\documentclass[12pt]{article}

  \usepackage{graphicx}
  \usepackage{epsfig}
  \usepackage{amssymb}
  \usepackage{subfigure}
  \usepackage{amsmath} 

\evensidemargin - .5truecm
\allowdisplaybreaks[1]

\def \lsim
{\raisebox{-3pt}{$\>\stackrel{<}{\scriptstyle\sim}\>$}}
\def \gsim
{\raisebox{-3pt}{$\>\stackrel{>}{\scriptstyle\sim}\>$}}

\begin{document}

\newcommand{\CC}{{\mathbb C}}
\newcommand{\RR}{{\mathbb R}}
\newcommand{\ZZ}{{\mathbb Z}}
\newcommand{\QQ}{{\mathbb Q}}
\newcommand{\NN}{{\mathbb N}}
\newcommand{\beq}{\begin{equation}}
\newcommand{\eeq}{\end{equation}}
\newcommand{\beal}{\begin{align}}
\newcommand{\eeal}{\end{align}}
\newcommand{\nn}{\nonumber}
\newcommand{\bea}{\begin{eqnarray}}
\newcommand{\eea}{\end{eqnarray}}
\newcommand{\ba}{\begin{array}}
\newcommand{\ea}{\end{array}}
\newcommand{\bfig}{\begin{figure}}
\newcommand{\efig}{\end{figure}}
\newcommand{\bc}{\begin{center}}
\newcommand{\ec}{\end{center}}

\newenvironment{appendletterA}
{
  \typeout{ Starting Appendix \thesection }
  \setcounter{section}{0}
  \setcounter{equation}{0}
  \renewcommand{\theequation}{A\arabic{equation}}
 }{
  \typeout{Appendix done}
 }
\newenvironment{appendletterB}
 {
  \typeout{ Starting Appendix \thesection }
  \setcounter{equation}{0}
  \renewcommand{\theequation}{B\arabic{equation}}
 }{
  \typeout{Appendix done}
 }

%%%%%%%%%%%%%%%%%%%%%%%%%%%%%%%%%%%%%%%%%%%%%%%%%%%%%%%%%%%%%%%%%%%%%%%%
%
%
%\begin{fmffile}{HIGAGA}
%
% 
%%%%%%%%%%%%%%%%%%%%%%%%%%%%%%%%%%%%%%%%%%%%%%%%%%%%%%%%%%%%%%%%%%%%%%%%

\begin{titlepage}
\nopagebreak

\renewcommand{\thefootnote}{\fnsymbol{footnote}}
\vskip 2cm
\begin{center}
\boldmath

{\Large\bf On the Unsolvability of Bosonic Quantum Fields}

\unboldmath
\vskip 1.cm
{\large  U.G.~Aglietti}
\vskip .2cm
{\it Dipartimento di Fisica, Universit\`a di Roma ``La Sapienza''}
\end{center}
\vskip 0.7cm

\begin{abstract}

Two general unsolvability arguments for interacting bosonic
quantum field theories are presented, based on Dyson-Schwinger 
equations on the lattice and cardinality considerations.
The first argument is related to the fact that, on a lattice
of size $N$, the system of lattice Dyson-Schwinger equations
closes on a basis of "primitive correlators"
which is finite, but grows exponentially with $N$.
By properly defining the continuum limit, one finds
for $N \to \infty$ a countably-infinite basis of the 
primitive correlators.
The second argument is that any conceivable exact analytic calculation
of the primitive correlators involves, in the continuum limit,
a linear system of coupled partial differential equations 
on an infinite number of unknown functions,
namely the primitive correlators, 
 evolving with respect to
an infinite number of independent variables. 

\vskip .4cm

\end{abstract}
\vfill
\end{titlepage}    

\setcounter{footnote}{0}

\newpage

\tableofcontents

\newpage

\section{Introduction}
\label{sect1}

"Hard" physics is often described by a Quantum Field
Theory (QFT), or by some generalization of it.
Originally created to describe the interaction of light
with atoms and, more
generally, to combine special relativity with ordinary
quantum mechanics, QFT was formulated
for the first time as Quantum Electrodynamics (QED),
a local relativistic quantum field theory.
Then it came the successful generalization to describe weak 
and  strong  interactions, finally giving rise to the well-known 
Standard Model (SM) of particle physics.

Euclidean versions of quantum field theories were
systematically applied to describe second-order 
phase transitions in statistical mechanics.
In the latter case, the problem was to calculate 
thermal fluctuations effects (temperature $T \ne 0$) 
in an interacting many-body system, rather than quantum 
fluctuations as in high-energy physics
(controlled by Planck's fundamental constant $\hbar \ne 0$).
It was also found, roughly speaking, that
Euclidean theories were simpler than the 
original Minkowski ones and could actually be used
for a rigorous mathematical formulation of
some Minkowski models, via functional analysis and
stochastic calculus \cite{BerrySimon}.
Furthermore, non-relativistic quantum field theories 
were introduced to describe the excitation spectra of many-body 
systems in condensed matter physics 
(phonons, quantum liquids, etc.).
In more recent years, field theories have been introduced
to describe turbulence in fluid mechanics --- an old
classical-physics problem involving (infinitely) many 
strongly-interacting degrees of freedom.

It seems that the fate of a system 
with many fluctuating degrees
of freedom, an-harmonically interacting with each other,
is that of being described, sooner or later, by some version of
a quantum field theory. 
Because of its generality, we may think that, in the near future, 
quantum field theory will invade engineers, biologists and geologic 
models. Remarkably enough, 
%for the profane, 
chaotic models are currently 
under attention to describe complex legislature systems and maybe QFT 
could be an alternative method. 

The understanding within Quantum Chromodynamics (QCD),
the QFT of strong interactions, of the observed striking properties of this 
fundamental force --- such as color confinement, mass gap generation,
spontaneous chiral symmetry breaking, string effects, etc. ---
turned out to be an extraordinarily difficult task.
Standard perturbative expansion in the interaction
coupling was not working and it seemed it was necessary to exactly
solve the theory to succeed --- or at least that it
was necessary to find a different kind of expansion.

Despite the ever-increasing range of applications, 
with different attempts by many great scientists
in the decades, no "realistic" quantum field theory
--- such as for example QED in four space-time dimensions ---
has ever been exactly solved.
Furthermore, interacting quantum field theories are defined {\it operationally},
i.e. are introduced by constructing some truncated formal expansion around 
the free theory.
That is in sharp contrast with mathematical tradition,
where models are defined through (after) some abstract existence theorem.
In QCD, for example, one calculates perturbative expansions in some small 
coupling, such as the strong coupling constant $\alpha_S \ll 1$ or 
$1/N_C$, with $N_C \gg 1$ the number of quark colors.
Non-perturbative QCD computations on an (euclidean) space-time lattice 
can be viewed as some sort of expansions in $1/N$,
where $N$ is the size of the lattice  
(the number of points) $\approx$ the number of degrees of freedom
of the system.

%
% In medias res:
%

According to a qualitative argument given by G. Preparata 
\cite{Preparata}, interacting quantum field theories will never
be exactly solved because of the threshold structure
of the correlation functions.
By pushing the perturbative expansion to progressively higher
orders, intermediate states with an arbitrarily large number 
of particles are created, producing an infinite
sequence of singularities in the exact correlators. 
All available higher-order pertubative calculations
fully confirm this argument.
Actually, reality seems to surpass imagination in QFT.
Let's just to give a few examples. 
The analytic structure of four-point functions
tremendously complicates in going from one loop to two loops 
\cite{RemiddiGehrmann}.
Anomalous thresholds already appear
in one-loop three-point functions and
additional singularities ---
not possessing any threshold interpretation --- do appear in 
the evolution equations of 
massive two-point and three-point functions at two loops
\cite{Remiddi, noi}. 

%%%%%%%%%%%%%%%%%%%%%%%%%%%%%%%%%%%%%%%%%%%%%%%%
%                      SI PARTE
%%%%%%%%%%%%%%%%%%%%%%%%%%%%%%%%%%%%%%%%%%%%%%%%

In this paper we present a quite different argument
with respect to the Preparata one,
in favor of unsolvability of bosonic theories, based on
lattice regularization \cite{Montvay}, 
Dyson-Schwinger (DS) equations \cite{DSeq,ItzZuber}
and general cardinality considerations.
As well known, DS equations never close in the formal
continuum, as an $n$-point correlator $G^{(n)}$
is always expressed, for any $n=2,3,4, \cdots$, 
in terms of higher-order correlators $G^{(n+1)}$, 
$G^{(n+2)} \cdots$ multiplied by some positive
power of the interaction coupling $\lambda^s$, $s\ge 1$.
The first-principle use of the DS equations
mainly involves the systematic generation of
Feynman diagrams ($\lambda \ll 1$), together with 
techniques to approximately resum the perturbative series to
all orders in $\lambda$.

The first point of our analysis is that,
unlike in the continuum, DS equations {\it do close}
on a lattice, of whatever size (i.e. number of points) $N<\infty$.
This is, in some sense, a good new.
It also explains why it was hopeless to try
to close the DS system in the formal
continuum, where one takes $N=\infty$
from the very beginning.
The bad new is that DS equations close {\it exponentially},
rather than power-like, with the lattice size $N$.
In the case of a $\lambda \, \phi^4$ theory,
for example, we find by explicit computation
that the number of correlators which need to be known, 
let's call them the {\it primitive} ones% 
%%%%%%%%%%%
\footnote{
The primitive correlators might equally well be called
{\it irreducible correlators} (or even {\it master correlators}).
},
%%%%%%%%%%%
in terms of which all correlators can be expressed, is 
\beq
\# \,\, \mathrm{of \,\, primitive \,\, correlators} \, = \, 
\mathcal{O}\left(3^N\right).
\eeq
That it is a huge growth with $N$.
If we consider for example a lattice in 
a four-dimensional space-time with $20$ points
along each direction 
--- well below current Monte-Carlo simulations ---
the number of primitive correlators to evaluate is
of order
\beq
3^{20^4} \, \simeq \, 2.5 \times 10^{763397}.
\eeq
More generally, on a lattice of size $N$,
\beq
\label{general_formula}
\# \,\, \mathrm{of \,\, primitive \,\, correlators} 
\, \approx \,
\left( m_{\mathrm{anh}} - 1 \right)^N ,
\eeq
where $m_{anh}$ is the maximal anharmonicity (or non linearity) 
of the theory, assumed to be finite, defined by
\beq
\qquad\qquad\qquad\qquad\qquad
\mathcal{L}_{int} \, = \, \sum_{i=2}^{m_{\mathrm{anh}}} c_i \, \phi^i ;
\qquad\qquad\qquad
2 \, \le \, m_{\mathrm{anh}} \, < \, \infty; \quad c_{m_{\mathrm{anh}}} \ne 0.
\eeq
For a cubic interaction, for example,
\beq
\mathcal{L}_{int} \, = \, c_3 \, \phi^3,
\eeq
we have
\beq
m_{\mathrm{anh}} \, = \, 3 : \qquad 
\eeq
and, at lattice size $N$,
\beq
\# \,\, \mathrm{of \,\, primitive \,\, correlators} 
\, = \, \mathcal{O}\left(2^N\right).
\eeq
Being exponential, that is also a huge growth with $N$, comparable to the previous
one:
\beq
2^{20^4} \, \simeq \, 6.3 \times 10^{48166}.
\eeq
For a gaussian theory, having
\beq
m_{\mathrm{anh}} \, = \, 2,
\eeq
eq.(\ref{general_formula}) gives a number of primitive correlators
of order one for any $N$, as $1^N \equiv 1$ uniformly in $N$:
\beq
\# \,\, \mathrm{of \,\, primitive \,\, correlators}  
\, = \, \mathcal{O}\left(1\right) ,
\eeq
as it should.
In general, eventual symmetries of the lattice theory
relate primitive correlators to each other, so
their total number is diminished.
In the case of a $\lambda \, \phi^4$ scalar theory,
however, lattice symmetries produce a mild, power-like
suppression of the above exponential growth with $N$,
implying that the behavior of the theory 
in the continuum limit $N \to \infty$ is 
not affected by the lattice symmetries.
We believe that this situation is the "normal"
or "generic" one, in the usual mathematical sense.
Of course, one can also imagine {\it exceptionally
symmetric theories}, possessing so many symmetries
--- once regularized on some lattice --- so as to kill the 
exponential growth above.
However, we have not been able to find a physically-sensible
model exhibiting such mechanism.

If we take the direct limit $N\to \infty$ in the above
formulas, we conclude that the number of primitive
correlators  in $\lambda\, \phi^4$ theory has the cardinality of the continuum, as
\beq
\mathrm{Card}\left(3^\NN\right) \, = \, \mathrm{Card}\left(2^\NN\right)
\, \equiv \, \aleph_1.
\eeq
It exists however the possibility of defining
the continuum limit in a weaker sense, which
is physically the right choice, in which
the number of primitive correlators is countable.

According to eq.(\ref{general_formula}), in the limit $N \to \infty$,
a strong discontinuity manifests itself in going from a 
free theory to any interacting theory. 
While in the free theory the number of primitive
correlators always remains finite and of order one, in a generic
interacting theory we obtain, with any
definition of the continuum limit, an infinite number of
primitive correlators;
Intermediate cardinalities, namely those of big finite sets,
do not appear.

Let us remark that we do not address existence problems
in quantum field theories, but only analytic solvability 
issues --- once existence has been proved or it is assumed. 
Indeed our arguments --- as we are going to show in detail ---
do not even depend on the dimension $d$ of the space-time
where the quantum fields live, which is instead a crucial parameter 
in existence proofs, as it controls the density of states at high energy.
In particular, we do not study the invariance of
the observable, low-energy physics under an unbounded increase 
of the ultraviolet cutoff on the energies,
\beq
\Lambda_{\mathrm{UV}} \, \approx \, \frac{N}{T},
\eeq 
where $T$ is the linear dimension of the lattice.
The above one is the well-known Renormalization-Group (RG) problem.
In this respect, our arguments are {\it meta arguments}.

The relevant phenomena we intend to show, can
already be understood by looking at a
quantum anharmonic oscillator, i.e. at a  
$\lambda \, \phi^4$ theory in space-time
dimension $d=1$ (which certainly exists!%
%%%%%%%%%%
\footnote{
The existence of the scalar $\lambda \, \phi^4$
theory in the continuum limit has been
proved for $d=2$ and $d=3$, where the coupling
constant $\lambda$ has a positive mass dimension
and the number of primitively divergent diagrams
is finite (super-renormalizable cases).
}).
%%%%%%%%%
In this case, the space-dimension $d_S=d-1$ vanishes,
\beq
d_S \, = \, 0 ,
\eeq
and the field $\phi(t)$ is actually a particle 
coordinate,
\beq
\phi(t) \, = \, x(t).
\eeq
The couplings of the fields $\left\{\phi\left(t_i\right)\right\}$ 
at different times $t_i$ (coming from the discretization of the
time-derivative term $d\phi/dt$ in the continuum action),
together with the anharmonic fluctuations, are already
responsible for all the effects we wish to describe%
%%%%%%%%%%
\footnote{
If often happens that a system exactly solvable in
classical mechanics is also solvable in the quantum
theory. Well-known cases are the harmonic oscillator
and the Kepler problem (the hydrogen atom).
Such correspondence is violated in this case:
while the free anharmonic oscillator is integrable
by quadrature in classical physics
(elliptic functions are obtained),
being an autonomous one-degree of freedom system, 
the quantum case is not. 
}
%%%%%%%%%%

The paper is organized as follows.
Since our arguments do not involve
explicit perturbative computations
and are, by necessity, rather implicit,
we devote the next four sections,
i.e. sect.$\,$\ref{sect2} to sect.$\,$\ref{sect5},
to an elementary discussion of the anharmonic
oscillator in the relevant continuum
and lattice spaces.
That also gives us the possibility
of discussing the relation of the
symmetries in the different spaces.
These sections can be skipped 
by a reader familiar with quantum field
theory on a lattice.
In sect.$\,$\ref{sect6} we consider the
continuum limit, i.e. the limit
of vanishing lattice spacing, of the
lattice theory.
In sect.$\,$\ref{sect7} we derive the Dyson-Schwinger
equations for the anharmonic oscillator
on the lattice --- hereafter Lattice Dyson-Schwinger
(LDS) equations --- and we solve them in
sect.$\,$\ref{sect8} by introducing the
primitive correlator basis.
In sects.$\,$\ref{sect9} and \ref{sect10} we present an algebraic and geometric formulation
of the resolution process of the LDS
equations respectively.
In sect.$\,$\ref{sect11} we discuss the
effects on the reduction to primitive
correlators of the Ward identities of the
lattice theory.
In sect.$\,$\ref{sect12} we discuss the
reduction to primitive correlators
in the continuum limit $N \to \infty$.
In sect.$\,$\ref{sect13} we present
a general method to evaluate
the primitive correlators by means
of systems of ordinary or partial differential
equations.
In sect.$\,$\ref{sect14} we derive the
form of the system of partial
differential equations on the
primitive correlators basis in the continuum
limit $N \to \infty$.
In sects.$\,$\ref{sect15} to \ref{sect17}
we discuss generalizations of the
results obtained for the anharmonic
oscillator, to space-time dimension
$d>1$, i.e. to true scalar
QFT's, and to theories
involving interacting bosons with non-zero 
spin.
Finally in sect.$\,$\ref{sect18}
we draw our conclusions and discuss
possible developments.

%%%%%%%%%%%%%%%%%%%%%%%%%%%%%%%%%%%%%%%%%%%%%%%%%

\section{Euclidean Anharmonic Oscillator on the Real Line}
\label{sect2}

Let's first consider a quantum anharmonic oscillator
in the continuum, with the euclidean time ranging on the
entire real line,
\beq
t_E \, \in \, \RR .
\eeq
As well known, the euclidean times are related to the 
(purely imaginary) Minkowski times by the relation
\beq
\label{analitic_cont}
t_M \, = \, e^{ - i ( \pi/2 - \epsilon ) } \, t_E,
\qquad
0 \, < \, \epsilon \, \ll \, 1.
\eeq
In general, the correlation functions to exactly compute, read
in configuration space
\beq
\label{basic_corr_cont}
\langle 
\phi\left(t_1\right) \, \phi\left(t_2\right) \cdots \phi\left(t_n\right) \rangle 
\, = \, \frac{1}{Z}
\int \mathcal{D}\phi \, 
\phi\left(t_1\right) \, \phi\left(t_2\right) \cdots \phi\left(t_n\right)
\exp\big(-S[\phi]\big); 
\qquad n \, = \, 1, 2, 3, \cdots,
\eeq
where $Z \equiv \langle 1 \rangle$,
the times $t_i \in \RR$ are not necessarily distinct%
%%%%%%%%%%%%%%%
\footnote{Correlators involving local composite operators of the form $\phi^n(t)$, $n \ge 2$,
can be obtained by taking some of the times $t_1, t_2,\cdots,t_n$ equal.
Operators containing time derivatives, such as $d\phi(t)/dt$,
$\phi(t) \, d\phi(t)/dt$, $\phi(t) \, d^2\phi(t)/dt^2$, etc.,
can be obtained by taking time derivatives on both sides of eq.(\ref{basic_corr_cont}) and then identifying some of the times.
}
%%%%%%%%%%%%%%%
and we have dropped, to have a lighter notation, 
the "$E$ "subscript $\left( t \equiv t_E \right)$.
To compute correlators, one needs:
\begin{enumerate}
\item
{\it An (euclidean) action}, which we take as
\beq
S[\phi] \, \equiv \, 
\int\limits_{-\infty}^{+\infty} 
\left[
        \frac{1}{2} \left( \frac{d\phi}{dt} \right)^2 
\, + \, \frac{1}{2} m_0^2 \, \phi(t)^2 
\, + \, \frac{\lambda_0}{4} \, \phi(t)^4 
\right] dt ,
\eeq
where $m_0>0$ and $\lambda_0>0$ are the bare mass
(frequency) and the bare coupling. 
Note that field histories $\phi(t)$, $t \in \RR$,
which do not vanish for $t\to\pm\infty$
have infinite action;
%%%%%
\item
{\it A path measure}, which we take as a formal infinite product
of Lebesgue measures on the real line%
%%%%%%%%%%%%%%%%%%%%%%%%%%%%%%%%%%%%%%
\footnote{In the Euclidean case, the functional measure 
\beq
d\mu_0 \, \equiv \, \mathcal{D} \, \Phi \, e^{-S_0[\Phi]},
\eeq
with $S_0 \, \equiv \, S(\lambda_0=0)$ the free 
(harmonic oscillator) action, is the standard
Ornstein-Uhlenbeck measure \cite{BerrySimon}.
}
%%%%%%
\beq
\mathcal{D} \Phi \, \equiv \, \prod_{t \in \RR} d\phi(t) .
\eeq
\end{enumerate}

%%%%%%%%%%%%%%%%%%%%%%%%%%%%%%%%%%%%%%%%%%%%

\subsection{Symmetries}

In this section we consider the symmetries
of the euclidean oscillator on the real line.
We study this problem in detail because
we will encounter "complications" of these
symmetries when we will define the oscillator
on a circle, on an infinite lattice immersed
in the real line and on a finite lattice 
immersed in a circle.
By looking at the action and at the integration measure, one finds
that the theory has the following symmetries:
\begin{enumerate}
\item
{\it Change of sign of the field $\phi$,}
\beq
\phi(t) \, \to \, - \, \phi(t), \qquad t \in \RR;
\eeq
%%%%%
\item
{\it Symmetries related to the time $t$:}
\begin{enumerate}
\item
{\it Continuous time translations, $t \to t + \alpha$, $\alpha \in \RR$}.
The symmetry is formally described by the group $H$ defined as the set
\beq
H \, \equiv \, \left\{ h_\alpha; \,\, \alpha \, \in \, \RR \right\},
\eeq
with the group operation, in additive notation, given by
\beq
h_\alpha \, + \, h_\beta \, \equiv \, h_{\alpha + \beta}.
\eeq
$H$ is a one-dimensional, abelian group, isomorphic
to the real line equipped with the ordinary sum,
\beq
H \, \sim \, \left( \RR , + \right).
\eeq 
Its action on the times $t \in \RR$ reads
\beq
h_\alpha(t) \, \equiv \, t \, + \, \alpha.
\eeq
%%%%%
\item
{\it Reflections of the time $t$ about any point
$a \in \RR$ of the time axis}.

This case is more complicated than the previous one,
so it is convenient to consider first the action
of the related symmetry group $G$ on the times $t$.
The reflection $s_a$ about the point $a$ of a time
$t \in \RR$ is defined as
\beq
s_a(t) \, \equiv \, 2a \, - \, t, \qquad t \, \in \, \RR. 
\eeq
The point $t=a$ goes into itself, 
\beq
s_a(a) \, = \, a,
\eeq
as it should.
For $a=0$, we obtain the usual time-reversal operation,
\beq
s_0(t) \, \equiv \, - \, t.
\eeq
As expected, the square of any reflection is the identity:
\beq
s_a^2(t) \, \equiv \, s_a\left(s_a(t)\right) \, = \,
s_a(2a-t) \, = \, 2a-(2a-t) \, = \, t \, = \, \mathrm{id}(t),
\quad \forall t \in \RR, \,\,\, \forall a \in \RR.
\eeq
Let's now compose two different reflections:
\beq
s_b \circ s_a(t) \, \equiv \, s_b\left(s_a(t)\right)
\, = \, s_b\left(2a-t\right) \, = \, 2b-(2a-t) \, = \, 2(b-a) \, + \, t
\, = \, h_{2(b-a)}(t).
\eeq
The composition of two reflections is then a translation.
The group $G$ therefore is not commutative, as
\beq
\left[s_a,s_b\right](t) \, = \, h_{2(a-b)}(t) - h_{2(b-a)}(t) 
\, = \, 4(a-b) 	\, \ne \, 0, \qquad (a \, \ne \, b).
\eeq
\end{enumerate}
Formally, the complete group of the symmetries related to
the time can be described as a 
semi-direct product of the group $H$ of the translations
and the order-two group generated by the time-reversal
$t \to -t$.
\end{enumerate}

%%%%%%%%%%%%%%%%%%%%%%%%%%%%%%%%%%%%%%%%%%%%%%%%

\subsection{Breaking of Symmetries}

The symmetries discussed in the previous section
can be explicitly broken by modifying the action
as below.
\begin{enumerate}
\item
{\it Change of sign of the field, $\phi\to-\phi$.}
The symmetry can be broken by adding to the action 
the following odd functional in $\phi$
\beq
\Delta S[\phi] \, = \, \, \equiv \, 
\int\limits_{-\infty}^{+\infty} 
\left[ 
a \, \phi(t) 
\, + \, 
\frac{1}{3} g \, \phi^3(t)
\right] dt,
\eeq
with $a$ and $g$ are constants.
The total action is then
\beq
S[\phi] \, + \, \Delta S[\phi];
\eeq
%%%%%
\item
{\it Translation and reflection of time.}
The symmetry can be broken by generalizing
the action as
\beq
\label{almost_complete_S}
S[\phi] \, \equiv \, 
\int\limits_{-\infty}^{+\infty} 
\left[
        \frac{1}{2} \eta(t) \, \left( \frac{d\phi}{dt} \right)^2 
\, + \, \frac{1}{2} m^2(t) \, \phi^2(t) 
\, + \, \frac{\lambda(t)}{4} \, \phi^4(t) 
\right] dt, 
\eeq
where $\eta(t), m(t), \lambda(t) > 0$ are given functions of time.
\end{enumerate}
Both symmetries above can be broken by adding
to the action in eq.(\ref{almost_complete_S}) the functional
\beq
\Delta S[\phi] \, = \, \, \equiv \, 
\int\limits_{-\infty}^{+\infty} 
\left[
a(t) \, \phi(t) 
\, + \, 
\frac{1}{3}g(t) \, \phi^3(t)
\right] dt,
\eeq
with $a(t)$ and $g(t)$ are given functions of $t \in \RR$.

%%%%%%%%%%%%%%%%%%%%%%%%%%%%%%%%%%%%%%%

\subsection{Free Propagator}

\begin{figure}[ht]
\begin{center}
\includegraphics[width=0.5\textwidth]{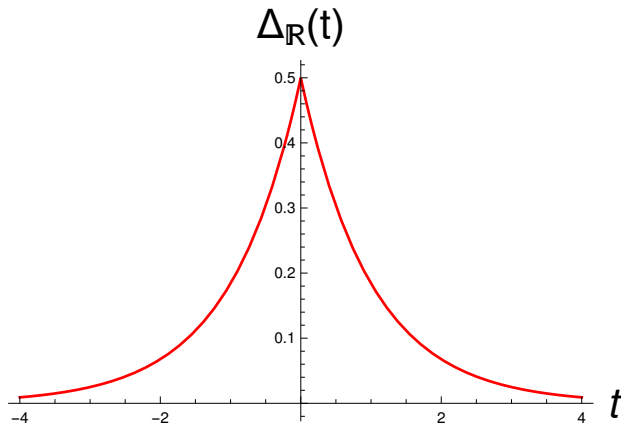}
\footnotesize
\caption{
\label{fig_plotDeltaR}
\it Propagator $\Delta_\RR(t)$ of an euclidean harmonic oscillator on the real line $\RR$ as a function of time $t\equiv t_E$ for $m=1$. 
An exact exponential decay with lifetime $\tau=1/m=1$ is observed for
any time.
}
\end{center}
\end{figure}

The euclidean propagator of the harmonic oscillator ($\lambda_0=0$)
reads in momentum (energy) space
\beq
S_\RR\left(E\right) \, = \, \frac{1}{E^2 \, + \, m^2}.
\eeq
In configuration (time) space, the propagator is given by 
(see fig.\ref{fig_plotDeltaR}):
\beq
\Delta_\RR(t) \, = \, 
\int\limits_{-\infty}^{+\infty} S_\RR(E) \, e^{-iEt} \, \frac{dE}{2\pi}
\, = \,
\frac{e^{ - m |t|}}{2m} \, .
\eeq
%
%%%%%%%%%%%%%%%%%%% FIGURA %%%%%%%%%%%%%%%%%%%%
%%%%%%%%%% FINE FIGURA %%%%%%%%%%%%%%%%%%%%%%%%
%
It exponentially decays with $|t|$ with
the characteristic time (or life-time)
\beq
\tau \, \equiv \, \frac{1}{m}.
\eeq
Let us remark that, since the time $t$ ranges
in the entire real line, by going to very large
times, 
\beq
|t| \, \gg \, \tau, 
\eeq
one can arbitrarily decorrelate the system,
as the exponential can become infinitesimal,
\beq
e^{-|t|/\tau} \, \to \, 0^+
\qquad \mathrm{for}
\,\,\, t \, \to \, \pm \, \infty .
\eeq
We will see that this possibility 
does not hold anymore when we define 
the theory on a circle.

%%%%%%%%%%%%%%%%%%%%%%%%%%%%%%%%%

\subsection{Renormalization}

It is not possible to exactly resum the correlators
to all orders in $\lambda_0$, because one does
not know exactly, for general $n$, the coefficient $c_n$
of $\lambda_0^n$ \cite{BenderWu}.
One has then to resort to perturbation theory in 
$\lambda_0$.
The most singular diagram is the tadpole one
(a one-loop diagram involving a single propagator), 
which is ultraviolet finite:
\beq
\approx \, \lambda_0 \int\limits_{-\infty}^{+\infty} 
\frac{dE}{2\pi} \, \frac{1}{E^2 \, + \, m_0^2}
\, = \, \frac{\lambda_0}{2 m_0}
\, < \, \infty.
\eeq
That implies that the renormalization of any parameter 
of the field normalization constant is finite:
\bea
m &=& m_0 \, + \, \delta m\left(\lambda_0\right) ;
\qquad 
\left| \delta m\left(\lambda_0\right) \right| \, < \, \infty ;
\nonumber\\
\lambda &=& \, \lambda_0 \,\, + \,\, \delta \lambda\left(\lambda_0\right) ;
\qquad \,\,
\left| \delta \lambda\left(\lambda_0\right) \right| \, < \, \infty;
\nonumber\\
Z &=& \, Z_0 \, + \,\, \delta Z\left(\lambda_0\right) ;
\qquad \,
\left| \delta Z \left(\lambda_0\right) \right| \, < \, \infty.
\eea
With the canonical normalization of the field $\phi(t)$
which we have chosen, $Z_0=1$.

Note that a power infrared divergence $\propto 1/m_0$ occurs in the
above tadpole for $m_0 \to 0$, related to the fact
that, in the massless limit, the motion of the particle
becomes a free one.

%%%%%%%%%%%%%%%%%%%%%%%%%%%%%%%%%%%%%%%%%%%%%%%%%%%%%

\subsection{Analytic continuation to Minkowski space}

By means of analytic continuation according to eq.(\ref{analitic_cont}),
the euclidean correlators become, in Minkowski space,
the $T$-ordered products of the field operators
averaged over the vacuum state:
\beq
G_M\left(t_1,t_2,\cdots,t_n\right) 
\, \equiv \, 
\langle 0 | T \phi\left(t_1\right) \phi\left(t_2\right) \cdots \phi\left(t_n\right)  | 0 \rangle .
\eeq
Because of eq.(\ref{analitic_cont}),
the Minkowski correlators are boundary values of the
Euclidean correlators for complex times.

We can obtain the propagator in Minkowski momentum (energy) space
by means of the following complex rotation%
%%%%%%%
\footnote{Note the change of sign in the rotation angle 
with respect to the relation between the corresponding times.}:
%%%%%%
\beq
k_0 \, = \, e^{ i ( \pi/2 - \epsilon) } \, E, 
\qquad
0 \, < \, \epsilon \, \ll \, 1,
\eeq
with $k_0$ the Minkowski (physical) energy.
It holds:
\beq
S_M\left(k_0\right) \, = \, \frac{1}{k_0^2 - m_0^2 + i \epsilon} .
\eeq
We can interpret the above formula as a relativistic propagator
in which the energy is replaced by the rest particle mass $m_0 > 0$,
\beq
\sqrt{\vec{k}^2 \, + \, m_0^2} \, \to \, m_0 .
\eeq
The latter is a static approximation for both the particle
and the antiparticle states.

%%%%%%%%%%%%%%%%%%%%%%%%%%%%%%%%%%%%%%%%%%%%%%%%%%%%%%%

\section{Euclidean Anharmonic Oscillator on a Circle $S^1$}
\label{sect3}

Since we want to define the theory on a finite lattice
with periodic boundary conditions,
let us first consider the simpler case of
a quantum anharmonic oscillator on  a circle, 
i.e. let us compactify the real line $\RR$ 
to a circle $S^1$,
\beq
\RR \, \to \, S^1.
\eeq
As regards to topology, we are making a
one-point compactification, by adding
one point at infinity.
As physics is concerned, 
by going from the real line to the circle, 
we are basically introducing an explicit time 
scale in the theory, namely the length $T$ of 
the circle $S^1_T$.
The parameter $T$ plays the role of the 
"largest possible time", in the sense 
that any time $t$ living in $S^1$ is subjected to
the limitation
\beq
|t| \, \lsim \, T \, < \, \infty.
\eeq
That implies, as we are going to explicitly show,
that the exponential decay of the propagator
\beq
\approx e^{-|t|/\tau}
\eeq 
cannot be observed for any time $t$, namely
up to infinitesimal values, unlike the case 
on the real line.
The inverse of the circle length,
\beq
\lambda \, \equiv \, \frac{1}{T} \, > \, 0
\eeq
is to be considered as an infrared cutoff
to the energies $E$:
\beq
E \, \gsim \, \lambda.
\eeq
In this case,
unlike the theory on $\RR$ where both time
and energy are continuous variables,
time is still continuous, while energy
is discrete because of the finite size of 
the circle. 
The circle $S^1$ is defined, as usual, as a closed segment with
its end-points identified
\beq
S^1_T \, \equiv \, 
\left\{ t_E \, \in \, \left[-\frac{T}{2},\frac{T}{2}\right] ; 
\,\, - \, \frac{T}{2} \, \sim \, + \, \frac{T}{2} \right\},
\eeq
with fixed
\beq
0 \, < \, T \, < \, \infty.
\eeq
$S^1_T$ is therefore our euclidean time domain. 
Note that the condition above (endpoint identification)
looses its meaning in the limit of an infinite $T$,
\beq
T \, \to \, + \infty.
\eeq 
This observation will become relevant when we
consider infinite lattices.
As well known in mathematics%
\footnote{
That is the dual characterization or functional
characterization of the circle.
}, we can forget end-point identification, i.e. take
\beq
t_E \, \in \, \left[ - \, \frac{T}{2}, \, + \, \frac{T}{2} \right], 
\eeq
but restrict to functions $\phi$ coinciding on the end-points
\beq
\phi\left( - \frac{T}{2} \right) \, = \, \phi\left( + \frac{T}{2} \right). 
\eeq
The correlation functions to exactly compute read:
\beq
\langle 
\phi\left(t_1\right) \, \phi\left(t_2\right) \cdots \phi\left(t_n\right) 
\rangle 
\, = \, \frac{1}{Z}
\int_{\RR^\infty} \mathcal{D}\phi \, 
\phi\left(t_1\right) \, \phi\left(t_2\right) \cdots \phi\left(t_n\right)
\exp\big(-S[\phi]\big),
\eeq
where we have dropped the "$E$" subscript $\left( t \equiv t_E \right)$.
The euclidean action has the expression
\beq
S[\phi] \, \equiv \, 
\int\limits_{-T/2}^{+T/2} 
\left[
        \frac{1}{2} \left( \frac{d\phi}{dt} \right)^2 
\, + \, \frac{1}{2} m_0^2 \, \phi(t)^2 
\, + \, \frac{\lambda_0}{4} \, \phi(t)^4 
\right] dt 
\eeq
and the measure reads
\beq
\mathcal{D}\phi 
\, \equiv \,
\prod_{t \in S^1_T } d\phi(t).
\eeq

%%%%%%%%%%%%%%%%%%%%%%%

\subsection{Symmetries}

The theory has a global $O(2)$ symmetry, related to
continuous angle rotations (the $SO(2)$ subgroup) and
discrete reflections of $S^1$ about any diameter%
\footnote{It is a "compact remnant" of the symmetry of
the theory in $\RR$ (see previous section).}.
If we represent the circle $S^1_T$ in the plane $\RR^2$ as
\bea
x_1(t) &=& \frac{T}{2\pi} \, \cos\left(\frac{2\pi t}{T}\right);
\nonumber\\
x_2(t) &=& \frac{T}{2\pi} \, \sin\left(\frac{2\pi t}{T}\right); 
\qquad t \, \in \, [0,T);
\eea
then the action of the $SO(2)$ group reads
\beq
t \, \to \, t \, + \, \alpha,
\eeq
while the reflection for example about the $x_1$ axis 
is given by
\beq
x_1 \, \to \, x_1; \qquad x_2 \, \to \, - \, x_2.
\eeq

%%%%%%%%%%%%%%%%%%%%%%%

\subsection{Free Propagator}

\begin{figure}[ht]
\begin{center}
\includegraphics[width=0.5\textwidth]{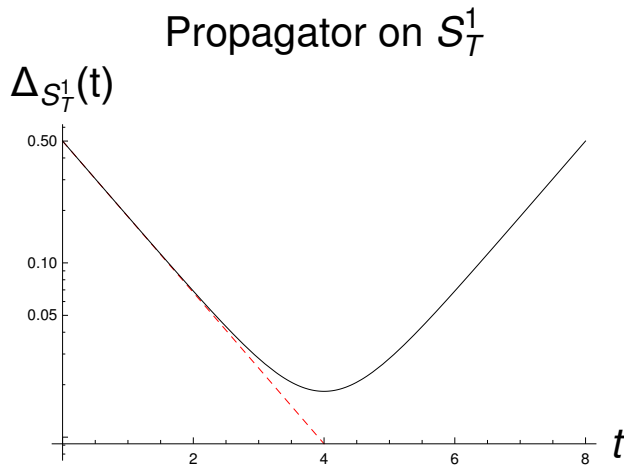}
\footnotesize
\caption{
\label{fig_ppropS1}
\it The black continuous line is the 
logarithmic plot (in the vertical scale) 
of the free propagator $\Delta_{S^1_T}(t)$
on a circle $S^1_T$ of length $T=8$ with mass $m=1$ 
as a function of time $t$ for an entire period, $t:0 \to 8$.
The red dashed line represents the propagator on the real line
$\Delta_\RR(t)$ with the same parameters, namely $m=1$ 
$(T=\infty$ because the real line has infinite length$)$, 
plotted for half of a period, $t:0 \to 4$.
The two propagators are very close to each other 
from $t=0$ up to $t \lsim T/2$.
While the propagator on the real line decays exponentially
for any time, the propagator on $S^1_T$ decays  
up to $T/2$ and then begins to rise up because of the 
circle periodic boundary condition.
In general, the comparison of these plots gives an idea
of the size of the finite-volume effects (finite $T$) a
various times $t$.
}
\end{center}
\end{figure}

The free propagator in momentum (energy) space is
obtained from the propagator on the real line by
replacing the continuous energies $E$ with
the allowed discrete energies $E_n$:
\beq
S^{1}_{\,n} \, \equiv \, S_\RR\left(E_n\right)
\, = \, 
\frac{ 1 }{ E_n^2 \, + \, m^2 }
\, = \, 
\frac{ 1 }{ (n \, 2\pi / T)^2 \, + \, m^2 } ,
\eeq
where
\beq
E_n \, = \, \frac{2\pi}{T} \, n, \qquad n \, \in \, \ZZ .
\eeq
Note that the discrete energies are evenly spaced:
\beq
\Delta E_n \, \equiv \, E_{n+1} \, - \, E_n \, = \, 
\frac{2\pi}{T} \, \equiv \, \Delta E .
\eeq
The propagator in configuration (time) space
is given by the following Fourier series
(see fig.\ref{fig_ppropS1}):
\bea
\label{S_S1}
\Delta_{S^1_T}(t) &=& \sum_{n=-\infty}^{+\infty} 
S^1_n \, \exp\left( - i E_n t \right) \,\, \frac{\Delta E_n}{2\pi} 
\, = \, 
\frac{1}{T} \sum_{n=-\infty}^{+\infty}
\frac{ \exp( - \, 2 \pi i \, n \, t / T ) }{ (n \, 2\pi / T)^2 \, + \, m^2 } \, =
\nonumber\\
&=& T \sum_{n=-\infty}^{+\infty}
\frac{ \exp( - \, 2 \pi i \, n \, t / T ) }{ (2\pi \, n )^2 \, + \, (m \, T)^2 } .
\eea
Note that it is a periodic function of the time $t$ with period $T$, 
\beq
\Delta_{S^1_T}(t+T) \, = \, \Delta_{S^1_T}(t),
\eeq
as it should.
%
%%%%%%%%%%%%%%%%%%% FIGURA %%%%%%%%%%%%%%%%%%%%
%%%%%%%%%% FINE FIGURA %%%%%%%%%%%%%%%%%%%%%%%%
%
Let us make a few observations:
\begin{enumerate}
\item Since the coefficients $c_n$ of the Fourier series above
behave asymptotically as
\beq
c_n \, \approx \, \frac{1}{n^2} 
\qquad 
\mathrm{for} \,\,\, n \, \to \, \pm \, \infty,
\eeq
the function $\Delta_{S^1_T}(t)$ has a discontinuous first
derivative at $t=0$, as can also be seen directly
by differentiating with respect to time 
the first and the last member in eq.(\ref{S_S1});
%%%%%
\item
In the formal limit $T \to \infty$, the energy spacing
$\Delta E = 2\pi/T \to 0$ and 
one recovers the euclidean propagator on the real line:
\beq
\sum_{n=-\infty}^{+\infty} f\left(E_n \, ;t\right) \, \Delta E_n
\to \int\limits_{-\infty}^{+\infty} f(E;t) \, dE
\qquad \mathrm{for} \,\, T \, \to \, + \infty,	
\eeq
where
\beq
f(E;\,t) \, \equiv \, S\left(E\right) \, e^{ - i E t } .  
\eeq
In practice, the propagator on $S^1$ is close
to the one on $\RR$ if the sum over the discrete energies
$E_n$, $n\in\ZZ$,
is a good approximation of the integral over the 
continuous energies $E\in \RR$.
For that to be true, the denominator on the last
member of eq.(\ref{S_S1}) must not vary much
when 
\beq
\label{nplusone}
n \, \to \, n + 1.
\eeq
This occurs if
\beq
m \, T \, \gg \, 1.
\eeq
Furthermore, also the oscillating exponential
at the numerator on the last member of eq.(\ref{S_S1})
must vary little under the
variation (\ref{nplusone}).
That implies its argument must be much 
less than one, i.e. that it must hold
\beq
\frac{|t|}{T} \, \ll \, 1;
\eeq
%%%%%
\item
The propagator on $S^1$ roughly decays
exponentially with the lifetime $\tau =1/m$
for half of the circle length
(see fig.\ref{fig_ppropS1}):
\beq
S(t) \, \approx \, e^{-|t| /\tau}
\qquad \mathrm{for} \,\, |t| \, \lsim \, \frac{T}{2}.
\eeq
The maximal decorrelation, i.e. the smallest
possible value of the exponential, is therefore
\beq
\min_{t \in S^1_T} S(t) \, \approx \, e^{ - \, m \, T / 2}.
\eeq
By requiring to be close to the theory on $\RR$,
we therefore obtain again the relation
\beq
m \, T \, \gg \, 1.
\eeq
If the condition above is not satisfied,
finite volume effects are substantial
and the theory on the circle has no resemblance to the
one in the continuum.
\end{enumerate}

%%%%%%%%%%%%%%%%%%%%%%%%%%%%%%%%%%%%%%%%%%%%%%%%%%%%

\section{Anharmonic Oscillator on an Infinite Lattice
$L \subset \RR$}
\label{sect4}
Compared to the theory on the real line, 
the theory on a lattice $L \subset \RR$ 
of infinite spatial extent 
contains an additional scale, namely the lattice spacing $a>0$.
Unlike the length "$T$" of the circle $S^1_T$
of the previous section, the lattice spacing "$a$"
has to be considered as the "shortest possible time"
in the theory,
\beq
|t| \, \gsim \, t_{\min} \, \equiv \, a \, > \, 0.
\eeq
Its inverse
\beq
\Lambda_{\mathrm{UV}} \, \equiv \, \frac{\pi}{a} \, < \, \infty
\eeq 
plays the role of an ultraviolet cutoff 
on the energies $E$:
\beq
E \, \lsim \, \Lambda_{\mathrm{UV}}. 
\eeq
Since the propagator decays by the factor
\beq
\approx \, e^{ - m \, a}
\eeq
when we move away from the origin by one lattice spacing,
i.e. by the smallest possible distance,
in order to be close to the exponential
decay on the real line, $e^{-m|t|}$, $t \in \RR$, 
the above factor has to be close to
one, so it must be
\beq
m \, a \, \ll \, 1 .
\eeq
The doubly-infinite lattice $L$ immersed in $\RR$ 
is written
\beq
L \, \equiv \, \left\{ n \, a; \,\,\, n \, \in \, \ZZ \right\}
\, \subseteq \, \RR.
\eeq
The theory on the lattice $L$ has "specular" properties 
with respect to the theory on the circle $S^1$:
time is discrete, while energy is continuous, because
the lattice $L$ has infinite spatial extent, so there
is no infrared cutoff. 

%%%%%%%%%%%%%%%%%%%%%%%%%%%%%%%%%%%%%%%%%%%%%%

\subsection{Symmetries}

The theory on the lattice $L$ has a discrete
symmetry group $G$ generated by the translation
of the time $t_n \equiv n a$ by one lattice spacing,
\beq
t_n \, \to \, t_n \, + \, a \, = \, (n+1)a \, \equiv \, t_{n+1},
\eeq
and the reflection about the point $t=0$,
\beq
t_n \, \to \, - \, t_n.
\eeq
$G$ is the crystallographic group of an infinite one-dimensional
lattice with a single lattice spacing $a$.

%%%%%%%%%%%%%%%%%%% FIGURA %%%%%%%%%%%%%%%%%%%%
\begin{figure}[ht]
\begin{center}
\includegraphics[width=0.5\textwidth]{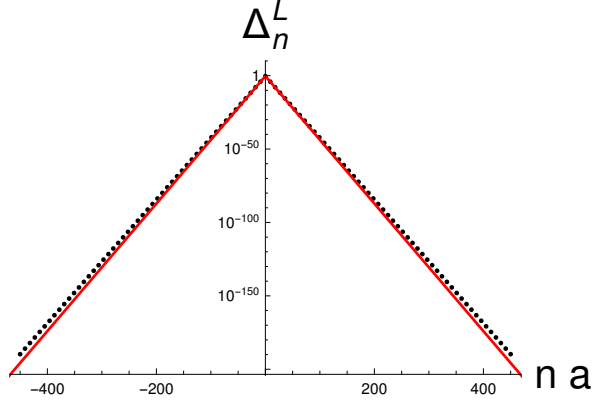}
\footnotesize
\caption{
\label{fig_platticeinf}
\it The black points represent a logarithmic plot 
of the propagator $\Delta^L_{\,n}$ on the infinite lattice 
$L$ in configuration space for mass $m=1$ and lattice
spacing $a=0.9$.
The lattice times $t_n=n a$ are plotted in
multiples of ten, i.e. $n=\cdots,-10,0,+10,+20,\cdots$.
The red continuous line represents the propagator
on the real line, $\Delta_\RR(t)$, for the same
parameters, namely $m=1$ ($a=0$ because we are in the
continuum).
The difference is barely visible because the lattice
corrections are very small (see text).
The pole residue on the lattice 
$Z_{\mathrm{eff}} = \Delta^L_{\,n=0} \simeq 0.456$,
while on the real line $Z=\Delta^\RR(t=0)=0.5$, 
i.e. there is a $-9\%$ difference
with respect to the continuum.
The effective mass is $m_{\mathrm{eff}} \simeq 0.969$,
so there is a $-3\%$ difference with respect to the
physical mass $m=1$.
}
\end{center}
\end{figure}
%%%%%%%%%% FINE FIGURA %%%%%%%%%%%%%%%%%%%%%%%%

%%%%%%%%%%%%%%%%%%%%%%%%%%%%%%%%%%%%%%%%%%%%%%%%%%%%

\subsection{Free Propagator}

The free propagator on the doubly-infinite lattice $L$ reads:
\beq
S_L(E) \, = \, \frac{a^2}{ 2 \left[1 - \cos(E a) \right] \, + \, (m a)^2 },
\eeq
Since the r.h.s. of the above equation is a periodic function of
the energy $E$ with period $2\pi/a$, the latter is typically restricted 
to the first Brillouin zone,
\beq
\label{first_Brillouin}
- \, \frac{\pi}{a} \, < \, E \, \le \, + \, \frac{\pi}{a}.
\eeq
In the continuous limit, i.e. in the limit of zero lattice spacing, 
$a \to 0^+$, one recovers the euclidean propagator on the real 
line $\RR$,
\beq
S_L(E) \,\, \to \,\, 
S(E) \, = \, \frac{1}{ E^2 \, + \, m^2 }
\qquad \mathrm{for} \,\, a \, \to \, 0^+ ,
\eeq
with the energy $E$ now ranging in the entire real line,
\beq
- \, \infty \, < \, E \, < \, + \, \infty.
\eeq
In configuration (time) space, the propagator reads
\beq
\label{propagL}
\Delta^L_{\,n} 
\, \equiv \, 
\int\limits_{-\pi/a}^{+\pi/a} 
S_L(E) \, e^{-i E n a} \, \frac{dE}{2\pi}
\, = \, Z_{\mathrm{eff}} \, \exp\left( - \, m_{\mathrm{eff}} \, |n| \, a \right),
\qquad n \, \in \, \ZZ,
\eeq
where we have defined the effective mass and the effective pole residue:
\bea
m_{\mathrm{eff}} &=& 
- \, \frac{1}{a} \ln\left(1+\eta-\sqrt{2\eta+\eta^2}\right) ;
\nonumber\\
Z_{\mathrm{eff}} &=& + \, \frac{a}{ 2 \sqrt{2 \eta + \eta^2} };
\eea
with
\beq
\eta \, \equiv \, \frac{1}{2} \, (m a)^2.
\eeq
The following remarks are in order.
\begin{enumerate}
\item
On the lattice $L$, the only "allowed" times are integer
multiples of the lattice spacing,
\beq
t \, = \, t_n \, \equiv \, n \, a,
\qquad n \,\in \, \ZZ,
\eeq
so that the argument of the exponent on the last member
of eq.(\ref{propagL}) contains the modulus
of the discrete times,
\beq
|n| \, a \, = \, \left|t_n\right|;
\eeq
%%%%%
\item
The effective mass $m_{\mathrm{eff}}$ 
and the effective pole-residue $Z_{\mathrm{eff}}$ 
of the propagator on $L$, unlike the corresponding quantities
$m$ and $Z$ in the continuum, do depend on the lattice spacing $a$
and have $\mathcal{O}\left(\left(m a\right)^2\right)$ corrections.
\bea
m_{\mathrm{eff}} &=& m_{\mathrm{eff}}(a) 
\, = \, m 
\left\{ 
1 \, - \, \frac{(m a)^2}{24} 
\, + \, \mathcal{O}\left[\left(m a \right)^4 \right] 
\right\};
\nonumber\\
Z_{\mathrm{eff}} &=& Z_{\mathrm{eff}}(a) \, = \, \frac{1}{2m}
\left\{
1 \, - \, \frac{(m a)^2}{8}
\, + \, \mathcal{O}\left[\left(m a \right)^4 \right] 
\right\}.
\eea
As already discussed, the theory has indeed an ultraviolet cutoff provided by 
$1/a$  (or $\pi/a$), the inverse of the lattice spacing, but no infrared cutoff,
because the lattice $L$ has an infinite spatial extent.
Note that the above corrections are very small.
\end{enumerate}

%%%%%%%%%%%%%%%%%%%%%%%%%%%%%%%%%%%%%%%%%%%%%%%%%%%%%%%%%%%%%%

\section{Anharmonic Oscillator on a Circular Lattice $\Lambda \subset S^1$}
\label{sect5}

To define an anharmonic oscillator on a finite lattice $\Lambda$
of size $N$ immersed in the circle $S^1_T$,
\beq
\Lambda \, \subset \, S^1_T,
\eeq
we discretize the euclidean time $t \to t_i \equiv i \, a$, 
with $i = 1, 2, \cdots, N$ and $a \equiv T/N$ the lattice spacing,
and define the scalar field $\phi(t)$ at each lattice point as
\beq
\phi_i \, \equiv \, \phi\left(t_i\right),
\qquad
i \, = \, 1, 2, \, \cdots, \, N.
\eeq
We will call the lattice $\Lambda$ a "circular lattice"
when boundary conditions become relevant.
The time derivative of the field $\phi(t)$ is discretized as
a nearest neighborhood interaction, 
\beq
\left. \frac{d\phi}{dt}\right|_{t=t_i} \,\, \to \,\,\, 
\frac{\phi_{i+1} \, - \, \phi_i}{a} .
\eeq
By omitting a trivial normalization, the correlators to compute read
\beq
\label{G_integral}
G\left(\nu \right) 
\, \equiv \,  
\int\limits_{\RR^N}  D \Phi \, \Phi^\nu \, 
\exp\left[ - \, S(\Phi) \right],
\eeq
where we have introduced the following compact notation:
\begin{enumerate}
\item
The string of fields at the lattice points,
\beq
\Phi \, \equiv \, 
\left(
\phi_1, \, \phi_2 ,
\, \cdots,
\, \phi_N 
\right);
\eeq
%%%%%%%%%%%
\item
The multi-index
\beq
\mathcal{\nu} 
\, \equiv \,  
\left( 
\nu_1, \, \nu_2, \cdots, \, \nu_N 
\right) ,
\eeq
with components, since
we are dealing with a bosonic theory, 
in the range
\beq
\qquad 0 \, \le \, \nu_i \, < \, \infty;
\eeq
%%%%%
\item
The product of the string of fields
with the chosen exponents
\beq 
\Phi^\nu \, \equiv \, 
\prod_{i=1}^N \phi_i^{\nu_i} \, = \, 
\phi_1^{\, \nu_1} \, \phi_2^{\, \nu_2}
\, \cdots \, 
\phi_N^{\nu_N} ;
\eeq
%%%%%%
\item
The integration measure given by an ordinary product
of Lebesgue measures on $\RR$,
\beq
D \Phi \, \equiv \, \prod_{i=1}^N d\phi_i . 
\eeq
We are indeed dealing with an ordinary multiple integral;
%%%%%
\item
The generalized lattice action
\beq
S[\Phi] 
\, \equiv \,
\sum_{i=1}^{N} 
\left(
\frac{1}{2} k_i \, \phi_i^2
\, - \, w_{i,i+1} \, \phi_i \, \phi_{i+1}
\, + \, \frac{\lambda_i}{4} \, \phi_i^4
\right),
\eeq
where the last coupling $w_{N,N+1}$ involves, because of
periodicity, the field
\beq
\phi_{N+1} \, \equiv \, \phi_1 .
\eeq
We have introduced a different coupling in each term of
$S[\Phi]$ to have more freedom in writing evolution
equations (see later).
In reality, as well known, a given discretization of the time derivative
in the continuum action, together with the choice of $m_0$, 
unambiguously fixes the couplings $k_i$ and $w_{i,i+1}$.
\end{enumerate}
%%%%%%%%%%%%%%%%%%%%%%
Let us end this section with a few remarks.
\begin{enumerate}
\item
Because of ultraviolet finiteness (finite tadpole), 
as well as infrared finiteness ($m_0 \ne 0$), 
we can assume the bare parameters $m_0$ and $\lambda_0$ entering $S[\Phi]$ 
to be constant in varying the lattice spacing $a$. In other words, the Renormalization Group flow is trivial:
we remain in the same physical (low-energy) theory by varying $a$
while keeping the bare couplings above fixed;
%%%%%
\item
We consider the correlator $G(\nu)$ as a function of the
multi-index $\nu$, i.e. as a discrete function of the
exponents of the fields at all lattice points.
Unlike classical field theory, where one looks
at the lattice fields $\phi_i$ individually, 
in the quantum case a "global" information is needed, involving
the simultaneous knowledge of the exponents of the fields at all 
points.
The quantum case is therefore radically more "correlated"
than the classical one, already at the level of formulation.
%%%%%
\item
By looking at the expression of $G(\nu)$ one finds that,
since the indices $\nu_i$ can be varied independently from each other
at each lattice point, from zero to infinity, we have to compute 
\beq
\chi_N 
\, \equiv \, 
\NN^N 
\, \equiv \, 
\left\{ f: \left\{ 1,2,\cdots,N \right\} \, \to \, \NN \right\}
\eeq
independent correlation functions, where $\NN$
is the set of the integers, zero included,
\beq
\NN \, \equiv \, \left\{ 0,1,2, \cdots, n, n+1, \cdots \right\}.
\eeq
The number of kinematically-independent correlators to compute 
on a finite lattice of whatever size $N$, is therefore
countable
\beq
\mathrm{Card}\left(\chi_N\right)
\, = \, \mathrm{Card}\left(\NN\right)
\, \equiv \, \aleph_0 ;
\eeq
%%%%%
\item
As already discussed, to solve a quantum field theory means to know all the correlators
$G(\nu)$'s, which are average values of monomials in the $N$ fields 
$\phi_1, \phi_2, \cdots, \phi_N$,
\beq
G(\nu) \, \equiv \,
\langle \phi_1^{\nu_1} \phi_2^{\nu_2} \cdots \phi_N^{\nu_N} \rangle. 
\eeq
Now, a polynomial $P$ in the $\phi_i$'s is a finite linear
combination of the monomials above, 
\beq
P\left(\phi_1,\phi_2,\cdots,\phi_N\right)
\, = \, \sum_{\mathrm{finite}} c_{\nu_1,\nu_2,\cdots,\nu_N}
\phi_1^{\nu_1} \, \phi_2^{\nu_2} \, \cdots \, \phi_N^{\nu_N} ,
\eeq
where $c_{\nu_1,\nu_2,\cdots,\nu_N}$ are arbitrary
coefficients. 
Therefore, we may also say that to solve a quantum field theory 
means to know the expectation values of any polynomial 
in the $\phi_i$'s,
\beq
\langle 
P\left(\phi_1,\phi_2,\cdots,\phi_N\right)
\rangle .
\eeq   
It is clear that this second formulation
of the theory
is completely equivalent to the first one.
\end{enumerate}

%%%%%%%%%%%%%%%%%%%%%%%%%%%%%%%%%%%%%%

\subsection{Free Propagator}

\begin{figure}[ht]
\begin{center}
\includegraphics[width=0.5\textwidth]{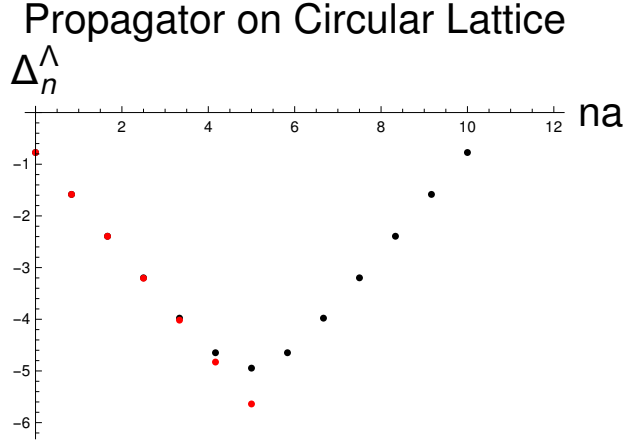}
\footnotesize
\caption{
\label{fig_pLambda1}
\it The black dots represent a logarithmic plot of the free propagator 
$\Delta^{\Lambda}_{\, n}$ on the circular lattice $\Lambda$,
as a function of the discrete times $t_n \equiv n \, a$.
We have taken the mass $m=1$, the physical length of the lattice
$T=10$ and the lattice size $N=12$, 
so that the lattice spacing is $a = T/N = 5/6$.
The red dots represent the propagator on the infinite lattice,
$\Delta_L(t)$, with the same parameters, namely $m=1$
and $a=5/6$ ($T=N \, a=\infty$ as $N=\infty$). 
In the latter case, finite-volume effects
are zero, so a comparison between the two plots
gives an idea of the size of these effects for various
times $t_n$.
}
\end{center}
\end{figure}

\begin{figure}[ht]
\begin{center}
\includegraphics[width=0.5\textwidth]{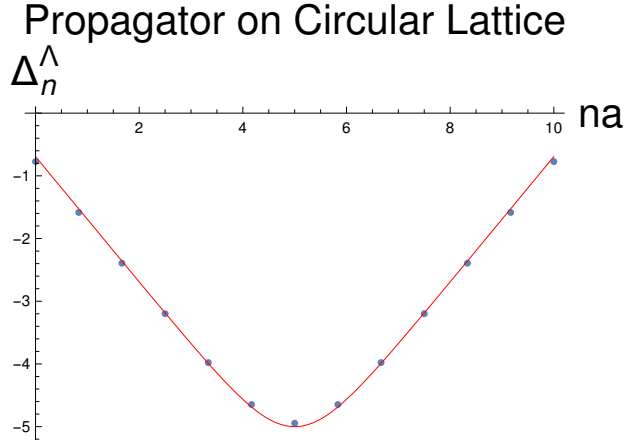}
\footnotesize
\caption{
\label{fig_pLambda2}
\it Logarithmic plot of the free propagator $\Delta^{\Lambda}_{n}$
as a function of $n a$ with the same parameters of the previous
plot.
The continuous red line is the propagator on the
circle $\Delta_{S^1_T}(t)$ with the same parameters
of the discrete plot, 
namely $m=1$ and $T=10$ ($a=T/N=0$ as $N=\infty$).
The finite-lattice spacing effects, i.e. the effects related to finite
instead of infinitesimal $a$, are barely visible
because, as discussed in the main text, they are very small
up to rather large values of $m a \lsim 1$.
}
\end{center}
\end{figure}

In this section, we derive the free propagator
of the theory on the circular lattice $\Lambda$.
Compared to the theory on the real line,
the theory on $\Lambda \equiv \Lambda_T^N \subset S^1_T$
contains {\it two} additional scales,
namely the lattice spacing $a$ and the length
$T$ of the embedding circle $S^1_T$.
These two scales are related to each other by the size $N$
of the lattice, i.e. by the number of its points,
by the relation
\beq
a \, = \, \frac{T}{N}.
\eeq
Combining the results of the previous two sections,
we derive that the times $t_n$ of the theory on $\Lambda_T^N$ 
are subjected to the double limitation
\beq
a \, \lsim \, \left|t_n\right| \, \lsim \, T.
\eeq
If we consider instead the energies $E$,
the limitations are "reversed" and read
\beq
\frac{1}{T} \, \lsim \, E \, \lsim \, \frac{1}{a}.
\eeq
In order to have a match of the range of the discrete energies
$E_n$ with the range of the continuum energies $E$ in the infinite
lattice, which we have taken in the first Brillouin 
zone (\ref{first_Brillouin}), 
let us write the points of $\Lambda \subset S^1$ 
in a symmetric way, as
\beq
\Lambda \, = \, 
\Big\{ 
\big([-N/2] +1\big) a, \cdots, -a, \, 0 \, , a \, , 2a \, , \, \cdots \, , [N/2] \, a 
\Big\}
\, \subseteq \, S^1,
\eeq
where $[\alpha]$ is the integer part of $\alpha$,
defined in such a way that the remainder is always positive
(e.g. $[-3/2]=-2$).
The $\Lambda$ propagator in the energy space is given by
\beq
S_{\,e}^{\Lambda} \, = \, S_L\left(E_e\right) \, = \,
\frac{a^2}{ 2 \left[1 - \cos\left(E_e a\right) \right] \, + \, (m a)^2 }
\, = \, \frac{a^2}{ 
2 \left[1 - \cos\left( 2\pi \, e / N \right) \right] \, + \, (m a)^2 },
\eeq
with the integer $e$ in the range
\beq
e \, = \, [-N/2] + 1, \, \cdots, \, - 1 \, , \, 0, 
\, 1, \, 2, \, \cdots, \, [N/2].
\eeq
In the last member of the above equation,
we have simply replaced, in the propagator on the
infinite lattice $L$, the allowed, discrete energies $E_e$
in place of the continuous energies $E$,
\beq
E \,\, \to \,\, E_e \, = \,  \frac{2 \pi \, e}{T} .
\eeq
The propagator on the circular lattice is obtained via the Discrete Fourier Transform (DFT) of the previous one
(see figs.\ref{fig_pLambda1} and \ref{fig_pLambda2}):
\bea
\label{prop_DFT}
\Delta^{\Lambda}_{\,n} &=& \frac{1}{N a}\sum_{e=[-N/2]+1}^{[N/2]} S_{\,e}^{\Lambda} 
\, \exp\left( - 2\pi i \frac{ n \, e}{N} \right) \, =
\nonumber\\
&=& \frac{1}{Na}\sum_{e=[-N/2]+1}^{[N/2]}  
\frac{a^2}{ 
2 \left[1 - \cos\left( 2 \pi e /N \right) \right] \, + \, (m a)^2 }
\, \exp\left( - 2\pi i \frac{n \, e}{N} \right) . 
\eea
%
%%%%%%%%%%%%%%%%%%% FIGURA %%%%%%%%%%%%%%%%%%%%
%%%%%%%%%% FINE FIGURA %%%%%%%%%%%%%%%%%%%%%%%%
%
Let us now discuss the relation of the above propagator
with the propagators derived in the previous sections.
\begin{enumerate}
%%%%%%
\item
{\it Continuum limit, i.e. limit of zero lattice spacing,
at finite volume, i.e. at finite and fixed $T$,}
\beq
a \, \to \, 0^+; \qquad T \,\, = \,\, \mathrm{const.}.
\eeq
In this case, the lattice size diverges, as
\beq
N \, = \, \frac{T}{a} \, \to \, \infty. 
\eeq
The energy spacing $\Delta E = 2\pi/T$ 
does not go to zero, but it is constant in the above limit.
The propagator on $\Lambda$ in the energy space has the limit
\beq
S^\Lambda_e  \, = \,
\frac{a^2}{ 2 \left[1 - \cos\left(E_e a\right) \right] \, + \, (m a)^2 }
\,\, \to \,\, \frac{1}{ E_e^2 \, + \, m^2 }
\qquad 
\mathrm{for} \,\,\, a \, \to \, 0^+,
\eeq
with
\beq
E_e \, = \,  \frac{2\pi e}{T}
\eeq
constant, with the integer $e$ now taking any integer value
\beq
e \, \in \, \ZZ.
\eeq
The exponent is naturally written as
\beq
\exp\left( - \, 2 \pi i \, \frac{ n \, e}{N} \right)
\, = \, \exp\left( - \, i \, E_e \, t_n \right)
\, = \,
\exp\left( - \, \frac{ 2 \pi i \, t_n \, e}{T} \right),
\eeq
where $t_n \equiv n \, a$.
Since $Na = T\,=\,$const., the propagator $\Delta^{\Lambda}_{\,n}$,
on the second member of eq.(\ref{prop_DFT}), 
tends to the propagator on the circle $S^1_T$
(see fig.\ref{fig_pLambda1}):
\beq
\Delta^{\Lambda}_{\,n}
\,\, \to \,\, 
T \sum_{e = - \infty}^{+\infty} 
\frac{\exp( - \, 2 \pi i \, t_n \, e / T)}{(2 \pi \, e)^2 \, + \,(m \, T)^2}
\, = \, \Delta_{S^1_T}\left(t_n\right)
\qquad \mathrm{for} \,\, a \, \to \, 0^+, \,\,\, \mathrm{fixed} \,\, T. 
\eeq
As already discussed, the propagator on the last
member of the above equation tends to the propagator in the real line 
for $T \to + \infty$;
%%%%%
\item
{\it Limit of infinite volume, i.e. of infinite time $T$,
at fixed lattice spacing $a$,}
\beq
T \, \to \, \infty; \qquad a \,\, = \,\, \mathrm{const.}. 
\eeq
The lattice size $N$ diverges in this limits as,
\beq
N \, = \, \frac{T}{a} \, \to \, \infty.
\eeq
Since
\beq
\Delta E_e \, \equiv \, E_{e+1} \, - \, E_e
\, = \, \frac{2\pi}{T} \, \equiv \, \Delta E \, \to \, 0
\qquad \mathrm{for} \,\, T \, \to \, \infty,
\eeq
we recover the propagator on the infinite lattice $L$
(see fig.\ref{fig_pLambda2}):
\beq
\Delta^{\Lambda}_{\,n} \, \to \, \Delta_n^L \, = \, \int\limits_{-\pi/a}^{+\pi/a}
\frac{a^2}{ 
2 \left[1 - \cos\left(E a\right) \right] \, + \, (m a)^2 } 
\, \exp\left(-i E n a\right) \, \frac{dE}{2\pi} 
\qquad \mathrm{for} \,\,\, T \, \to \, \infty.
\eeq
In turn, the above propagator tends, in the limit of vanishing lattice spacing,
as we have seen above, to the propagator on the real line:
\beq
\Delta_n^L 
\, \to \,
\Delta_\RR(t) \, = \,
\int\limits_{-\infty}^{+\infty}
\frac{\exp\left(-i E t\right)}{ E^2 \, + \, m^2 } \, \frac{dE}{2\pi} 
\, = \,
\frac{1}{2m} \exp(-m|t|)
\qquad \mathrm{for} \,\, a \, \to \, 0^+.
\eeq
\end{enumerate}
%
%%%%%%%%%%%%%%%%%%% FIGURA %%%%%%%%%%%%%%%%%%%%
%%%%%%%%%% FINE FIGURA %%%%%%%%%%%%%%%%%%%%%%%%
%
The pattern of the limits considered above can be
summarized by the following diagram:
\beq
\begin{array}{ccccc}
\, &\, & \Delta^\Lambda_{\,n} & \, & \, 
\\
T\to\infty; \, a=\mathrm{const.} & \,\,\, \swarrow & \, & 
\!\!\!\!\!\!\!\!\!\!\!\!\!\!\!\!\!\! \searrow & a \to 0; \, 
T=\mathrm{const.} 
\\
\, &\Delta^L_n & \, & \Delta_{S^1_T}(t) & \,
\\
\qquad \qquad a \to 0 & \,\,\, \searrow & \, & 
\!\!\!\!\!\!\!\!\!\!\!\!\!\!\!\!\!\! \swarrow & T \to \infty \qquad\qquad
\\
\, & \, & \Delta_\RR(t) & \, & \,
\end{array}
\eeq
Let us end this section with a few comments.
\begin{enumerate}
\item
All dimensionfull parameters have disappeared
from the propagator in eq.(\ref{prop_DFT}), 
as $m a$ is adimensional and the length $T$ 
of the circle $S^1$ does not explicitly appear;
%%%%%
\item
As already discussed, since the circular lattice 
$\Lambda \subset S^1_T$ is finite, 
the theory has both an infrared cutoff $T$
and an ultraviolet cutoff $a=T/N$ on the lengths.
In order to make finite-volume effects small,
one has to take
\beq
m \, T \, = \, m \, a \, N \, \gg \, 1,
\eeq
while, in order to render finite lattice-spacing
effects small, one has to take
\beq
m \, a \, \ll \, 1.
\eeq
At a given lattice size $N$, one has to compromise
between the two requests, 
a possibility being given by the "symmetric choice"
\beq
m \, a \, \approx \, \frac{1}{\sqrt{N}} \, \ll \, 1,
\eeq
giving
\beq
N \, m \, a \, \approx \, \sqrt{N} \, \gg \, 1;
\eeq
%%%%%
\item
According to convenience, we have indexed the lattice 
points $i$ in different ways,
\beq
i \, = \, 1, 2, \cdots, N,
\eeq
or
\beq
i \, = \, [-N/2] + 1, \cdots, -1, 0, 1, 2, \cdots, [N/2].
\eeq
Since the circular lattice $\Lambda$ is clearly invariant
under rotations of $\theta=2\pi$, we can add
to any index $i$ an integer multiple of $N$,
\beq
i \, \to \, i' \, = \, i \, + \, s \, N, \qquad s \, \in \, \ZZ,
\eeq
without changing the lattice point.
That is equivalent to say that we can identify the above
integers
\beq
i \, \sim \, i' \qquad \mathrm{iff} \qquad i'\, - \, i \, \in \, 
N \, \ZZ \, \equiv \, \left\{ \cdots,-N,0,N,2N,\cdots\right\}.
\eeq
It is therefore natural to think to the
circular-lattice indices as elements
of the quotient ring
\beq
\ZZ_N \, \equiv \, \frac{\ZZ}{N \, \ZZ}
\eeq
i.e. to take
\beq
i \, \in \, \ZZ_N.
\eeq
\end{enumerate}

%%%%%%%%%%%%%%%%%%%%%%%%%%%%%%%%%%%%%%%%%%%%%%%%%%%%%%

\subsection{Symmetries}

In this section we consider the classical
and quantum symmetries of the anharmonic
oscillator defined on the circular lattice $\Lambda$
of size $N$.
In general, by regularizing the theory on a lattice,
the symmetries of the original continuum
theory are, roughly speaking, drastically reduced.
One goes from continuous groups, having the cardinality
of the continuum, such as the orthogonal group 
$O(2)$, to finite groups.

%%%%%%%%%%%%%%%%%%%%%%%%%%%%%%%%%%%%%

\subsubsection{Classical Case}

Symmetries of the classical field theory
are given, as well known, by the
invariance group of the action.
In the general case, the lattice action,
\beq
S[\Phi] 
\, \equiv \,
\sum_{i\in\ZZ_N}
\left(
a_i \, \phi_i \, + \, 
\frac{1}{2} \, k_i \, \phi_i^2
\, - \, w_{i,i+1} \, \phi_i \, \phi_{i+1}
\, + \, \frac{g_i}{3} \, \phi_i^3
\, + \, \frac{\lambda_i}{4} \, \phi_i^4
\right),
\eeq
has not any linear symmetry.
Relevant symmetries emerge in the following
two particular cases:
\begin{enumerate}
\item
{\it The coefficients of the terms involving odd powers of the fields $\phi_i$
vanish:}
\beq
a_i \, = \, 0; \qquad g_i \, = \, 0;
\qquad i \, \in \, \ZZ_N.
\eeq
In this case, the action is even under change of sign of all the fields,
\beq
S[-\Phi] \, = \, S[\Phi].
\eeq
The symmetry group is then the group with two elements
\beq
\ZZ_2 \, = \, \left\{ +1, -1 \right\};
\eeq
%%%%%
\item
{\it The coefficients do not depend on the index $i$,
i.e. on the lattice point,} 
\beq
a_i \, = \, a; 
\quad k_i \, = \, k;
\quad w_{i,i+i} \, = \, w; 
\quad g_i \, = \, g;
\quad \lambda_i \, = \, \lambda;
\qquad i \, \in \, \ZZ_N.
\eeq
so that the action reads:
\beq
\label{action_symm}
S[\Phi] \, = \,
a \sum_{i\in\ZZ_N} \phi_i 
\, + \, 
\frac{k}{2} \sum_{i\in\ZZ_N} \phi_i^2
\, - \, 
w \sum_{i\in\ZZ_N} \phi_i \, \phi_{i+1} \, + \,
\frac{g}{3} \sum_{i\in\ZZ_N} \phi_i^3
\, + \, 
\frac{\lambda}{4} \sum_{i\in\ZZ_N} \phi_i^4 .
\eeq
For $N \le 3$, all the above sums
are invariant under the symmetric group $S_N$
acting on the set of all the lattice points
$\left\{1,2,\cdots,N\right\}$.
For $N \ge 4$, the third sum on the r.h.s. of the
above equation, having $w$ as coefficient,
is invariant under the dihedral group $D_N$,
while the other sums are still invariant under 
$S_N$.
The group $D_N$, defined for any $N \ge 3$, 
is the group of the symmetries of a regular
polygon with $N$ sides, often called $N$-gone for 
brevity \cite{Serre} (see fig.$\,$\ref{fig_ptriangle} 
for the case $N=3$, an equilater triangle, 
and fig.$\,$\ref{fig_psquare} for $N=4$, a square).
%
%%%%%%%%%%%%%%%%%%% FIGURA %%%%%%%%%%%%%%%%%%%%
\begin{figure}[ht]
\begin{center}
\includegraphics[width=0.5\textwidth]{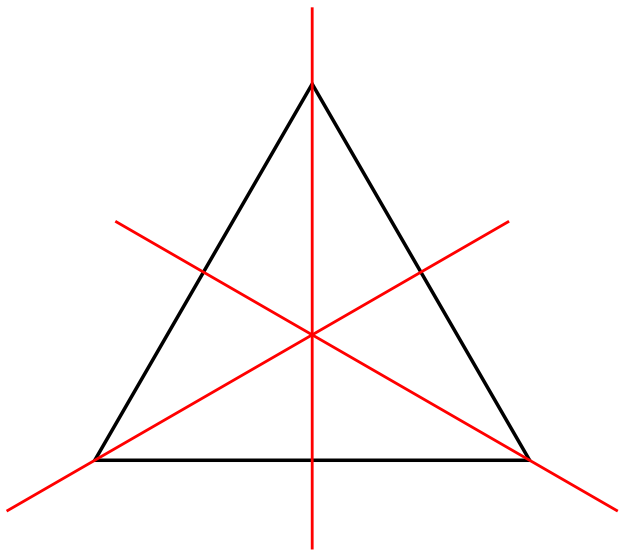}
\footnotesize
\caption{
\label{fig_ptriangle}
\it Equilateral triangle (in black), i.e. regular polygon with $N=3$
vertices (or sides), together with its three symmetry axes (in red).
The latter pass through the center $C$ of the triangle
and anyone of its vertices.
An equilater triangle is invariant under rotations about 
$C$ of integer multiples of the angle $2\pi/3$.
}
\end{center}
\end{figure}
%%%%%%%%%% FINE FIGURA %%%%%%%%%%%%%%%%%%%%%%%%
%
Presented in terms of generators and relations,
the dihedral group reads:
\beq
D_N \, = \,
\langle
x, \, a; \,\, x^N \, = \, 1; \,\, a^2 \, = \, 1; \,\, x a x \, = \, a
\rangle .
\eeq
%
%%%%%%%%%%%%%%%%%%% FIGURA %%%%%%%%%%%%%%%%%%%%
\begin{figure}[ht]
\begin{center}
\includegraphics[width=0.5\textwidth]{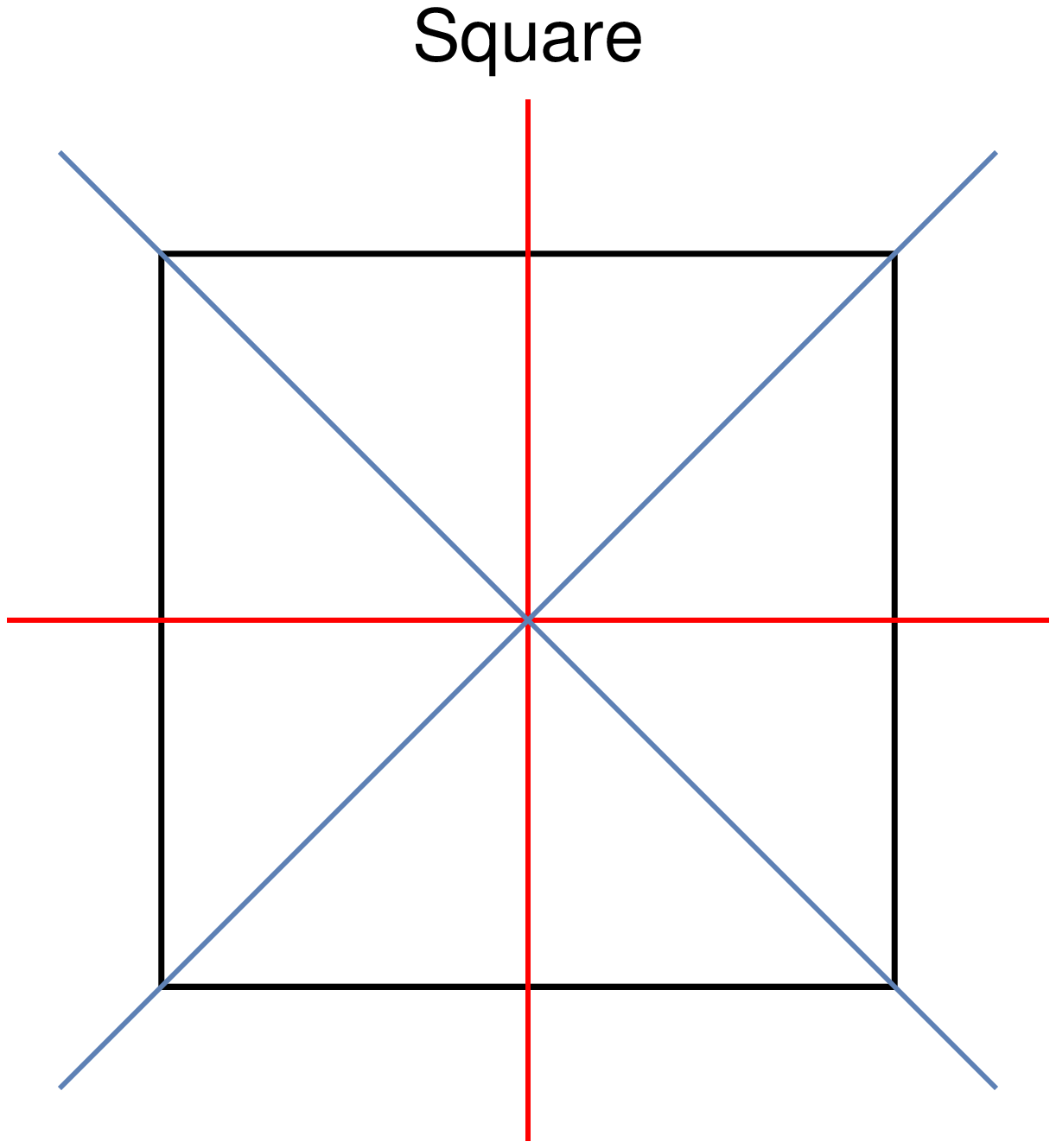}
\footnotesize
\caption{
\label{fig_psquare}
\it Square (in black), i.e. regular polygon with $N=4$
vertices (or sides), together with its four symmetry axes.
Two of them (in blue) coincide with the diagonals, i.e.
pass through opposite vertices, while the other
two (in red) pass through the midpoints of opposite 
sides.
The square is invariant under rotations about its
center of integer multiples of a right angle 
($\theta=\pi/2$).
}
\end{center}
\end{figure}
%%%%%%%%%% FINE FIGURA %%%%%%%%%%%%%%%%%%%%%%%%
%
The subgroups of $D_N$  can be obtained as
follows.
\begin{enumerate}
\item
By dropping the generator "$a$" from the above formula, 
we obtain the subgroup $C_N$, the cyclic group of order $N$,
\beq
C_N \, \equiv \, 
\langle
x; \,\, x^N \, = \, 1
\rangle \, \subseteq \, D_N.
\eeq
A regular $N$-gone $P_N$ can be inscribed into a
unit circle $S^1$ centered in zero in
the complex plane $\CC$:
\beq
S^1 \, \equiv \, 
\left\{
z \, \in \, \CC; \,\, |z| \, = \, 1 
\right\} .
\eeq
By putting one vertex on the positive axis,
let's call it $v_0$, the vertices of $P_N$ form the set
\beq
P_N \, \sim \,
\left\{ 
v_k \, \equiv \, \exp\left( i \frac{2\pi k}{N}\right); \,\, k=0,1,2,\cdots,N-1
\right\},
\eeq
where we have identified $P_N$ with its vertices.
The group $C_N$ can be written as
\beq
C_N \, = \, 
\left\{ 
h_j \, \equiv \, \exp\left( i \frac{2\pi j}{N}\right); 
\,\, j = 0,1,2,\cdots,N-1
\right\}.
\eeq
It holds
\beq
h_j \circ h_l \, = \, h_{j+l}. 
\eeq
The action of the cyclic group $C_N$
on $P_N$ consists in rotating it rigidly
by integer multiples of the angle $2\pi/N$.
Such an action is induced on the set of the
vertices of $P_N$:
\beq
h_j \cdot v_k \, \equiv \, v_{j+k}.
\eeq
It is intended that indices are defined modulo $N$;
%%%%%
\item
By dropping instead the generator "$x$", we obtain the
order-two subgroup
\beq
\ZZ_2 
\, \equiv \, 
\langle a; \,\, a^2 \, = \, 1 \rangle,
\eeq
which represents a reflection of the $N$-gone
about one of its symmetry axes;
%%%%%
\item
Given the reflection above "$a$", other $N-1$ reflections
are obtained as
\beq
a \, \to \, z \, a \, z^{-1}, \qquad z \, \in \, C_N.
\eeq
The elements on the r.h.s. of the above relation
indeed have order two:
\beq
\left( z \, a \, z^{-1} \right)^2
\, \equiv \, 
z \, a \, z^{-1} z \, a \, z^{-1}
\, = \, z \, a \, a \, z^{-1}
\, = \, z \, z^{-1} \, = \, 1.
\eeq
Geometrically, for $N$ odd, the reflections
are about lines passing through the center
$C$ of the polygon and one of its vertices
(see fig.\ref{fig_ptriangle}).
For $N$ even, there are $N/2$ reflections
with respect to axes passing through
opposite vertices and $N/2$ reflections
with respect to axes passing through the
midpoints of opposite sides 
(see fig.\ref{fig_psquare}).
\end{enumerate}
From the above calculations, it follows that
$D_N$ is a group of order $2N$, i.e. it contains $2N$ elements.
For any $N \ge 3$, it clearly holds
\beq
\label{S_N_largest}
D_N \, \subseteq \, S_N.
\eeq
Since for $N=3$,
\beq
\left|D_3\right| \, =  \, \left|S_3\right| \, = \, 6,
\eeq
it follows that in this case these groups coincide:
\beq
D_3 \, = \, S_3.
\eeq
Since, for $N \ge 4$,
\beq
\left|D_N\right| \, = \, 2 N \, < \, \left|S_N\right| \, = \, N! 
\qquad (N \, \ge \, 4) ,
\eeq
it follows instead the dihedral group is a proper
subgroup of the symmetric group,
\beq
D_N \, \subsetneq \, S_N.
\eeq
Roughly speaking, we may say that the dihedral group 
$D_N$ is the "discrete remnant" of the orthogonal symmetry group $O(2)$ 
of the continuous theory on $S^1$.
To be more specific, we may say that:
\begin{enumerate}
\item
The cyclic group $C_N$ is the "discrete remnant" of 
the special orthogonal group $SO(2)$ because, while
in the continuous case we can rotate a circle around its center 
by any real angle, in the discrete case we can rotate an $N$-gone 
only by integer multiples of $2\pi/N$;
%%%%%
\item
The $N$ reflections in $D_N$ are the discrete
remnant of the reflections about any diameter of $S^1$.
\end{enumerate}
Because of eq.(\ref{S_N_largest}),
we conclude that the action $S[\Phi]$ in eq.(\ref{action_symm}) has
for any $w \ne 0$ the symmetry group
\beq
G_N \, = \, D_N \qquad (w \ne 0).
\eeq
Explicitly,
\beq
S\left( \phi_{g\cdot 1}, \phi_{g\cdot 2}, \cdots, \phi_{g\cdot N} \right)
\, = \,
S\left( \phi_1, \phi_2, \cdots, \phi_N \right),
\qquad \forall g \, \in \, G_N,
\eeq
where by $g\cdot i$ we denote the action of the group element 
$g$ on the $i^{\mathrm{th}}$ lattice point.

%%%%%%%%%%%%%%%%%%%%%%%%%%%%%%%%%%%%%%

\vskip 0.3truecm

\noindent 
{\bf Random Field}

\vskip 0.2truecm

We may ask ourself what happens to the action in eq.(\ref{action_symm})
if we set to zero all the coupling between different fields,
i.e. if we take
\beq
w \, = \, 0.
\eeq
In this case there is no more correlation between fields at different
points and the theory describes a random field.
For $N=2$, the term under consideration explicitly reads
\beq
\mathcal{P}_2[\Phi] \, = \, w \, \phi_1 \, \phi_2
\eeq
and is symmetric under exchange of the indices $1$ and $2$,
like all the other terms in $S[\Phi]$; the symmetry group is
$S_2 \sim C_2 \sim Z_2$.
For $N=3$, the term proportional to $w$ is
\beq
\mathcal{P}_3[\Phi] \, = \, w \left( \phi_1 \, \phi_2 \, + \, \phi_2 \, \phi_3 
\, + \, \phi_3 \, \phi_1 \right)
\eeq
and has again the same symmetry $D_3=S_3$ of the other
terms in $S[\Phi]$.
For $N=4$, one has a polynomial in the $\phi_i$'s
given by
\beq
\mathcal{P}_4[\Phi] \, = \,
w \left( \phi_1 \, \phi_2 \, + \, \phi_2 \, \phi_3 
\, + \, \phi_3 \, \phi_4 \, + \, \phi_4 \, \phi_1
\right),
\eeq
which is invariant only under $D_4 \subsetneq S_4$
and not under $S_4$.
That occurs because $\mathcal{P}_4[\Phi]$
is not a symmetric polynomial, as it does not
contains the monomials
\beq
\phi_1 \, \phi_3 \quad \mathrm{and} \quad \phi_2 \, \phi_4.
\eeq
A similar phenomenon to the latter one also holds for any $N \ge 4$.
The conclusion is that, if we take
\beq
w \, = \, 0, 
\eeq
then the action $S[\Phi]$ becomes invariant under
the full symmetric group,
\beq
G_N \, = \, S_N \qquad (w = 0),
\eeq
which is a much larger group than the dihedral one
$D_N$ for $N \gg 1$. 
%%%%%
\item
{\it Both symmetries above.} 
The action in this case reads
\beq
S[\Phi] 
\, = \,
 \frac{k}{2} \sum_{i\in\ZZ_N} \phi_i^2
\, - \, 
w \sum_{i\in\ZZ_N} \phi_i \, \phi_{i+1}
\, + \, 
\frac{\lambda}{4}  \sum_{i\in\ZZ_N} \phi_i^4 .
\eeq
The theory then has both a $\ZZ_2$
and a $D_N$ symmetry.
Since $\ZZ_2$ acts on the values of all the fields 
$\phi_i \to \pm \phi_i$, $i=1,2,\cdots,N$,
without touching the indices,
while $D_N$ acts on the indices of the fields,
without touching their values,
these two groups commute with each other and the
resulting symmetry is given by their direct product:
\beq
G_N \, = \, D_N \times \ZZ_2 .
\eeq
\end{enumerate}

%%%%%%%%%%%%%%%%%%%%%%%%%%%%%%%%%%%%%%%%%%

\subsubsection{Quantum Case}

In looking at the symmetries at the quantum level,
we have also to look at the integration measure
\beq
D\Phi \, \equiv \, \prod_{i \in \ZZ_N} d\phi_i.
\eeq
The measure\ above is invariant under:
\begin{enumerate}
\item
{\it Change of sign of all the fields,}
\beq
\phi_i \, \to \, - \, \phi_i, \qquad i \, \in \, \ZZ_N,
\eeq
because the jacobian of the transformation has modulus
equal to one;
%%%%%
\item
{\it Action under the symmetric group $S_N$,} because
\beq
\prod_{i \in \ZZ_N} d\phi_{\sigma\cdot i} 
\, = \, 
\prod_{i \in \ZZ_N} d\phi_i, \qquad \forall \sigma \, \in \, S_N.
\eeq
The differentials of the fields $d\phi_i$ indeed commute with
each other.
\end{enumerate}
As a consequence, the classical symmetries
go over into the corresponding quantum symmetries or,
in other words, there is no anomaly.
Let us then reconsider the classical symmetries of the previous
section, which now take the form of Ward identities
relating different correlators among each other.
\begin{enumerate}
\item
$\ZZ_2$ {\it Symmetry.} It implies that 
all the correlators with an odd sum of all the exponents 
$\nu_i$ identically vanish:
\beq
G(\nu) \, = \, 0 
\quad \mathrm{if} \quad
\sum_{i\in\ZZ_N} \nu_i \, = \, \mathrm{(odd)}.
\eeq
%%%%%%
\item
$D_N$ {\it Symmetry}. It implies on correlation functions that:
\beq
G\left( 
\nu_{1}, \, \nu_{2}, \cdots \nu_{N} 
\right)
\, \equiv \, 
\langle \phi_1^{\nu_1} \, \phi_2^{\nu_2} \cdots \phi_N^{\nu_N} \rangle
\, = \, 
\langle \phi_{g\cdot 1}^{\nu_1} \, \phi_{g\cdot 2}^{\nu_2} 
\cdots \phi_{g\cdot N}^{\nu_N} \rangle,
\qquad \forall g \, \in \, G_N ,
\eeq
For a given $g \in G_N$, let us now consider lattice indices 
$k_1,k_2,\cdots,k_N \in \ZZ_N$ (which exist and are uniquely
defined) such that:
\beq
g\cdot k_1 \, = \, 1;
\qquad
g\cdot k_2 \, = \, 2;
\quad
\cdots
\quad
g\cdot k_N \, = \, N;
\eeq
so that
\beq
k_1 \, = \, g^{-1} \cdot 1;
\quad
k_2 \, = \, g^{-1} \cdot 2;
\quad
\cdots
\quad
k_N \, = \, g^{-1} \cdot N;
\eeq
It follows that
\beq
G\left( 
\nu_{g^{-1}\cdot 1}, \, \nu_{g^{-1}\cdot 2}, \cdots \nu_{g^{-1}\cdot N} 
\right)
\, = \, 
G\left( 
\nu_{1}, \, \nu_{2}, \cdots \nu_{N} 
\right), \qquad \forall g \in G_N.
\eeq
Let us define
\beq
h \, \equiv \, g^{-1}.
\eeq
Since $h$ takes values in the whole group 
as we vary $g$ in all $G_N$,
the above equation is rewritten as
\beq
G\left( 
\nu_{h \cdot 1}, \, \nu_{h \cdot 2}, \cdots \nu_{h \cdot N} 
\right)
\, = \, 
G\left( 
\nu_{1}, \, \nu_{2}, \cdots \nu_{N} 
\right), \qquad \forall h \, \in \, G_N.
\eeq
It expresses the action of the symmetry $D_N$ on the 
correlators.

Let us define the sum of all the exponents (indices) of a given
correlator
\beq
\mathcal{R}[G(\nu)] \, = \, \mathcal{R}(\nu)
\, \equiv \, \sum_{i=1}^N \nu_i.
\eeq
Since
\beq
\sum_{i=1}^N \nu_{g\cdot i} \, = \, \sum_{i=1}^N \nu_i ,
\qquad \forall g \, \in \, G_N,
\eeq
the symmetry group $G_N$ relates correlators
with the same weight.
In other words, $\mathcal{R}$ is an invariant,
as it commutes with the action of $G_N$.
That implies, for example, that a 2-point correlator with $\mathcal{R}=2$ 
(i.e. a propagator) will never mix with a 
4-point function of the elementary fields, having $\mathcal{R}=4$.

As already discussed, the action of $C_N \equiv \langle x; \,\, x^N = 1\rangle$, 
the cyclic group of order $N$,
on a lattice point $i \in \ZZ_N$ reads
\beq
x^k \cdot i \, = \, k \, + \, i , \qquad i \, \in \, \ZZ_N.
\eeq
Invariance under $C_N$ implies 
\beq
G\left( 
\nu_{1+k}, \, \nu_{2+k}, \, \cdots, \, \nu_{N+k}
\right) 
\, = \, 
G\left( 
\nu_{1}, \, \nu_{2}, \, \cdots, \, \nu_{N}
\right) ,
\eeq
where $k$ is any integer.
As usual, it is intended that indices in the last equation are defined
modulo $N$.
\end{enumerate}
In general, we will call equations of the above form
{\it Lattice Symmetry Equations} (LSE).
We will use the above symmetries in the next sections.
The idea is that, as far as the counting of
independent correlators is concerned,
we go from individual correlators to orbits of correlators.

%%%%%%%%%%%%%%%%%%%%%%%%%%%%%%%%%%%%%%%%%%%%%%%%%%

\section{Continuum Limit of the Theory on $\Lambda$}
\label{sect6}

Suppose we want to describe hadron dynamics by
means of an euclidean space-time lattice.
Since the typical dimension of a hadron, such as a proton $P$ or a $\rho$ meson, 
is of the order of one Fermi,
\beq
d_P \, \approx \, d_\rho \, \approx \, 1 \, \mathrm{fm} \, = \, 10^{-13} \, \mathrm{cm},
\eeq
we expect that a box of a linear dimension $T$ of, let's say,
ten Fermi's should be reasonably good,
\beq
T \, = \, 10 \, \mathrm{fm}.
\eeq
Let's then imagine to fix $T$ to the above value.
The lattice spacing $a$ is given by
\beq
\label{alN}
a \, = \, \frac{T}{N},
\eeq
where $N$ is the lattice size.
The continuum limit is defined as the limit of vanishing lattice 
spacing $a$%
%%%%%%%%%%
\footnote{
If the theory contains for example
a particle with mass $m \ne 0$, as in our case,
that physically means to send to zero the 
adimensional quantity $m \, a$,
\beq
m \, a \, \to \, 0^+.
\eeq
If only massless particles are involved, then
one may require for example
\beq
E_1 \, a \, \to \, 0^+,
\eeq
where $E_1$ is the lowest non-zero energy.}, 
%%%%%%%%%%
\beq
a \, \to \, 0^+.
\eeq
According to eq.(\ref{alN}), it implies that the
lattice size $N$ diverges:
\beq
N \, \to \, + \, \infty.
\eeq
Actually, while increasing $N$, one can also increase
$T$ in order to render finite-volume effects
vanishing small, so long as the lattice spacing
still goes to zero.
Let us make a few observations.
\begin{enumerate}
\item
Taking the continuum limit while keeping the
lattice size $N$ fixed implies
\beq
T \, \to \, 0,
\eeq
i.e. one goes to the continuum, but in an infinitesimal
box, which is not what physics usually requires;
\item
The limit of infinite lattice size 
does not necessarily implies the continuum limit. 
Indeed, one can take the limit
$N \to \infty$ while keeping $a$ constant,
implying that one is taking the dimension
of the box $T=Na$ growing exactly as $N$.
\end{enumerate}	
To solve the theory means to obtain analytical expression 
of the correlators
\beq
G\left(\cdots, \nu_{-2},\nu_{-1},\nu_0,\nu_1,\nu_2, \cdots, \nu_n, \cdots\right)
\, \equiv \, \int
\prod_{s=-\infty}^{+\infty} d\phi_s  
\, 
\prod_{i=-\infty}^{+\infty} \phi_i^{\nu_i}  
\, e^{-S[\Phi]} ,
\eeq
where the lattice action is given by the following series:
\beq
S[\Phi] 
\, \equiv \,
\sum_{i=-\infty}^{+\infty} 
\left(
\frac{k_i}{2} \, \phi_i^2
\, - \, w_{i,i+1} \, \phi_i \, \phi_{i+1}
\, + \, \frac{\lambda_i}{4} \, \phi_i^4
\right).
\eeq
We have used an explicit notation in order to stress the
difference with respect to the finite $N$ case.
While for any finite lattice size, $N<\infty$, there is only one 
possibility in defining what we mean by analytic solution, 
in the continuum limit there are instead two different possibilities, 
which we call the {\it strong limit} and the {\it weak limit}.
\begin{enumerate}
\item
{\it Strong Limit.}
That is the first possibility that comes to mind.
One considers all the possible values
of the multi-index 
\beq
\nu \, \equiv \, \left( \cdots,\nu_{-1},\nu_0,\nu_1,\nu_2,\cdots,\nu_n,\cdots\right); 
\qquad \nu_i \, = \, 0,1,2,\cdots; \quad i \, \in \, \ZZ;
\eeq
which can be thought of as a map from the ring of the integers to the
natural numbers:
\bea
\nu: \ZZ &\to& \NN
\nonumber\\
i &\mapsto& \nu_i,
\eea
where the set of the natural numbers is defined as including the zero,
\beq
\NN \, \equiv \, \left\{ 0,1,2,\cdots \right\}.
\eeq
According to this definition of solution of the theory,
the knowledge of correlators such as for example the one with
$\nu_i=1$ for any $i \in \ZZ$,
\beq
G(\cdots,1,\cdots,1,\cdots),
\eeq
is then required.
The cardinality of the correlators to evaluate is then
\beq
\mathrm{Card}\left[\left\{G(\nu)\right\}\right] 
\, = \, \mathrm{Card}\left(\NN^\ZZ\right),
\eeq
which is the cardinality of the continuum.
Along this route, we find an intractable theory.
However, at this point, physics comes to our help: 
we know that only correlators $G$ with a
{\it finite} number of points
--- though arbitrarily large --- 
are needed:%
\footnote{
For example, at the $P+P$ Large Hadron Collider (LHC),
presently operating at the European Center
for Nuclear Physics (CERN) at a Center-of-Mass
energy of 13 $\mathrm{TeV}$, processes
with up to $\mathcal{O}(10)$ partons in 
the final state are studied.
}
%%%%%%%%%%%
\beq
G\left(x_1,x_2,\cdots,x_n\right)
\, \equiv \,
\left\langle 0 \right| 
T \phi\left(x_1\right) \phi\left(x_2\right)
\cdots \phi\left(x_n\right)
\left| 0 \right\rangle, 
\qquad n = 2 ,3, 4, \cdots.
\eeq
Indeed, if we consider scattering processes
of particles with a mass $m \ne 0$
at any given energy $E<\infty$, then
the number of particles in the initial
state, $n_{\mathrm{in}}$,
and in the final state, $n_{\mathrm{in}}$, 
is subjected to the upper kinematic bound
\beq
n_{\mathrm{in}}, \,\, n_{\mathrm{out}} 
\, \le \, \frac{E}{m} \, < \, \infty.
\eeq
The above bound is empty in the case
massless particles ($m=0$), such as photons or gluons,
are involved in the process.
In such cases, as well known, if infrared divergences
are present, 
one is forced to consider correlators
with an arbitrarily large number $n$ of particles,
yet still finite.   
To summarize, as far as physics
is concerned, correlators with an 
infinite number of points are not 
relevant.
This observation offers us the possibility
to define a simpler continuum limit.
%%%%%
\item
{\it Weak Limit.}
A much simpler theory, with the same
physical content as the above one,
is obtained if we require
the (infinite) sum of all the occupation
numbers $\nu_i$ to be {\it finite}:
\beq
\label{basic_assunz}
\sum_{i=-\infty}^{+\infty} \nu_i \, < \, \infty.
\eeq
We then require all the infinite occupation numbers $\nu_i$,
$i \in \ZZ$,
to be zero, except for a finite number of them.%
\footnote{
It's like to have an infinite number of distinguishable boxes,
but only a finite number of balls to put inside them.
Only finitely many boxes are not empty.
}
%%%%%%%
In other words, the $\nu_i$'s become definitively zero 
going in the positive direction, $i \to + \infty$, 
as well as going in the negative direction, 
$i \to - \infty$.
That is to say that we restrict to maps
\bea
\nu: \ZZ &\to& \NN
\nonumber\\
i &\mapsto& \nu_i
\eea
which are definitively zero.
Let us denote the set of such maps with the
symbol
\beq
\left(\NN^\ZZ\right)_0
\, \equiv \,
\left\{
\nu: \ZZ \, \to \, \NN;
\,\, \sum_{i=-\infty}^{+\infty} \nu_i \, < \, \infty
\right\}.
\eeq
Correlators such as the one considered
in the previous point, with
$\nu_i=1$ for any $i \in \ZZ$,
\beq
G(\cdots,1,\cdots,1,\cdots),
\eeq
are then excluded  from the definition of
solution of the theory.

As already discussed, for finite lattice sizes,
$N < \infty$,
the theory can also be formulated in terms
of average values of polynomials in the $\phi_i$'s,
$i=1,2,\cdots,N$: 
\beq
\langle
P\left(\phi_{1}, \phi_2, \cdots, \phi_N, \right)
\rangle.
\eeq
In the continuum case,
we may also say that solving the theory
means to know
the average value of an arbitrary polynomial
in the infinitely many variables $\phi_i$,
\beq
\langle
P\left(\cdots, \phi_{-1}, \phi_0, \phi_1, \cdots, \phi_n, \cdots \right)
\rangle,
\eeq
where the dots $\cdots$ at the beginning and at the end
of the string denote that there is neither
a first variable nor a last variable.

Going back to correlators, we can assign to each $G(\nu)$
two characteristic numbers:
\begin{enumerate}
\item
{\it The sum of all the indices,}
\beq
\mathcal{R}(\nu) 
\, \equiv \,
\sum_{i=-\infty}^{+\infty} \nu_i.
\eeq
The vacuum correlator, having all its indices
equal to zero,
\beq
G(\cdots,0,\cdots,0,\cdots),
\eeq
for example, has 
\beq
\mathcal{R} \, = \, 0.
\eeq
Correlators with one non-zero index $\nu_i$
at any point $i \in \ZZ$,
\beq
G\left(\cdots; 0 ; \cdots; 0 ; \, \nu_i > 0 \, ; \cdots; 0 ; \cdots \right),
\eeq
have 
\beq
\mathcal{R} \, = \, \nu_i ,
\eeq
and so on.
We will see that the above integer number will
play an important role when we will discuss
the dynamics in the continuum limit;
%%%%%
\item
{\it The number specifying how many indices $\nu_i$ are not zero,
but strictly positive.}
Given
\beq
\nu \, \in \, \left( \NN^\ZZ \right)_0,
\eeq
let us define
\beq
\tau(\nu)
\, \equiv \,
\mathrm{Card}\left[\left\{ \nu_i > 0; \,\, i \in \ZZ \right\}\right]
\, = \, 
\sum_{ i \in \ZZ \,\, \mathrm{s.t.} \, \nu_i > 0 } 1 .
\eeq
For any correlator $G(\nu)$, it clearly holds
\beq
0 \, \le \, \tau(\nu) \, \le \, \mathcal{R}(\nu) \, < \, \infty.
\eeq
Correlators with one non-zero index $\nu_i$
at any point $i \in \ZZ$,
\beq
G\left(\cdots; 0;\cdots; 0; \, \nu_i = 1,2,\cdots \, ; 0 ; \cdots; 0; \cdots \right),
\eeq
for example, have $\tau=1$.
Note that, since the non-zero index can be put at any place,
it follows that
\bea
&&\mathrm{Card}
\left[
\left\{
G(\cdots; 0; \cdots; 0; \, \nu_i = 1,2,\cdots \, ; 0; \cdots; 0; \cdots)
\right\}
\right]
\, = \, \mathrm{Card}\left( \ZZ \times \NN_+ \right) \, = \,
\nonumber\\
&=& \mathrm{Card}(\NN) \, \equiv \, \aleph_0,
\eea
since the direct product of a finite family of countable sets is countable.
$\NN_+$ is the set of strictly positive integers,
\beq
\NN_+ \, \equiv \, \left\{ 1, 2, 3, \cdots \right\}.
\eeq
Correlators with exactly two strictly-positive
indices
\beq
G\left(\cdots; 0; \, \nu_i=1,2,\cdots \, ; 0;  \cdots; 0; 
\, \nu_j=1,2,\cdots \, ;  0; \cdots \right),
\qquad i \, \ne \, j,
\eeq
have $\tau=2$.
For the first index $\nu_i \in \NN_+$ we can take any 
position $i \in \ZZ$, while for the second index
$\nu_j\in \NN_+$ we can take any position $j \in \ZZ$
with $j \ne i$.
The cardinality of the correlators with $\tau=2$
is therefore given by
\beq
\mathrm{Card}
\left[\left\{ G(\nu ); \,\, \tau=2 \right\}\right]
\, = \, \mathrm{Card}[(\ZZ \times \NN_+)^2] \, = \, \aleph_0.
\eeq
Since each correlator has a well-defined value of $\tau$,
the set of all $G(\nu)$'s can be written in the form:
\beq
\big\{ G(\nu) \big\}
\, = \, \cup_{n=0}^\infty \big\{ G(\nu); \,\, \tau(\nu) \, = n \big\},
\eeq
where the union is disjoint, i.e. it involves pair-wise disjoint sets.

Since the union of a countable family of countable sets is countable, we
conclude that the set of all correlators is countable
in the weak limit:
\beq
\mathrm{Card}[\left\{ G(\nu) \right\}] \, = \, \aleph_0.
\eeq
That has to be compared with the strong-limit case,
in which the correlators have instead the cardinality 
of the continuum.
Therefore, by requiring the indices $\mathrm{R}(\nu)$ or $\tau(\nu)$ 
to be finite for any $G(\nu)$, we reduce the cardinality of the correlator set
from $\aleph_1$ to $\aleph_0$.
\end{enumerate}

\end{enumerate}
The difference between the strong and the weak limits can be 
easily understood by means of the following simple example.
The weak limit is the analog of the vector space of all 
the polynomials in one indeterminate $x$, let's say on the
real field,
\beq
\RR[x] \, \equiv \, 
\left\{
\sum_{k=0}^N a_k \, x^k; \,\, N \, = \, 0,1,2,\cdots, \,\, a_k \, \in \, \RR
\right\}.
\eeq
The latter is an infinite-dimensional space, having
as algebraic (or Hamel) basis, for example,
the countable set of all the monomials
\beq
\mathcal{B} \, \equiv \, \left\{ x^n; \,\, n \, = \, 0,1,2,3, \cdots \right\}.
\eeq
The vector spaces of all the
polynomials with degree up to $n=0,1,2,\cdots$,
\beq
\mathrm{Pol}_n[x]
\, \equiv \,
\left\{
\sum_{k=0}^n a_k \, x^k; \,\, a_k \, \in \, \RR
\right\},
\qquad \dim\left(\mathrm{Pol}_n[x]\right) \, = \, n + 1,
\eeq
form a strictly-increasing sequence (filtration) of
finite-dimensional vector spaces 
\beq
\cdots \, \subsetneq \,\mathrm{Pol}_{n-1}[x] \, \subsetneq \,\mathrm{Pol}_n[x] \, 
\subsetneq \, \mathrm{Pol}_{n+1}[x] \, \subsetneq \, \cdots,
\eeq
whose union is the space under consideration:
\beq
\RR[x] \, = \, \cup_{n=0}^\infty \mathrm{Pol}_n[x].
\eeq
The strong limit is the analog 
of the vector space of all the formal power series:
\beq
\RR[[x]] \, \equiv \, 
\left\{
\sum_{k=0}^\infty a_k \, x^k; \,\, a_k \, \in \, \RR
\right\}.
\eeq
The latter is a much bigger vector space, 
\beq
\RR[x] \, \subsetneq \, \RR[[x]],
\eeq
which cannot be invaded by sequences
of finite-dimensional spaces,
with an uncountable algebraic basis.

%%%%%%%%%%%%%%%%%%%%%%%%%%%%%%%%%%%%%%%%%%%%%%%%%%%%%%%%%%%%%

\subsection{Symmetries}

We expect the lattice theory to acquire, in the continuum limit 
$N \to \infty$, symmetries described by finite or infinite discrete 
groups.
\begin{enumerate}
\item
{\it $\ZZ_2$ Symmetry.}
The invariance of the theory under change of sign of the fields
on a finite lattice of size $N$, 
\beq
\phi_i \, \to \, - \, \phi_i; \qquad i \, \in \, \ZZ_N;
\eeq
simply becomes, in the limit $N \to \infty$:
\beq
\phi_i \, \to \, - \, \phi_i; \qquad i \, \in \, \ZZ;
\eeq
\item
{\it $D_N$ Symmetry.}
The "limit" $N \to \infty$ of the dihedral group $D_N$
can be simply defined by removing the condition $x^N=1$
(which becomes meaningless) from the definition:
\beq
D_\infty \, \equiv \, 
\langle
x, \, a; \,\, a^2 \, = \, 1; \,\, x a x \, = \, a
\rangle .
\eeq
An infinite abelian subgroup of $D_\infty$ is the free group
generated by the element $x$:
\beq
\langle x \rangle \, = \, \left\{ x^n; \, \,\, n \, \in \, \ZZ \right\}
\, \subsetneq \, D_\infty.
\eeq
As well known, the above group is isomorphic to 
the additive group of the ring of the integers,
\beq
\langle x \rangle \, \sim \, (\ZZ,+).
\eeq
By acting on the reflection $a$ as
\beq
a \, \to \, x^n \, a \, x^{-n}, \qquad n = 1, 2, 3,\cdots, 
\eeq
one generates a countable number of reflections.
It is straightforward to check that the $N\to \infty$
symmetric lattice action 
\beq
\label{S_infinito}
S[\Phi] 
\, \equiv \,
\sum_{i=-\infty}^{+\infty} 
\left(
\frac{k}{2} \, \phi_i^2
\, - \, w \, \phi_i \, \phi_{i+1}
\, + \, \frac{\lambda}{4} \, \phi_i^4
\right).
\eeq
is invariant under $D_\infty$, so let us just sketch the proof.
The action above is invariant under any shift of the index
\beq
i \, \to \, i \, + \, j, \qquad j \, \in \, \ZZ, 
\eeq
i.e. it is invariant under $(\ZZ,+)$.
If we identify the symmetry "$a$" with the reflection
about the lattice point $j=0$, 
\beq
i \, \to \, - \, i,
\eeq
then $S[\Phi]$ is invariant under such transformation.
Since we have already proved the invariance of the action 
under the group $(\ZZ,+)$, it follows that $S[\Phi]$ is also 
invariant under the action of $x^n \, a \, x^{-n}$, 
$n \in \ZZ$,
and then under the complete $N=\infty$ dihedral group $D_\infty$.

If we set to zero the coupling $w$ in the action in eq.(\ref{S_infinito}), 
then the theory becomes invariant under the symmetric group $S_\infty$,
the symmetric group acting on the infinite set $\NN$ of the natural numbers,
\beq
S_\infty
\, \equiv \,
\left\{ f:\NN \, \to \, \NN; \,\,\, f \,\, \mathrm{invertible} \right\}.
\eeq
\end{enumerate}

%%%%%%%%%%%%%%%%%%%%%%%%%%%%%%%%%%%%%%%%%%%%%%%%%%%%%%%%%%%%%

\section{Dyson-Schwinger Equations on the Lattice}
\label{sect7}

Linear relations between the correlators $G(\nu)$'s
are obtained by means of the Lattice Dyson-Schwinger (LDS) equations:
\beq
\int\limits_{\RR^{N}}  D \Phi \, \frac{\partial}{\partial \phi_i} 
\Big\{
\Phi^\nu \, \exp\left( - S\left[\Phi\right] \right)
\Big\}
\, = \, 0 ,
\qquad
i \, \in \, \ZZ_N.
\eeq
By explicitating the derivatives, one obtains
\beq
\int\limits_{\RR^{N}}  D \Phi 
\left(
\nu_i \, \Phi^{\nu \, - \, e_i} 
\, - \, \Phi^\nu \, \frac{\partial S}{\partial \phi_i}
\right)
\exp\left( - S\left[\Phi\right] \right)
\, = \, 0 ,
\eeq
with $\left(e_i\right)_j \equiv \delta_{ij}$ or, more explicitly,
\beq
e_i \, \equiv \, \left(0, \cdots, 0_{i-1}, 1_i, 0_{i+1}, \cdots, 0 \right).
\eeq
In a more compact notation,
\beq
\nu_i \left\langle \Phi^{\nu \, - \, e_i} \right\rangle
\, = \, 
\left\langle \Phi^\nu \, \frac{\partial S}{\partial \phi_i} \right\rangle;
\qquad \nu_i \, = \, 0,1,2,\cdots; \quad i \, \in \, \ZZ_N.
\eeq
Going back to the index notation, one obtains the set of equations:
\bea
\label{LDS}
&+& \quad \nu_i \,\,\,\,\, G\left[\nu_{[-N/2]+1}; \, \cdots; \, \nu_{i-1}; 
\,\, \left\{ - \, 1 \, + \, \nu_i \right\} \,\, ; \,\, \nu_{i+1}; \, \cdots; \, \nu_{[N/2]} \right] \, +
\nonumber\\
&-& \quad k_i \,\,\,\,\, G\left[\nu_{[-N/2]+1}; \, \cdots; \, \nu_{i-1}; 
\,\, \left\{ + \, 1 \, + \, \nu_i\right\} \,\, ; \,\, \nu_{i+1}; \, \cdots, \, \nu_{[N/2]} \right] \, +
\nonumber\\
&+& w_{i-1, \,i} \, G\left[\nu_{[-N/2]+1}; \, \cdots; \, \nu_{i-2}; 
\,\, \left\{ + \,1 \, + \, \nu_{i-1} \right\} 
\,\, ; \, \nu_i, \, \cdots, \, \nu_{[N/2]} \, \right] \, +
\nonumber\\
&+& w_{i, \,i+1} \, G\left[\nu_{[-N/2]+1}; \, \cdots; \, \nu_i; 
\,\, \left\{ + \, 1 \, + \, \nu_{i+1} \right\} 
\,\, ; \,\, \nu_{i+2}; \, \cdots, \, \nu_{[N/2]} \right] \, +
\nonumber\\
&-& \quad \lambda_i \,\,\,\,\, G\left[ \nu_{[-N/2]+1}; \, \cdots; \, \nu_{i-1} ; 
\,\, \left\{ + \, 3 \, + \, \nu_i \right\} \,\, ; 
\,\, \nu_{i+1}; \, \cdots, \, \nu_{[N/2]} \right] \, = \, 0 ,
\eea
with
\beq
\qquad\qquad
0 \, \le \, \nu_i \, < \, \infty,
\qquad
i \, \in \, \ZZ_N.
\eeq
The term on the first line above vanishes for $\nu_i \, = \, 0$.
The shifted indices have been put inside curly brackets for clarity purposes.
Let's make a few comments.
\begin{enumerate}
\item
If we had discretized the time derivative $\phi'(t)$ 
of the field $\phi(t)$ by using also
next-to-nearest differences, also a term
of the form
\bea
&& w_{i-2, \,i} \, G\left[\nu_{[-N/2]+1}; \, \cdots; \, \nu_{i-3}; 
\,\, \left\{ + \,1 \, + \, \nu_{i-2} \right\} 
\,\, ; \, \nu_{i-1}; \nu_i; \, \cdots, \, \nu_{[N/2]} \, \right] \, +
\nonumber\\
&& w_{i,i + 2} \, G\left[\nu_{[-N/2]+1}; \, \cdots; \, \nu_i; \, \nu_{i + 1}; 
\,\, \left\{ + \, 1 \, + \, \nu_{i + 2} \right\} 
\,\, ; \,\, \nu_{i + 3}; \, \cdots, \, \nu_{[N/2]} \right]
\eea
would have appeared in the LDS equation above;
\item
In a theory with a cubic interaction,
such as for example a $g \, \phi^3$
theory,%
%%%%%%%%%
\footnote{It is well known that the $g \, \phi^3$
theory does not exist because the cubic potential
is unbounded from below for any real $g \ne 0$, so the
functional integral is divergent.
By giving up unitarity, we can formally overcome this 
difficulty by taking $g$ purely imaginary.
}
%%%%%%%%
the last term of the LDS equation
would have the index $\nu_i + 3$ replaced
by $\nu_i + 2$.
If the theory under study involves 
different interactions, the sum
of the corresponding contributions
does appear.
\end{enumerate}
%%%%%%%%%%%%%%%
If we define the observables of the quantum field
theory by means of average values of polynomials
in the $\phi_i$'s,
\beq
\langle P(\Phi) \rangle,
\eeq
with
\beq
P(\Phi) \, = \, P\left(\phi_1,\phi_2,\cdots,\phi_N\right),
\eeq
then the lattice Dyson-Schwinger equations are written as
\beq
\int\limits_{\RR^{N}}  D \Phi \, \frac{\partial}{\partial \phi_i} 
\Big\{
P(\Phi) \, \exp\left( - S\left[\Phi\right] \right)
\Big\}
\, = \, 
\int\limits_{\RR^{N}}  D \Phi 
\Big\{
\frac{\partial P(\Phi)}{\partial \phi_i} 
\, - \, P(\Phi) \, \frac{\partial S(\Phi)}{\partial \phi_i}  
\Big\}
\exp\left( - S\left[\Phi\right] \right)
\, = \, 0.
\eeq
In a more compact notation,
\beq
\left\langle \frac{\partial P(\Phi)}{\partial \phi_i} \right\rangle 
\, = \, 
\left\langle
P(\Phi) \, \frac{\partial S(\Phi)}{\partial \phi_i} 
\right\rangle , 
\qquad i \, \in \, \ZZ_N ,
\eeq
where $P(\Phi)$ is any polynomial in the $\phi_i$'s.

%%%%%%%%%%%%%%%%%%%%%

\subsection{Examples}

To get some intuition concerning the LDS systems,
let us see a few examples on small lattices:
\begin{enumerate}
\item
For $N=1$ there is the LDS equation
\beq
\nu \, G(\nu-1) - k \, G(\nu+1) - \lambda \, G(\nu+3) \, = \, 0 ;
\qquad \nu \ge 0
\eeq
Being there only one field, one can only study self-correlations
of the single field $\phi$ with itself.
In the gaussian case,
\beq
\lambda = 0,
\eeq
the LDS equation becomes a simple two-step, two-term equation
\beq
\nu \, G(\nu-1) \, - \, k \, G(\nu+1) \, = \, 0 ;
\qquad \nu \ge 0;
\eeq
%%%%%
\item
For $N=2$ one has the two LDS equations:
\bea
\label{N2gaussian}
\nu_1 \, G(\nu_1-1,\nu_2) - k_1 \, G(\nu_1+1,\nu_2) + w \, G(\nu_1,\nu_2+1)
- \lambda_1 \, G(\nu_1 +3 ,\nu_2) &=& 0;
\nonumber\\
\nu_2 \, G(\nu_1,\nu_2-1) - k_2 \, G(\nu_1,\nu_2+1) + w \, G(\nu_1+1,\nu_2)
- \lambda_2 \, G(\nu_1,\nu_2+3) &=& 0.
\eea
This is, in some sense, the lowest-order non-trivial case,
involving correlations at two different points.
The gaussian theory corresponds to 
\beq
\lambda_1 \, = \, \lambda_2 \, = \, 0.
\eeq
\end{enumerate}

%%%%%%%%%%%%%%%%%%%%%%%%%%%%%%%%%%%%%%%%%%%%%%%%%%%%%%%

\section{Solution of Lattice Dyson-Schwinger (LDS) Equations}
\label{sect8}

In general, in order to solve a model,
one has to combine all the information available.
In our case, a theory on a lattice, one has to combine together
the Lattice Dyson-Schwinger Equations (LDS) 
with the Lattice Symmetry Equations (LSE).
However, in order to understand the general "philosophy"
of our study, let's neglect the symmetry equations to begin with
--- they will be included later.

%%%%%%%%%%%%%%%

We have one LDS equation for each lattice point $i \in \ZZ_N$
and for each possible choice of the index $\nu_i=0,1,2,\cdots$.
So, by substituting all possible numerical values
for all the indices,
we get a huge linear homogeneous
system on all the correlators $G(\nu)$'s.
In physical language, the quantum equations of motion 
produce linear relations between the kinematically
independent correlators.
There is an infinite number of unknown correlators $G(\nu)$'s
and an infinite number of equations, so the
solution of the system is, a priori, not trivial at all.
Since the system is homogeneous in the correlators,
one has to {\it arbitrarily decide} which correlators are to be
considered as known --- i.e. to be put on the right
hand sides of the final solutions --- and which are to be considered 
as unknown --- and then kept on the left hand sides.
It is somewhat natural to solve the above system by expressing 
correlators with large values of the indices in terms of correlators 
with smaller values of the indices.
This idea can be formalized by defining a recursive weight, defined individually for each correlator, which involves, for example,
the sum of all the indices in $G(\nu)$,
\beq
\mathcal{R}\left[G(\nu)\right]
\, = \, 
\mathcal{R}\left( \nu \right)
\, \equiv \, 
\sum_{i=[-N/2]+1}^{[N/2]} \nu_i
\, = \,
\nu_{[-N/2]+1} \, + \, \cdots \, + \, \nu_i \, + \, \nu_{i+1} 
\, + \, \cdots \, + \, \nu_{[N/2]} .
\eeq
The correlator with the lowest possible recursive weight,
for example, is the vacuum amplitude $G(0,0,\cdots,0)$,
having weight zero:
\beq
\mathcal{R}\left[G(0,0,\cdots,0)\right] \, = \, 0 .
\eeq
The propagators,
\beq
G\left(0; \, \cdots; \, 0; \, \nu_i \, = \, 1; 0; \, \cdots; \, 0; \, \nu_j \, = \, 1; \, 0 ; \, \cdots; \, 0 \right) ,
\qquad\qquad i < j ,
\eeq
have $\mathcal{R} = 2$ and so on.
The terms on the symbolic LDS eq.(\ref{LDS}), in the same order
in which they are written, have the following weights:
\bea
\mathcal{R}\left( \propto \nu_i \right) \quad \,\,\, &=& \sigma \, - \, 1 ;
\nonumber\\
\mathcal{R}\left( \propto k_i \right) \quad \,\,\, &=& \sigma \, + \, 1 ;
\nonumber\\
\mathcal{R}\left( \propto w_{i-1,i} \right) &=& \sigma \, + \, 1  ;
\nonumber\\
\mathcal{R}\left( \propto w_{i,i+1} \right) &=& \sigma \, + \, 1  ;
\nonumber\\
\mathcal{R}\left( \propto \lambda_i \right) \quad \,\,\, &=& \sigma \, + \, 3  ;
\eea
where $\sigma$ is the symbolic sum of all the $\nu_i$'s:
\beq
\sigma \, \equiv \, \sum_{i=[-N/2]+1}^{[N/2]} \nu_i \,\, < \, \infty . 
\eeq
Note that all weights $\mathcal{R}$ above differ from each other
by an even number (zero,two and four),
because the action $S[\Phi]$ is even in $\Phi$.%
\footnote{If we were to add to $S[\Phi]$ terms linear or cubic in the $\phi_i$'s, then odd differences between the $\mathcal{R}$'s would appear.
The parity of $\mathcal{R}$ would not be respected in correlator 
decomposition.
}

The term with the highest weight is then the last one in eq.(\ref{LDS}), the quartic one, 
which represents the interaction in our model.
In the interacting case $\left( \lambda_i \ne 0 \right)$,
eq.(\ref{LDS}) is a 4-step recurrence equation.
As expected from experience, there are three regimes in which equation (\ref{LDS}) drastically differ, the first two being infinitely simpler than 
the third one.
%%%%%%%%%%%%%%%%%
\begin{enumerate}
\item
{\it Random Field},
\beq
w_{i,i+i} \, = \, 0 \qquad \forall\, i \, \in \, \ZZ_N,
\eeq
in which the LDS equations loose the couplings between
different lattice points.
The fields $\phi_i$ at various lattice points
fluctuate independently on each other, in non-gaussian way
for $\lambda_i \ne 0$.
As a consequence, there is not any wave propagation
and correlations at different points;
%%%%%
\item
{\it Gaussian or Free Theory}, 
\beq
\lambda_i \, = \, 0 \qquad \forall\, i \, \in \, \ZZ_N,
\eeq
in which the action $S[\Phi]$ become quadratic in the fields $\phi_i$
and the LDS equations reduce to 2-step recurrence equations 
--- note the discontinuity.
We recover in this case the standard gaussian theory (Wick theorem);
%%%%%
\item
{\it Fully-interacting Theory,}
\beq
w_{i,i+i} \, \ne \, 0; \quad \lambda_i \, \ne \, 0  
\qquad \forall\, i \, \in \, \ZZ_N ;
\eeq
in which both propagating and anharmonic effects
are fully retained. This is our main concern.
\end{enumerate}

%%%%%%%%%%%%%%%%%%%%%%%%%%%%%%%%%%%%%%%%%%%%%%%%

\subsection{Random Field}

The LDS equations reduce in this case to the following ones.
\bea
\label{LDS_Random}
&+& \quad \nu_i \,\,\,\,\, G\left[\nu_{[-N/2]+1}; \, \cdots; \, \nu_{i-1}; 
\,\, \left\{ - \, 1 \, + \, \nu_i \right\} \,\, ; \,\, \nu_{i+1}; \, \cdots; \, \nu_{[N/2]} \right] \, +
\nonumber\\
&-& \quad k_i \,\,\,\,\, G\left[\nu_{[-N/2]+1}; \, \cdots; \, \nu_{i-1}; 
\,\, \left\{ + \, 1 \, + \, \nu_i\right\} \,\, ; \,\, \nu_{i+1}; \, \cdots, \, \nu_{[N/2]} \right] \, +
\nonumber\\
&-& \quad \lambda_i \,\,\,\,\, G\left[ \nu_{[-N/2]+1}; \, \cdots; \, \nu_{i-1} ; 
\,\, \left\{ + \, 3 \, + \, \nu_i \right\} \,\, ; 
\,\, \nu_{i+1}; \, \cdots, \, \nu_{[N/2]} \right] \, = \, 0 .
\eea
In the non-gaussian (anharmonic) case
\beq
\lambda_i \, \ne \, 0 ,
\eeq
the solution reads:
\bea
\label{solve_Random}
&& G\left[ \nu_{[-N/2]+1} ; \, \cdots ; \, \, \nu_{i-1} ; 
\, \left\{ \nu_i \, \ge \, 3 \right\}; \, \nu_{i+1}; \, \cdots; \, \nu_{[N/2]} \right]
\,\, \mapsto \,\, 
\nonumber \\
&& \qquad + \, \frac{\nu_i \, - \, 3}{\lambda_i} \, G\left[ \nu_{[-N/2]+1}; \, \cdots; \, \nu_{i-1}; \, \left\{ - \, 4 \, + \, \nu_i \right\} ; \, \nu_{i+1}; \cdots; \nu_{[N/2]} \right] \, +
\nonumber \\
&& \qquad - \quad \frac{k_i}{\lambda_i} \quad \, G\left[ \nu_{[-N/2]+1}; \, \cdots; \, \nu_{i-1}; \,  \left\{ - \, 2 \, + \, \nu_i \right\} ; \, \nu_{i+1}; \, \cdots; \, \nu_{[N/2]} \right] .
\eea
The shifted index $\nu_i$ (we have not changed the symbol) has the restriction $\nu_i \, \ge \, 3$, while the other indices have the usual range: 
\beq
\nu_i \, \ge \, 3; 
\qquad 
\nu_j \, \ge \, 0 \qquad \, j \, \ne \, i, \quad j \, \in \, \ZZ_N .
\eeq
The first term on the r.h.s. vanishes for $\nu_i=3$.
It is a lucky circumstance that there is only
one correlator with maximal weight $\mathcal{R}$,
so it has not been necessary to take linear
combinations of different LDS equations.

%%%%%%%%%%%%%%%%%%%%%%%%%%%%%%%%%%%%%

\subsubsection{Random Gaussian Field}

In the Gaussian case the LDS equations reduce to
two-terms equations
\bea
\label{LDS_Random_Gauss}
&+& \nu_i \, G\left[\nu_{[-N/2]+1}; \, \cdots; \, \nu_{i-1}; 
\,\, \left\{ - \, 1 \, + \, \nu_i \right\} \,\, ; \,\, \nu_{i+1}; \, \cdots; \, \nu_{[N/2]} \right] \, +
\nonumber\\
&-& k_i \, G\left[\nu_{[-N/2]+1}; \, \cdots; \, \nu_{i-1}; 
\,\, \left\{ + \, 1 \, + \, \nu_i\right\} \,\, ; \,\, \nu_{i+1}; \, \cdots, \, 
\nu_{[N/2]} \right] 
\, = \, 0 .
\eea
The solutions read:
\bea
&& G\left[\nu_{[-N/2]+1}; \, \cdots; \, \nu_{i-1}; 
\,\, \left\{ \nu_i \, \ge \, 1 \right\} \,\, ; \,\, \nu_{i+1}; \, \cdots, \, 
\nu_{[N/2]} \right] 
\, \to \, 
\nonumber\\
&& \qquad\qquad
\frac{\nu_i - 1}{k_i} 
G\left[\nu_{[-N/2]+1}; \, \cdots; \, \nu_{i-1}; 
\,\, \left\{ - \, 2 \, + \, \nu_i \right\} \,\, ; \,\, \nu_{i+1}; \, \cdots; \, \nu_{[N/2]} \right].
\eea
For $\nu_i=1$ the r.h.s. of the above equation vanishes.

%%%%%%%%%%%%%%%%%%%%%%%%%%%

\subsection{Gaussian Theory}

In this case, the $i$-th (symbolic) LDS equation
contains three correlators,
\beq
G\left( + \, 1 \, + \, \nu_{i-1} \right);
\quad
G\left( + \, 1 \, + \, \nu_{i}   \right);
\quad
G\left( + \, 1 \, + \, \nu_{i+1} \right);
\eeq
having the same (maximal) weight,
\beq
\mathcal{R} \, = \, \sigma \, + \, 1,
\eeq
and one correlator,
\beq
G\left( \, - \, 1 \, + \, \nu_i \right),
\eeq
with a weight smaller by two units,
\beq
\mathcal{R} \, = \, \sigma \, - \, 1 .
\eeq
The LDS equations are then naturally written in this case as
\beq
\label{reduce_Gauss}
        w_{i-1} \, G\left( + \, 1 \, + \, \nu_{i-1} \right)
\, - \,   k_i   \, G\left( + \, 1 \, + \, \nu_{i}   \right) 
\, + \, w_{i+1} \, G\left( + \, 1 \, + \, \nu_{i+1} \right)
\, = \, - \, \nu_i \, G\left( \, - \, 1 \, + \, \nu_i \right);
\eeq
with $i = 1 ,2 , \cdots, N$. 
Note that we have slightly simplified the notation:
\beq
w_i \, \equiv \, w_{i,i+1} .
\eeq
By varying the index equation $i$ in the entire range
$\{1,2,\cdots,N\}$,
we obtain $N$ linearly independent equations,
which can be solved for the $N$ correlators, each one
having one index increased by one unit, namely:
\beq
G\left( + \, 1 \, + \, \nu_{1} \right); 
\quad
G\left( + \, 1 \, + \, \nu_{2} \right); 
\quad
\cdots;
\quad
G\left( + \, 1 \, + \, \nu_{N} \right).
\eeq
The known terms are linear combinations of
all the correlators with one index lowered
by one unit, namely:
\beq
G\left( - \, 1 \, + \, \nu_{1} \right); 
\quad
G\left( - \, 1 \, + \, \nu_{2} \right); 
\quad
\cdots;
\quad
G\left( - \, 1 \, + \, \nu_{N} \right).
\eeq
The solutions are then of the form
\beq
\label{sol_Poisson}
G\left( + \, 1 \, + \, \nu_{i} \right)
\, = \, \sum_{j=1}^N c_{ij}(\nu) \,
\, G\left( - \, 1 \, + \, \nu_j \right);
\qquad i = 1, 2, \cdots, N.
\eeq
The coefficients can be explicitly
calculated by inverting the tri-diagonal matrix
\beq
T \, \equiv \, \left(
\begin{array}{cccccc}
- k_1 & w_1 & 0 & \cdots & 0 & w_N
\\
w_1 & - k_2 & w_2 & 0 & \cdots & 0
\\
0 & w_2 & - k_3 & w_3 & \cdots & 0
\\
0 & \cdots & \cdots & \cdots & \cdots & 0
\\
w_N & 0 & \cdots & 0 & w_{N-1} & - k_N 
\end{array}
\right).
\eeq
We are indeed solving the Poisson equation
with a mass term added, on a one-dimensional 
lattice immersed in a circle, with known terms 
given by lower-weight correlators.

For computer applications, it is natural to have
the l.h.s. of the equations (\ref{sol_Poisson}) to contain
unshifted indices. By means of the shift
\beq
\nu_i \, \to \, \nu_i \, - \, 1,
\eeq
the solution (\ref{sol_Poisson}) is rewritten as
\bea
G\left( \nu_{i} \, \ge \, 1 \right)
&=& \sum_{1 < j < i} d_{ij}(\nu) \,
\, G\left( - \, 1 \, + \, \nu_j; \,  - \, 1 \, + \, \nu_{i} \right) 
\, + \, d_{ii}(\nu) \, G\left( - \, 2 \, + \, \nu_i \right) \, +
\nonumber\\
&& \quad + \, \sum_{i<j<N} d_{ij}(\nu) \,
   G\left( - \, 1 \, + \, \nu_{i} ; \, - \, 1 \, + \, \nu_j \right) ,
\qquad i=1,2,\cdots,N;   
\eea
where we now have the restriction on the indices 
\beq
\nu_i \ge 1; \qquad \nu_j \ge 0, 
\quad j \, \ne \, i \, \in \, \ZZ_N.
\eeq
By using the above set of equations, one is able
to express an arbitrary correlator $G(\nu)$,
with $\mathcal{R} > 0$, as a linear combination,
with known coefficients, of correlators
with $\mathcal{R} - 2$.
By means of an iteration, one can express $G(\nu)$
in terms of correlators with $\mathcal{R} - 4$
and so on.
By iterating this procedure up to the boundary values
for the indices, one can then reduce
any correlator to a combination of correlators
with weight
\beq
\mathcal{R} \, = \, 0, 1,
\eeq
namely the vacuum correlator ($\mathcal{R}=0$)
\beq
G(0,\cdots,0) ,
\eeq
and the $N$ tadpoles ($\mathcal{R}=1$)
\beq
G\left( \cdots, 0, 1_i, 0, \cdots \right), 
\qquad i \, = \, 1, 2, \cdots, N.
\eeq
In a symmetrical theory for $\phi \to - \phi$, 
such as our reference $\lambda \, \phi^4 $ theory,
odd $\mathcal{R}$ correlators vanish, so that
all correlators $G(\nu)$ can be expressed
as a multiple of the vacuum correlator,
\beq
G\left(\nu_1,\cdots,\nu_N\right) 
\, = \, \alpha\left(\nu_1,\cdots,\nu_N\right) \, G(0,\cdots,0),
\eeq
where $\alpha\left(\nu_1,\cdots,\nu_N\right)$ is a known function
depending, in addition to the indices, also on the
couplings $k_i$ and $w_i$ of the model.

%%%%%%%%%%%%%%%%%%%%%%%%%%%%%%%%%%%%%%%%%%%%%

\subsubsection{Examples}

Let us consider in this section a few examples
of reduction of Gaussian correlators.
\begin{enumerate}
\item
{\it Propagator.} By that we mean the $N(N+1)/2$ correlators% 
%%%%%%%%%%
\footnote{We assume that $i$ and $j$ may coincide ($j=i$),
giving rise in this case to the correlator with $\nu_i=2$, $\nu_{k\ne i} = 0$.}
%%%%%%%%%%
with $\mathcal{R} = 2$
\bea
&& G\left(0,\cdots, 0, 1_i, 0, \cdots, 0, 1_j, 0, \cdots,0 \right),
\qquad i \, < \, j ,
\nonumber\\
&& G\left(0, \cdots, 0, 2_i, 0, \cdots, 0 \right),
\qquad i \in \ZZ_N,
\eea
which can be expressed in terms of the vacuum correlator only,
\beq
G(0,\cdots,0),
\eeq
in just one step;
%%%%%%%%%%%%%%%%%
\item
{\it Explicit Solution at $N=2$.}
Let us discuss the explicit solutions of LDS equations in the free 
(or gaussian) case for $N=2$.
By solving the two symbolic equations (\ref{N2gaussian})
with respect to the two higher-weight corrrelators,
we obtain, after trivial shifts of the indices:
\bea
G\left(\nu_1 \ge 1; \, \nu_2 \ge 0 \right)
&=& \frac{k_2}{k_1 k_2 - w^2}(\nu_1 - 1)
G\left(\nu_1 - 2; \nu_2 \right) \, +
\nonumber\\
&+& \frac{w }{k_1 k_2 - w^2} \, \nu_2 \,
G\left(\nu_1 - 1; \nu_2 - 1 \right);
\nonumber\\
G\left(\nu_1 \ge 0; \, \nu_2 \ge 1 \right)
&=& \frac{k_1  }{k_1 k_2 - w^2} (\nu_2 - 1)
G\left(\nu_1 ; \nu_2 - 2 \right) \, +
\nonumber\\
&+& \frac{w }{k_1 k_2 - w^2} \, \nu_1 \,
G\left(\nu_1 - 1; \nu_2 - 1 \right).
\eea
Let us make a few comments.
\begin{enumerate}
\item
The singularity for
\beq
w \, \to \, \sqrt{k_1 k_2}^{\,\,-}
\eeq
originates from a massless zero mode;
%%%%%
\item
By replacing the numerical values
\beq
\nu_1 \, = \, \nu_2 \, = \, 1,
\eeq
one obtains the reduction of the propagator
to the vacuum amplitude
\beq
G(1,1) \, = \, \frac{w}{k_1 k_2 - w^2} G(0,0);
\eeq
%%%%%
\item
For $w\to 0$, the couplings between
different indices disappear.
\end{enumerate}
\end{enumerate}

%%%%%%%%%%%%%%%%%%%%%%%%%%%%%%%%%%%

\subsection{Fully-interacting Case}

In the fully-interacting case,
\beq
w_{i,i+1} \, \ne \, 0; 
\quad
\lambda_i \, \ne \, 0; \qquad i \, \in \, \ZZ_N;
\eeq
we solve eq.(\ref{LDS}), as explained, in terms of the last amplitude on the rhs of (\ref{LDS}).
It is then natural to make the shift on the $i$-th index only
\beq
\nu_i \, \to \, \nu_i \, - \, 3;
\qquad
\nu_j \, \to \, \nu_j , \quad j \, \ne \, i \, \in \, \ZZ_N .
\eeq
The symbolic solution of the LDS equation reads:
\bea
\label{solve}
&& G\left[ \nu_{[-N/2]+1} ; \, \cdots ; \, \, \nu_{i-1} ; 
\, \left\{ \nu_i \, \ge \, 3 \right\}; \, \nu_{i+1}; \, \cdots; \, \nu_{[N/2]} \right]
\,\, \mapsto \,\, 
\\
&& \qquad + \, \frac{\nu_i \, - \, 3}{\lambda_i} \, G\left[ \nu_{[-N/2]+1}; \, \cdots; \, \nu_{i-1}; \, \left\{ - \, 4 \, + \, \nu_i \right\} ; \, \nu_{i+1}; \cdots; \nu_{[N/2]} \right] \, +
\nonumber \\
&& \qquad - \quad \, \frac{k_i}{\lambda_i} \quad G\left[ \nu_{[-N/2]+1}; \, \cdots; \, \nu_{i-1}; \,  \left\{ - \, 2 \, + \, \nu_i \right\} ; \, \nu_{i+1}; \, \cdots; \, \nu_{[N/2]} \right] \, + 
\nonumber\\
&& \qquad + \, \frac{w_{i-1,i}}{\lambda_i} \, G\left[ \nu_{[-N/2]+1}; \, \cdots; \, \nu_{i-2}; \, \left\{ + \, 1 \, + \, \nu_{i-1} \right\}; 
\, \left\{ - \, 3 \, + \, \nu_i \right\}; \, \nu_{i+1}; \, \cdots; \, \nu_{[N/2]} \right] \, +
\nonumber\\
&& \qquad + \, \frac{w_{i,i+1}}{\lambda_i} \, G\left[ \nu_{[-N/2]+1}; \, \cdots; \, \nu_{i-1}; \, \left\{ - \, 3 \, + \, \nu_i \right\} ; \, \left\{ + \, 1 + \nu_{i+1} \right\} ; \, \nu_{i+2}; \, \cdots; \, \nu_{[N/2]} \right] .
\nonumber
\eea
The shifted index $\nu_i$ (we have not changed the symbol) has the restriction $\nu_i \, \ge \, 3$, while the other indices have the usual range: 
\beq
\nu_i \, \ge \, 3; 
\qquad 
\nu_j \, \ge \, 0 \quad \mathrm{for} \, j \, \ne \, i .
\eeq
If more than one correlator with maximal weight $\mathcal{R}$
had occurred in the LDS equations, it would have been necessary 
to take linear combinations of them.
Let us comment upon the various terms on the r.h.s.
of the reduction equation.
\begin{enumerate}
\item[I {\it term}:] 
It reduces the current occupation number $\nu_i$
by {\it four} units;
\beq
\nu_i \, \to \, \nu_i \, - \, 4. 
\eeq
Since it does not touch
any index $\nu_j$ with $j \ne i$, it is 
"diagonal" in index space.
It is the only term with a coefficient
depending on $\nu_i$
and it vanishes for $\nu_i = 3$;
%%%%%
\item[II {\it term}:]
It reduces the current occupation number $\nu_i$
by {\it two} units:
\beq
\nu_i \, \to \, \nu_i \, - \, 2. 
\eeq
It is also a diagonal term, in the sense specified
at the previous point;
%%%%%
\item[III {\it term}:]
It decreases the current index $\nu_i$ by {\it three} units
and at the same time it increases by {\it one} unit 
the index $\nu_{i-1}$ to the left of $\nu_i$,
\bea
\nu_i &\to& \nu_i \, - \, 3 ;
\nonumber\\
\nu_{i-1} &\to& \nu_{i-1} \, + \, 1 \quad \mathrm{(LM)}.
\eea
It is non diagonal in index space
and will be called a {\it Left Mover} (LM),
as it transfers part of the value of $\nu_i$ to $\nu_{i-1}$;
%%%%%
\item[IV {\it term:}]
It is the last one and it acts analogously to the previous one,
\bea
\nu_i &\to& \nu_i \, - \, 3 ;
\nonumber\\
\nu_{i+1} &\to& \nu_{i+1} \, + \, 1 \quad \mathrm{(RM)}.
\eea
It will be called a {\it Right Mover} (RM).
\end{enumerate}
The fundamental point is that, by repeatedly using the above equation, for different indices $i$ and for different values of the 
occupation numbers $\nu_i$,
one can reduce an arbitrary correlator $G(\nu)$ to
a (finite) linear combination, with known coefficients, of correlators having all indices less than or equal to two:
\beq
G(\nu) \, = \, 
\sum_{0 \le \mu_j \le 2; \, j \in \ZZ_N} c_\nu (\mu) \, P(\mu) ,
\eeq
where we have defined the multi-index
\beq
\mu \, \equiv \,  
\left( 
\mu_{[-N/2]+1}, \, \mu_{[-N/2]+2}, \, 
\cdots, \, \mu_0, \, \mu_1, \, \mu_2, \, 
\cdots, \, \mu_{[N/2]-1}, \, \mu_{[N/2]} 
\right) .
\eeq
We call the correlators $G(\mu)$ which appear on the right hand sides of the complete reductions {\it primitive correlators} 
and we have denoted them by $P(\mu)$,
\beq
G(\mu) \,\, \to \,\, P(\mu); \qquad 0 \, \le \, \mu_i \, \le \, 2;
\quad i \, \in \, \ZZ_N.
\eeq
The irreducible correlator with the highest weight has all its
indices equal to two and
\beq
\mathcal{R}\left[G(2,2,\cdots,2)\right] \, = \, 2 N ,
\eeq
where $N$ is the lattice size.
Let us make a few remarks:
\begin{enumerate}
\item
Many different reduction paths are possible in index ($\nu$) space.
Since the final result is well defined,
it follows by consistency that they are all equivalent;
\item
Each time the equation above is used an over-all factor
\beq
\frac{1}{\lambda_i}
\eeq
enters the decomposition.
Large (but finite) inverse powers of the couplings $\lambda_i$ then enter the
decomposition of any correlator;
\item
In general, correlators with many large indices, 
i.e. with a high weight $\mathcal{R} \gg 1$, involve a massive 
reduction process before primitive correlators are finally reached.
Let us remark however that
the number of reduction steps is finite in any case, because at each step
we generate correlators with weight reduced at least by two units.
However, it may happen that also correlators with relatively small weights may undergo a long reduction
process, with a large number of primitive correlators appearing in the final formula.
We will see explicit examples of these phenomena in a moment.
\end{enumerate}

%%%%%%%%%%%%%%%%%%%%

\subsubsection{Examples}

In this section we present some simple explicit examples 
of solutions of LDS equations.
\begin{enumerate}
\item
{\it Long Reduction Chain}.
\label{ex_long_chain}
A long reduction chain is obtained
by considering for example the correlator
\beq
G\left[ 
\nu_{[-N/2]+1} \, = \, 2 ; \, \cdots ; \, \nu_{-1} \, = \, 2 ; \, \nu_0 \, = \, 0  ; \, \nu_1 ; \, = \, 3 
; \, \nu_2 ; \, = \, 2\, \cdots ; \, \nu_{[N/2]} \, = \, 2 \right],
\eeq
having one index equal to zero, the adjacent one 
to the right equal to three, 
with all the other indices equal to two.
Since only the first index is greater than two,
\beq
\nu_1 \, = \, 3,
\eeq
the first reduction step necessarily involves eq.(\ref{solve}) for $i=1$.
Let analyze in turn the generated terms on the r.h.s.:
\begin{enumerate}
\item[I {\it term:}]
It vanishes, as already noted;
\item[II {\it term:}]
It produces a primitive corrrelator, 
\beq
- \frac{k_1}{\lambda_1} \,
P\left[ \nu_{[-N/2]+1} \, = \, 2 ; \, \cdots ; \, \nu_{-1} \, = \, 2 ; \, \nu_0 \, = \, 0 ; \, \nu_1 ; \, = \, 1 
; \, \nu_2 ; \, = \, 2\, \cdots ; \, \nu_{[N/2]} \, = \, 2 \right],
\eeq
because it just shifts 
\beq
\nu_1 \, = \, 3  \,\, \to \,\, 1 ,
\eeq
by keeping the remaining indices unchanged;
\item[III {\it term,}]
the $LM$: it also produces the primitive correlator
\beq
\frac{w_{0,1}}{\lambda_1} \,
P\left[ \nu_{[-N/2]+1} \, = \, 2 ; \, \cdots ; \, \nu_{-1} \, = \, 2 ; \, \nu_0 \, = \, 1 ; \, \nu_1 ; \, = \, 0 
; \, \nu_2 ; \, = \, 2\, \cdots ; \, \nu_{[N/2]} \, = \, 2 \right]. 
\eeq
It reduces to zero the index equal to three,
\beq
\nu_1 \, = \, 3 \,\, \to \,\, 0,
\eeq
and at the same time increases by one unit the index
to the left 
\beq
\nu_0 \, = \, 0 \,\, \to \,\, 1.
\eeq
Index increase is "potentially dangerous",
in the sense that it produces in general reducible
correlators, but this is not the case because
we have chosen an initial small value of $\nu_0=0$.
%%%%%
\item[IV {\it term,}] the RM: it produces a reducible correlator,
\beq
\frac{w_{1,2}}{\lambda_1} \,
G\left[ \nu_{[-N/2]+1} \, = \, 2 ; \, \cdots ; \, \nu_{-1} \, = \, 2 ; 
\, \nu_0 \, = \, 0 ; \, \nu_1 ; \, = \, 0 
; \, \nu_2 \, = \, 3 ; \, \cdots ; \, \nu_{[N/2]} \, = \, 2 \right]. 
\eeq
That is because, as in the previous case, the current index $\nu_1$
has been reduced below the critical value $\nu_{cr}=3$,
\beq
\nu_1 \, = \, 3 \,\, \to \,\, 0,
\eeq
but the index to the right, $\nu_2$, was initially set to two ,
so by increasing it by one unit we go above the critical value:
\beq
\nu_2 \, = \, 2 \,\, \to \,\, 3.
\eeq  
\end{enumerate}
%
%%%%%%%%%%%%%%%%%%% FIGURA %%%%%%%%%%%%%%%%%%%%
\begin{figure}[ht]
\begin{center}
\includegraphics[width=0.5\textwidth]{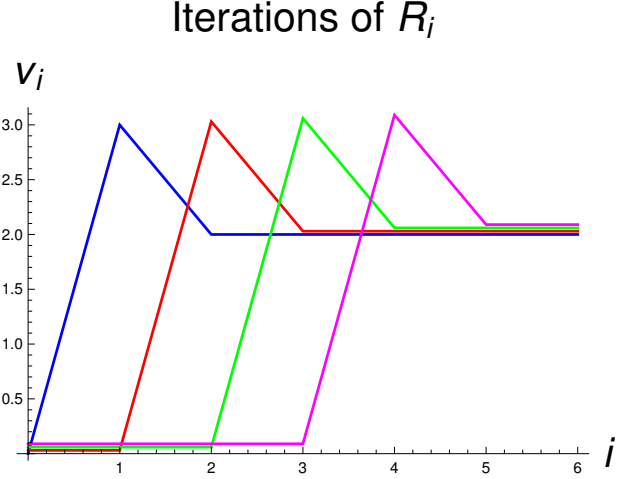}
\footnotesize
\caption{
\label{fig_pwave}
\it Effects of the operators $R_i$ for $i=1,2,3$,
in order, on the string of indices
$(0,3,2,2,2,2,2) \to (0,0,3,2,2,2,2) \to (0,0,0,3,2,2,2) \to (0,0,0,0,3,2,2)$.
The peak corresponding to $\nu_i=3$ moves to the right
by leaving behind him trailing zeros.
}
\end{center}
\end{figure}
%%%%%%%%%% FINE FIGURA %%%%%%%%%%%%%%%%%%%%%%%%
%
In summary, with the first reduction step,
we have produced three (non zero) correlators:
two primitive correlators and one reducible correlator.
The second reduction step then only involves
\beq
\frac{w_{1,2}}{\lambda_1} \,
G\left( \cdots ; \, \nu_{-1} \, = \, 2 \, ; \, \nu_0 \, = \, 0 \, ; 
\, \nu_1 ; \, = \, 0 \, ; \, \nu_2 \, = \, 3 \, 
; \, \nu_3 = 2, \, \cdots \right). 
\eeq
Since $\nu_2=3$, one needs the LDS equation with $i=2$.
The analysis is similar to the previous one:
only the RM produces a reducible correlator, because
it makes the transition
\beq
\left(\nu_2,\nu_3\right) \, = \, (3,2) \,\, \to \,\, (0,3).
\eeq
At this point, the mechanism should be clear:
when an index equal to two is to the right of an index
equal to three, then the RM increases it, producing
a reducible correlator.
By considering the RM only (see fig.\ref{fig_pwave}),
\beq
\left(\nu_1,\nu_2,\nu_3,\nu_4\right) \, = \, 
(3,2,2,2) \,\, \to \,\, (0,3,2,2) \,\, \to \,\, (0,0,3,2) \,\, \to \cdots .
\eeq
Since to the right of the index equal to three
there are only indices equal to two, by iterating, one finds that 
the index ``three'' propagates through all the lattice,
by leaving zeroes behind him.
A sort of "wave peak" propagates along the whole lattice,
resembling those games with sequences of Lego elements
falling progressively on each other.
This example can be generalized, in the sense that
basically nothing changes, by taking
\beq
\nu_0 \, = \, 0,1 \,\,\, \mathrm{and} \,\,\, \nu_1 \, = \, 3,4,5.
\eeq
It is remarkable that we are not solving some wave equation,
but we are {\it building up} the r.h.s. of an equation.
The complexity inherent {\it in solving} a wave equation 
is transferred, in some sense, to the equation itself.
The variables $w_{i,i+1}$ seem to represent a
current $J_{i,j}$ coupling to a pair of neighboring fields
$\phi_i$ and $\phi_j$ --- 
a generalization of the well-known Schwinger current $J_i$
coupled to the local field $\phi_i$.
%%%%%
\item
{\it Propagation on both directions of the lattice.} 
Let us consider a correlator with
all its indices equal to two, with the exception
of one index equal to three, such as
\beq
G\left( \nu_{[-N/2]+1} \, = \, 2 ; \, \cdots ; \, \nu_{-1} \, = \, 2 ; \, \nu_0 \, = \, 2  ; \, \nu_1 ; \, = \, 3 
; \, \nu_2 ; \, = \, 2\, \cdots ; \, \nu_{[N/2]} \, = \, 2 \right),
\eeq
In this case, more symmetrical than the previous one, 
both LM and RM generate recursions chains along all
the lattice;
%%%%%
\item
{\it Cutting a Long Reduction Chain}.
The reduction of example \ref{ex_long_chain} propagates 
an index equal to three on the whole lattice, 
because in any reduction step, an index equal to three only finds
indices equal to two to its right.
To "break" such a propagation, one just needs
to insert an index equal to zero or equal
to one "along the way" of the path of the index
equal to three,
\beq
G\left( \cdots ; 
\, \nu_0 \, = \, 0  ; \, \nu_1 ; \, = \, 3 ; 
\, \nu_2 ; \, = \, 2\, \cdots ; \, \nu_{k-1} \, = \, 2 ;
\, \nu_k \, < \, 2 ;
\, \cdots \right),
\eeq
where $k$ is some selected lattice point.
The above correlator has a reduction chain
to primitive correlators which stops
at $k$ because
\beq
\nu_k \, + \, 1 \, < \, 3 .
\eeq
\end{enumerate}

%%%%%%%%%%%%%%%

\subsection{Generalizations}

By a similar analysis, in a model with
a cubic interaction, such as for example 
a $g \, \phi^3$ theory,
one would be able to shift any index $\nu_i$
to
\beq
0 \, \le \, \nu_i \, \le \, 1; \qquad i \, \in \, \ZZ_N.
\eeq
The number of primitive correlators
on a lattice of size $N$ would then be
\beq
\mathcal{O}\left(2^N\right)
\eeq
again with the power of the continuum  
in the strong limit $N \to \infty$.
With a general interaction Lagrangian with
maximal anharmonicity $m_{\mathrm{anh}}$, by solving
with respect to the highest weight term,
having the shifted index
\beq
\nu_i \, + \, m_{\mathrm{anh}} \, - \, 1,
\eeq
one would be able to shift $\nu_i$ inside the range
\beq
0 \, \le \, \nu_i \, \le \, m_{\mathrm{anh}} \, - \, 2.
\eeq

%%%%%%%%%%%%%%%%%%%%%%%%%%%%%%%%%%%%%%%%%%%%%%

\section{Operator Algebra of LDS Equations}
\label{sect9}

In operator language, the solution of the $i^{\mathrm{th}}$ symbolic
LDS equation can be written as:
\bea
\label{solveO}
&&G\left(\nu_{[-N/2]+1},\, \cdots, \, \nu_{i-1},\, \, \nu_i \ge 3, \, \nu_{i+1}, \, \cdots, \, \nu_{[N/2]} \right) 
\mapsto 
\nonumber\\
&& \quad \quad\quad \quad\quad \quad\quad \quad\quad \quad\quad
O_i \,\, G\left(\nu_{[-N/2]+1},\, \cdots, \, \nu_{i-1},\, \, \nu_i , \, \nu_{i+1}, \, \cdots, \, \nu_{[N/2]} \right) , 
\eea
where the operator $O_i$ is defined as the following sum of linear operators:
\beq
O_i \, \equiv \, N_i \, + \, D_i \, + \, L_i \, + \, R_i , 
\eeq
having the expressions
\bea
N_i &\equiv& + \, \frac{1}{\lambda_i} \,\, i^{-4} \, \left( \hat{\nu}_i - 3 \right) ;
\\
D_i &\equiv& - \, \frac{k_i}{\lambda_i} \,\, i^{-2} ; 
\\
L_i &\equiv& + \, \frac{w_{i-1,i}}{\lambda_i} \,\, (i-1)^+ \, i^{-3} ;
\\
R_i &\equiv& + \, \frac{w_{i,i+1}}{\lambda_i} \,\, i^{-3} \, (i+1)^+ .
\eea
The operators on the r.h.s. of the above equations, in turn,
are defined as follows:
\begin{itemize}
\item
$\hat{\nu}_i$ is the $i^{\mathrm{th}}$ state 
occupation-number operator i.e., when applied to a correlator $G(\nu)$, 
it returns the occupation number of the $i^{\mathrm{th}}$ state, namely $\nu_i$,
\bea
&& 
\hat{\nu}_i \, G\left(\nu_{[-N/2]+1},\, \cdots, \, \nu_{i-1},\, \, \nu_i, \, \nu_{i+1}, \, \cdots, \, \nu_{[N/2]} \right) 
\, \equiv \, 
\nonumber\\
&& \qquad\qquad\qquad\qquad
\equiv \, \nu_i \,
G\left(\nu_{[-N/2]+1},\, \cdots, \, \nu_{i-1},\, \nu_i, \, \nu_{i+1}, \, \cdots, \, \nu_{[N/2]} \right).
\eea
In more compact notation, by writing only
the relevant indices,
\beq
\hat{\nu}_i \, G\left(\nu_i\right) \, = \, \nu_i \, G\left(\nu_i\right),
\qquad i \, \in \, \ZZ_N;
\eeq
%%%%%
\item
$i^{\pm}$ is the raising/lowering
operator of the occupation number $\nu_i$ for the
$i^{\mathrm{th}}$ state, i.e. the operator $i^{\pm}$
raises/lowers the index $\nu_i$ by one unit:
\bea
&& i^{\pm} \, G\left(\nu_{[-N/2]+1}; \, \cdots; \, \nu_{i-1}; \, \, \nu_i; \, \nu_{i+1}, \, \cdots, \, \nu_{[N/2]} \right)
\, \equiv \,
\nonumber\\
&& \qquad\qquad \qquad \equiv \, 
G\left(\nu_{[-N/2]+1}; \, \cdots;  \, \nu_{i-1}; \, \nu_i  \pm \, 1 ; \, \nu_{i+1}; \, \cdots; \, \nu_{[N/2]} \right),
\eea
with $i \, \in \, \ZZ_N$.
More briefly,
\beq
i^{\pm} \, G\left(\nu_i\right)
\, \equiv \,
G\left( \nu_i \, \pm \, 1 \right),
\qquad i \, \in \, \ZZ_N .
\eeq
Even more briefly,
\beq
i^{\pm} \, : \, \nu_i \, \mapsto \, \nu_i \, \pm \, 1,
\qquad i \, \in \, \ZZ_N .
\eeq
Powers are defined in straightforward way:
\beq
i^{++} \, \equiv \, i^+ \circ i^+ ,   
\eeq
so that
\beq
i^{++}: \, \nu_i \, \mapsto \, \nu_i \, + \, 2 ,  
\eeq
and so on ($i^{\pm} = i^{\pm 1}$, $i^{++} = i^{+2}$, etc.).
The zero power is the identity operator,
\beq
i^0 \, \equiv \, \mathrm{id},
\eeq
with
\beq
\mathrm{id} \, G(\nu) \, \equiv \, G(\nu).
\eeq
\end{itemize}
The main properties of these operators are given in the
following.
\begin{itemize}
\item
$i^{s}$ and $j^{t}$ commute for any lattice indices
$i$ and $j$ and any powers $s$ and $t$:
\beq
\label{ijdiff}
\left[ \, i^s, \, j^t \, \right] \,= \, 0 ;
\qquad i, \, j \, \in \, \ZZ_N; \quad s,t \, \in \, \ZZ;
\eeq
%%%%%
\item
$\hat{\nu}_i$ and $i^s$, $s \ne 0$, do not commute with each other,
as well known from elementary quantum mechanics:
\beq
\label{inui}
\left[ \hat{\nu}_i, \, i^s\right] \, = \, s \, i^s,
\qquad i \, \in \, \ZZ_N; \quad s \, \in \, \ZZ.
\eeq
For $s=\pm1$, for example:
\bea
\left[ \hat{\nu}_i, \, i^+\right] &=& + \, i^+ ;
\nonumber\\
\left[ \hat{\nu}_i, \, i^-\right] &=& - \, i^- ;
\qquad i \, \in \, \ZZ_N.
\eea
Commutativity holds instead for different indices
\beq
\qquad
\left[ \hat{\nu}_i, \, j^s \right] \, = \, 0;
\qquad i \, \ne \, j \, \in \, \ZZ_N, \quad s \, \in \, \ZZ.
\eeq
\end{itemize}

%%%%%%%%%%%%%%%%%%%%%%%%%%%%%%%%%%%%%

\subsection{Examples}

Let's then see a few examples of application of the
above rules.
\begin{enumerate}
%%%%%
\item
It follows from eq.(\ref{ijdiff}) that the operators $D_i$, $L_i$ and $R_i$ 
entering $O_i$ commute with each other even for different indices:
\bea
\left[ D_i, \, D_j \right] &=& 
\left[ L_i, \, L_j \right] \, = \, \left[ R_i, \, R_j \right] \, = \, 0;
\nonumber\\
\left[ D_i, \, L_j \right] &=& 
\left[ D_i, \, R_j \right] \, = \, \left[ L_i, \, R_j \right] \, = \, 0;
\qquad i, j \, \in \, \ZZ_N;
\eea
%%%%%
\item
According to eq.(\ref{inui}), the operator $N_i$ can also be written 
with the occupation-number operator $\hat{\nu}_i$
acting after $i^{-4}$:
\beq
N_i \, = \, \frac{1}{\lambda_i}  
\left( \hat{\nu}_i \, + \, 1  \right) \, i^{-4} ; 
\eeq
%%%%%
\item
The square of the operator $N_i$ reads:
\beq
N_i^2 \, = \, \frac{1}{\lambda_i^2} \, i^{-8} \left( \hat{\nu}_i - 7 \right)\left( \hat{\nu}_i - 3 \right),
\qquad \nu_i \, \ge \, 7.
\eeq
It vanishes for $\nu_i=7$;
%%%%%
\item
The commutation rules of $N_i$ with the other 
operators entering $O_i$ read:
\bea
\label{comm_to_use}
\left[N_i, D_i\right]
&=& + \, 2 \, \frac{\left(-k_i\right)}{\lambda_i^2} \, i^{-6};
\nonumber\\
\left[N_i, L_i\right]
&=& - \, 3 \, \frac{w_{i-1,i}}{\lambda_i^2} \, (i-1)^+ \, i^{-7};
\nonumber\\
\left[N_i, R_i\right]
&=& - \, 3 \, \frac{w_{i,i+1}}{\lambda_i^2} \,  i^{-7} (i+1)^+ ;
\eea
%%%%%
\item
Since $L_i$ commutes with $R_i$,
\beq
\left(L_i \, + \, R_i\right)^k
\, = \, \frac{ i^{-3k} }{ \lambda_i^k } 
\sum_{s=0}^k 
\left(
\begin{array}{c}
k 
\\
s
\end{array}
\right)
w_{i-1,i}^s \, w_{i,i+1}^{k-s} \, 
(i-1)^s \, (i+1)^{k-s};
\qquad k \, = \, 1, 2, 3, \cdots :
\eeq
This formula can be used, for instance, to 
understand the form of the
reduction of a correlator having all indices
equal to zero, except one index equal to a large value,
let's say $\nu_1 \gg 1$:
\beq
G(0,\cdots; \nu_0=0 \, ; \nu_1 \gg 1 \, ; \nu_2 = 0 \, ; \cdots,0).
\eeq
The indices $\nu_0$ and $\nu_2$ increase their value
from zero up to $[\nu_1/3]$, i.e. up to the integer
part of one-third of the initial large index value.
If $\nu_0 \ge 3$ or $\nu_2 \ge 3$, one has then to apply
$O_0$ or $O_2$ respectively, which in turn increase
neighboring indices of $i=0$ and $i=2$
up to the integer part of one-third of their values.
In total, some sort of diffusive behavior manifests
itself, with the initial peak value of $\nu_0$
spreading out, with "some loss", in a 
neighborhood of the lattice point $i=0$; 
%%%%%
\item
Since also $D_i$ commutes with $L_i$ and $R_i$,
the simple generalization of the above formula holds:
\beq
\left(D_i \, + \, L_i \, + \, R_i\right)^k
\, = \, \frac{ 1 }{ \lambda_i^k } 
\sum_{h+j+l=k}
\frac{k!}{h! \, j! \, l!}
\left(-k_i\right)^h \, w_{i-1,i}^j \, w_{i,i+1}^l \, 
i^{-2h-3(j+l)}(i-1)^j \, (i+1)^l;
\eeq
where the dummy indices are non-negative,
$h,j,l \, \ge \, 0$, and $k \, = \, 1, 2, 3, \cdots$;
%%%%%
\item
The calculation of the general power of the operator
$O_i$ is not easy because the operator $N_i$
does not commute with $D_i$, $L_i$ and $R_i$.
By defining
\beq
X_i \, \equiv \, D_i \, + \, L_i \, + \, R_i,
\eeq
the most explicit formula we could find is simply
\bea
O_i^k \, = \, \left( N_i \, + \, X_i\right)^k
&=& N_i^k \, + \, \sum_{a+b=k-1} N_i^a \, X_i \, N_i^b
\, + \, \sum_{a+b+c=k-2} N_i^a \, X_i \, N_i^b \, X_i \, N_i^c 
\, + \, \cdots 
\nonumber\\
&& \cdots \, + \, X_i^k ;
\eea
with $a,b,c,\cdots \ge 0$.
One has then to use the commutation rules in eq.(\ref{comm_to_use});
%%%%%
\item
As regards the composition of operators entering the $O_i$'s
for different indices, for $\nu_i, \, \nu_{i+1} \, \ge \, 3$,
the following formula holds:
\beq
L_{i+1} \, R_i \, = \, R_i \, L_{i+1} \, = \, 
\frac{w_{i,i+1}^2}{\lambda_i \, \lambda_{i+1}}
i^{\, --} \, (i+1)^{--};
\eeq
%%%%%
\item
Concerning the product of RM operators with
adjacent indices, the following expression is found: 
\beq
\prod_{s=0}^{k} R_{i+s}
\, = \, \left( \prod_{s=0}^{k} \frac{w_{i+s,i+s+1}}{\lambda_{i+s}} \right)
i^{-3} \left[ \prod_{s=1}^{k} (i+s)^{-2} \right]
(i+k+1)^{+1} .
\eeq
This formula simply accounts for the form 
of the long reduction chain considered in
an example in a previous section;
%%%%%
\item
Since the reduction of any correlator
to primitive correlators
is well defined, i.e. it does not depend
on the reduction path, by consistency
it must be true that
\beq
\left[ O_i,O_j \right] \, = \, 0; \qquad i,j \, \in \, \ZZ_N.
\eeq
The above result can also be checked by explicit
calculation;
%%%%%
\item
The individual operators entering $O_i$
commute with the operators entering $O_j$
if the indices differ at least by two units, i.e.
$|i-j| \ge 2$.
\end{enumerate}

%%%%%%%%%%%%%%%%%%%%%%%%%%%%%%%%%%%%%%%%%%%%%%%%%%%%%%%%%%%

\section{Geometry of LDS equations}
\label{sect10}

The rules to solve the LDS equations can be formulated
in the following geometric setting, giving rise
to a discrete geometry, of high dimension for $N \gg 1$.

\begin{figure}[ht]
\begin{center}
\includegraphics[width=0.5\textwidth]{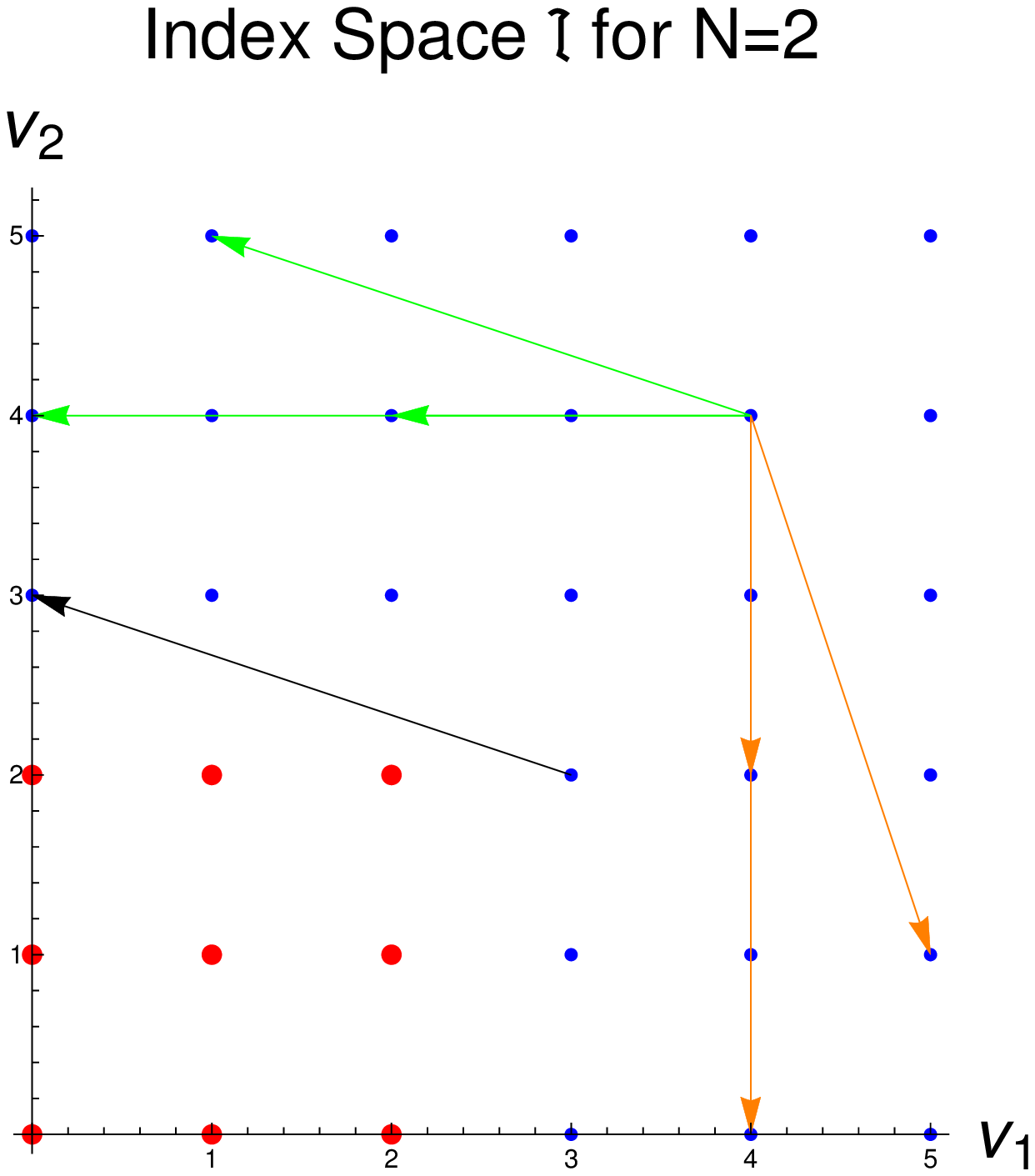}
\footnotesize
\caption{
\label{fig_pindex2}
\it Index space $\mathcal{I}$ for lattice size $N=2$.
The big red dots form the hypercube $\mathrm{Hc}$
of the set of the primitive correlators $\left\{ P(\nu)\right\}$, 
while the small blue dots represent kinematically-independent 
correlators $G(\nu)$'s.
The vertex of $\mathrm{Hc}$ outside the coordinate
lines $\nu_1=0$ and $\nu_2=0$ is the point
$(2,2)$, corresponding to the highest-weight 
primitive correlator.
The black vector, taking the point $(\nu_1,\nu_2)=(3,2) \to (0,3)$,
represent the Right-Mover Operator $R_i$ or the
Left Mover Operator $L_i$ for $i=1$;
Since in this case there are only two lattice points,
the Right Mover coincides indeed with the Left Mover.
The two horizontal vectors together with the tilted one, 
applied to the point $(4,4)$, all pointing to the left and 
plotted in green, represent the operators 
$N_1$, $D_1$ and $L_1=R_1$ respectively, entering $O_1$, 
while the vertical vectors and the tilted one,
still applied in $(4,4)$ and pointing down, in orange, 
represent the three operators in $O_2$ (see text).
}
\end{center}
\end{figure}

\begin{figure}[ht]
\begin{center}
\includegraphics[width=0.5\textwidth]{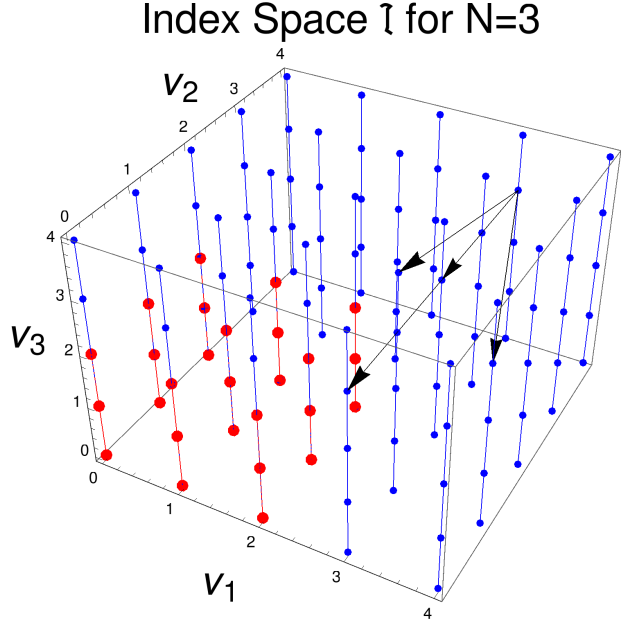}
\footnotesize
\caption{
\label{fig_psp3}
\it Index space $\mathcal{I}$ for $N=3$.
The big red dots form the hypercube $\mathrm{Hc}$
of the set of the primitive correlators $\left\{ P(\nu)\right\}$, 
while the small blue dots represent kinematically-independent 
correlators $G(\nu)$'s.
The vertex of $\mathrm{Hc}$ outside the coordinate
planes $\nu_i=0$, $i=1,2,3$, is the point
$(2,2,2)$, corresponding to the highest-weight 
primitive correlator.
To render the visualization of the 3-dimensional lattice easier,
vertical lines connecting the lattice points have been drawn.
The black vectors, applied to the point 
$\left(\nu_1,\nu_2,\nu_3\right)=(3,4,3)$, 
represent the operators $N_2$, $D_2$, $L_2$ and
$R_2$ entering $O_2$ (see text).
}
\end{center}
\end{figure}

We introduce the index space or $\nu$-space 
\beq
\mathcal{I} \, \equiv \, \NN^N \, \equiv \, \left\{ (\nu_1, \nu_2,\cdots, \nu_N); 
\,\,\, \nu_i \, = \, 0, 1, 2, 3, \cdots; \,\,\, 
i \, = \, 1, 2, \cdots, N \right\}.
\eeq
The space $\mathcal{I}$ is an $N$-dimensional lattice space,
infinite on one side of each of its $N$ dimensions
(see fig.$\,$\ref{fig_pindex2} for $N=2$ and fig.$\,$\ref{fig_psp3}
for $N=3$). 
An arbitrary correlator $G(\nu)$ is represented in this space
by a point $Q$, 
\beq
G(\nu)
\quad
\leftrightarrow
\quad
Q \, \equiv \, \left(\nu_1, \nu_2,\cdots, \nu_N\right) .
\eeq
Since one can make linear combinations
of the correlators $G(\nu)$'s, $\mathcal{I}$
has the structure of a  vector space.
%
%%%%%%%%%%%%%%%%%%% FIGURA %%%%%%%%%%%%%%%%%%%%
%%%%%%%%%% FINE FIGURA %%%%%%%%%%%%%%%%%%%%%%%%
%
%
%%%%%%%%%%%%%%%%%%% FIGURA %%%%%%%%%%%%%%%%%%%%
%%%%%%%%%% FINE FIGURA %%%%%%%%%%%%%%%%%%%%%%%%
%
\noindent
A norm for a vector $Q\in \mathcal{I}$ can be defined as
\beq
\left\| Q \right\|_\infty 
\, \equiv \,
\max_{i=1,2,\cdots,N} \left| \nu_i \right| .
\eeq
As well known, the above norm induces in $\mathcal{I}$
the following homogeneous and translation-invariant metric:
\beq
d_\infty\left(Q,Q'\right) \, \equiv \, \left\| Q' \, - \, Q \right\|_\infty , 
\eeq
where
\beq
Q' \, \equiv \, \left(\nu_1' \, ; \nu_2' \, ; \cdots \, ; \nu_N' \right).
\eeq
%
%%%%%%%%%%%%%%%%%%% FIGURA %%%%%%%%%%%%%%%%%%%%
\begin{figure}[ht]
\begin{center}
\includegraphics[width=0.5\textwidth]{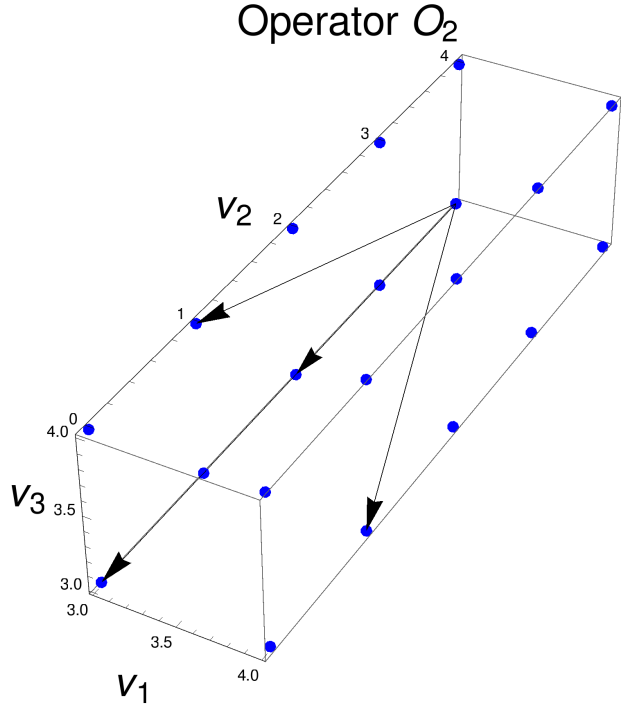}
\footnotesize
\caption{
\label{fig_plotO2}
\it Portion of the previous 3-dimensional plot
containing the vectors associated 
to the operators entering $O_2$, applied to the point 
$\left(\nu_1,\nu_2,\nu_3\right)=(3,4,3)$.
}
\end{center}
\end{figure}
%%%%%%%%%% FINE FIGURA %%%%%%%%%%%%%%%%%%%%%%%%
%
The set of all primitive correlators 
is represented in the space $\mathcal{I}$ by a hypercube $\mathrm{Hc}$ 
of edge size equal to two, with one vertex at the origin:
\beq
\left\{ P(\nu) \right\} 
\,\, = \,\, 
\mathrm{Hc} 
\, \equiv \, 
\left\{ \left( \nu_1, \nu_2,\cdots, \nu_N \right) \, \in \, \mathcal{I}; 
\,\, \nu_i \, = \, 0, 1, 2; \,\,\, i \, \in \, \ZZ_N \right\}.
\eeq
All the vertices of $\mathrm{Hc}$ belong to some
coordinate hyperplane
\beq
\pi_i \equiv \left\{ 
\left( \nu_1, \nu_2,\cdots, \nu_N \right) \, \in \, \mathcal{I}; 
\,\, \nu_i \, = \, 0
\right\},
\qquad i \, \in \, \ZZ_N ,
\eeq
with the exception of the highest-weight correlator
\beq
P(2,2,\cdots,2).
\eeq
We may also write
\beq
\mathrm{Hc} \, = \, \left\{ Q \, \in \, \mathcal{I}; 
\,\, \left\| Q \right\|_\infty \, \le \, 2 \right\} .
\eeq
Note that, for example,
\beq
d_\infty[G(2,2,\cdots,2), \,  G(3,2,\cdots,2)]
\, = \, d_\infty[G(2,2,\cdots,2), \, G(3,3,\cdots,3)] \, = \, 1,
\eeq
while the "distance", as far as the reduction to primitive
correlators is concerned, is clearly larger in the second case
compared to the first one.
To describe the difference between the above cases, it is convenient 
to introduce a second norm on $\mathcal{I}$:
\beq
\left\| Q \right\|_1
\, \equiv \,
\sum_{i=1}^N \left| \nu_i \right| .
\eeq
The inequality holds
\beq
\frac{1}{N} \, \left\| Q \right\|_1 
\, \le \,
\left\| Q \right\|_\infty 
\, \le \, \left\| Q \right\|_1
\qquad \forall Q \, \in \, \mathcal{I}. 
\eeq
The induced distance from the one-norm reads
\beq
d_1\left(Q,Q'\right) \, \equiv \, \left\| Q' \, - \, Q \right\|_1. 
\eeq
With the new distance,
\beq
d_1[G(2,2,\cdots,2), \,  G(3,2,\cdots,2)] \, = \, 1,
\eeq
while
\beq
d_1[G(2,2,\cdots,2), \, G(3,3,\cdots,3)] \, = \, N.
\eeq
We may also define the distance of a generic correlator $G(\nu)$,
identified by the point $Q$, from the hypercube $\mathrm{Hc}$:
\beq
d_1(Q,\mathrm{Hc})
\, \equiv \,
\min_{Q'\in \mathrm{Hc}} d_1\left(Q,Q'\right).
\eeq
Such a distance clearly vanishes if $Q$ is a primitive
correlator.

%%%%%%%%%%%%%%%%%%%%%%%%%%%%%%%%%%%%%%%%%%%%%%%

\subsection{Reduction to Primitive Correlators}

\begin{figure}[ht]
\begin{center}
\includegraphics[width=0.5\textwidth]{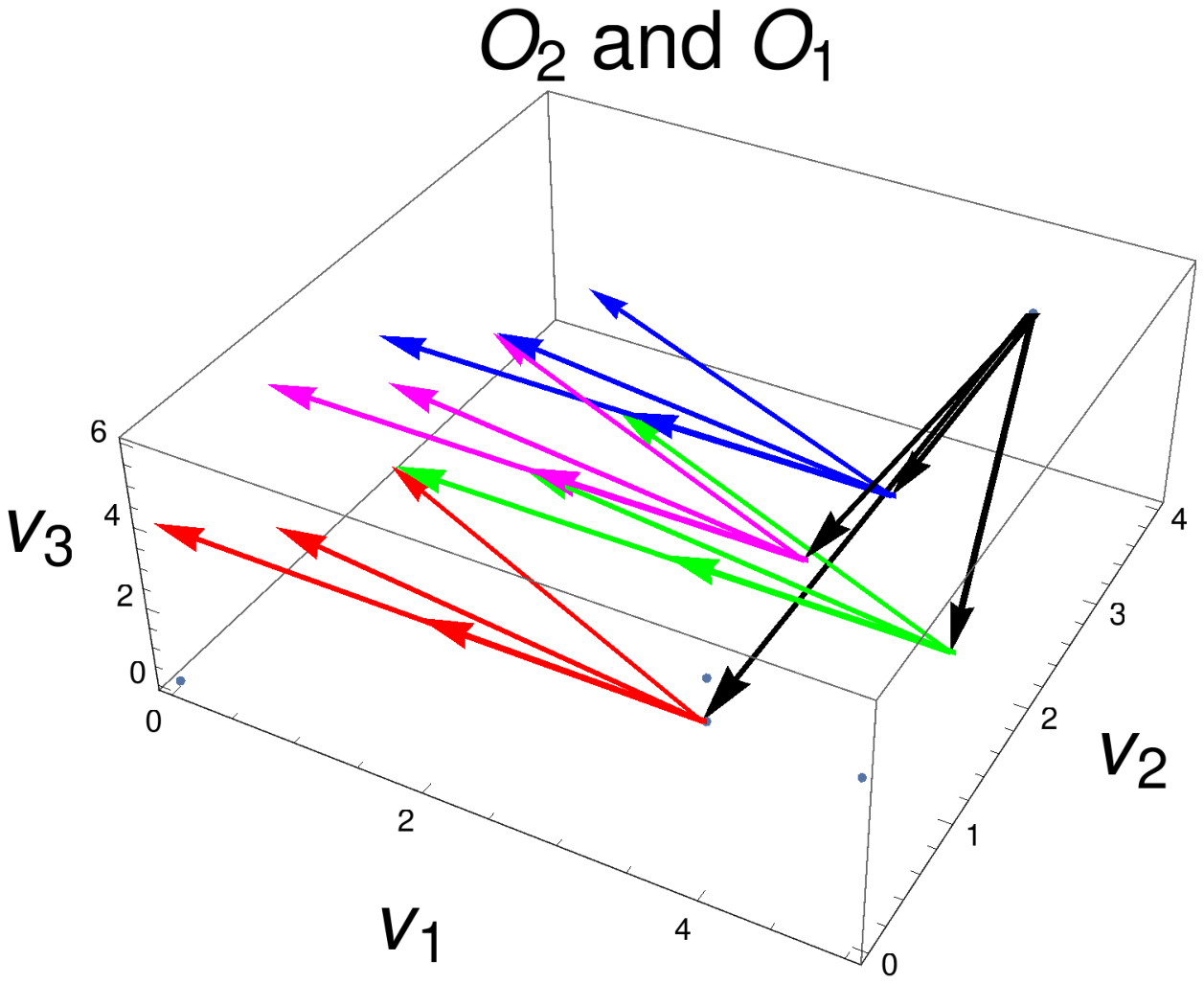}
\footnotesize
\caption{
\label{fig_pO1O2}
\it 
Action on the point $Q=(4,4,4)$,
in the index space $\mathcal{I}$ for $N=3$,
of the operator $O_{i=2}$ (black arrows), 
followed by the action, on each image point,
of $O_{i'=1}$ (colored arrows).
Also in this case, different paths lead sometimes
to the same final point.
}
\end{center}
\end{figure}

Let us now discuss the reduction of an arbitrary correlator $G(\nu)$
to primitive correlators.
This reduction involves paths along a tree in the space
$\mathcal{I}$, with the trunk 
beginning at $Q$ and with all the branches
ending inside $\mathrm{Hc}$.
As already observed, there is no a canonically-defined path,
but many equivalent paths.
%
%%%%%%%%%%%%%%%%%%% FIGURA %%%%%%%%%%%%%%%%%%%%
\begin{figure}[ht]
\begin{center}
\includegraphics[width=0.5\textwidth]{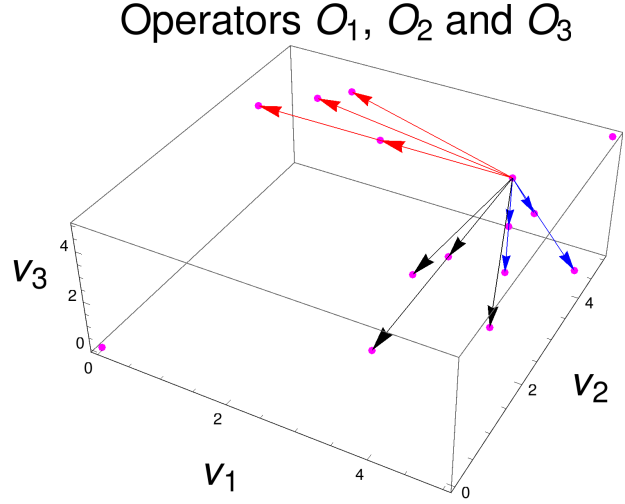}
\footnotesize
\caption{
\label{fig_pO1O2O3}
\it Representation of the operators $O_1$, $O_2$ and $O_3$ in index space $\mathcal{I}$
for $N=3$.
Red vectors: $O_1$;
Black vectors: $O_2$;
Blue vectors: $O_3$.
All the vectors are applied to the point
$\left(\nu_1,\nu_2,\nu_3\right)=(4,4,4)$.
}
\end{center}
\end{figure}
%%%%%%%%%% FINE FIGURA %%%%%%%%%%%%%%%%%%%%%%%%
%
The four operators $N_i$, $D_i$, $L_i$ and $R_i$ 
entering the operator $O_i$ solving the $i^{th}$ LDS equation
can be represented to the following four vectors in the space
$\mathcal{I}$ respectively:
\bea
u^{(i)} &=& \,\,\, \left( \, 0; \,\,\, - \, 4_i; \, 0 \, \right) ;
\nonumber\\
v^{(i)} &=& \,\,\, \left( \, 0; \,\,\, - \, 2_i; \, 0 \, \right) ;
\nonumber\\
\xi^{(i)} &=& \left( 1_{i-1}; - \, 3_i; \,\, 0 \, \right) ;
\nonumber\\
\eta^{(i)} &=& \,\,\, \left( \, 0 \,\, ; \, - \, 3_i ; 1_{i+1} \right) ;
\eea
where only the relevant vector components have been written
(see figs.$\,$\ref{fig_plotO2} and \ref{fig_pO1O2O3}).
To apply the operator $O_i$ to a correlator is equivalent 
to adding to $Q$ each one of the above vectors:
\beq
O_i \,\,\, \left(i^{th} \,\,\mathrm{LDS \,\, eq.}\right):
\,\,\, 
Q \,\, \to \,\, 
\left\{
\begin{array}{c}
Q \, + \, u^{(i)} ;
\\
Q \, + \, v^{(i)} ;
\\
Q \, + \, \xi^{(i)} ;
\\
Q \, + \, \eta^{(i)} .
\end{array}
\right. 
\eeq
Four points in the space $\mathcal{I}$
out of one point are then generated.
If some of the above points fall within the cube
$\mathrm{Hc}$, they are not transformed any more.
%
%%%%%%%%%%%%%%%%%%% FIGURA %%%%%%%%%%%%%%%%%%%%
\begin{figure}[ht]
\begin{center}
\includegraphics[width=0.5\textwidth]{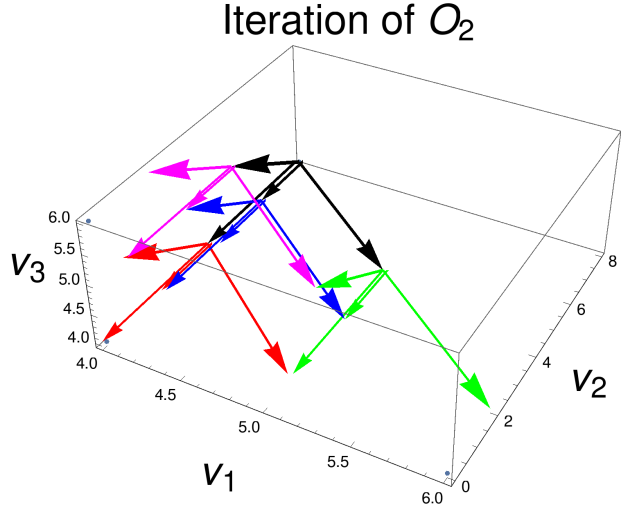}
\footnotesize
\caption{
\label{fig_pO2O2}
\it Iteration of the operator $O_2$ in the index space
$\mathcal{I}$ for $N=3$ ($i=i'=2$, see text).
The initial point is $Q=(4,8,4)$.
In some cases, different paths lead to the same final point.
}
\end{center}
\end{figure}
%%%%%%%%%% FINE FIGURA %%%%%%%%%%%%%%%%%%%%%%%%
%
Next, we iterate the above procedure, by applying the
operator $O_{i'}$ with the new index $i'$,
to each one of the four points above
(see figs.$\,$\ref{fig_pO2O2} and \ref{fig_pO1O2}).
%
%%%%%%%%%%%%%%%%%%% FIGURA %%%%%%%%%%%%%%%%%%%%
%%%%%%%%%% FINE FIGURA %%%%%%%%%%%%%%%%%%%%%%%%
%
Any index $i'$ is in principle good,
so long as $\nu_{i'} \ge 3$ for the current term,
because LDS equations with different indices
commute with each other and each one of them
reduces the weight $\mathcal{R}$.
In general, each of the four points above generates
four terms on his own, so that, after two
reduction steps, we have in general
$4^2$ points.
By iterating the process $k$ times, one obtains up to
$4^k$ points.
The iteration process above can be described as 
the branching of a quaternary tree.
The root of the tree, the trunk, is placed in the
initial point $Q$. At each step, each branch
produces four new secondary branches and so on.
The branching terminates when all the current
branches fall inside the cube $\mathrm{Hc}$. 
The branching is in general quite heavy
on the computational side, because
of the exponential increase described.

\begin{figure}[ht]
\begin{center}
\includegraphics[width=0.5\textwidth]{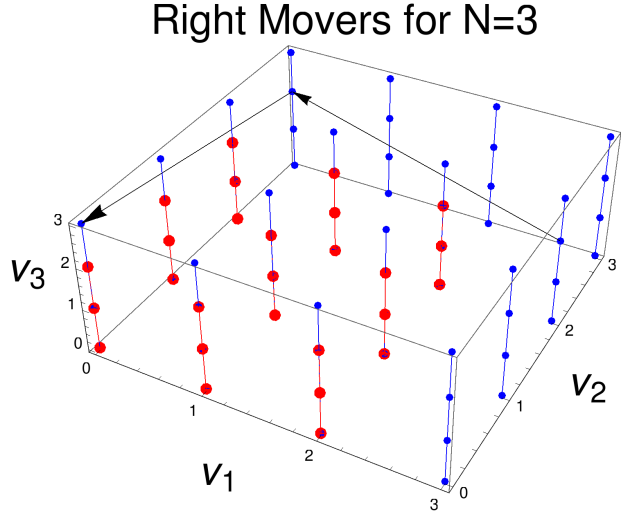}
\footnotesize
\caption{
\label{fig_pRMRM}
\it Graphical representation of the Right Mover 
Operators $R_i$ for $i=1,2$ in the index space
$\mathcal{I}$ for $N=3$, acting, in the order, 
on the initial point $Q=(3,2,2) \to (0,3,2) \to (0,0,3)$.
}
\end{center}
\end{figure}

The generation of long reduction chains
for correlators with an index equal to
three adjacent to a long sequence of indices
equal to two, such as for example
\beq
G\left(\nu_1=2,\nu_2=3,\nu_3=2,\cdots,\nu_N=2\right),
\eeq
can be viewed geometrically as follows.
The one-distance of the above correlator
from the hypercube of the primitive correlators
is only one,
\beq
d_1\left[G(\cdots),\mathrm{Hc}\right] \, = \, 1.
\eeq
With the first reduction, with $i=2$,
we get closer to $\mathrm{Hc}$ along the second direction, as
\beq
\nu_2 \, \to \, \nu_2 \, - \, 3,
\eeq
but we become more distant to $\mathrm{Hc}$ in the neighboring
directions, as
\bea
\nu_1 &\to& \nu_1 \, + \, 1 ;
\nonumber\\
\nu_3 &\to& \nu_3 \, + \, 1 .
\eea
We then have to reduce the distance along the above directions,
by applying the LDS equations for $i=1$ and $i=3$.
In general, by getting closer to $\mathrm{Hc}$,
by three units, in the $i^{\mathrm{th}}$ direction,
\beq
\nu_i \, \to \, \nu_i \, - \, 3,
\eeq
we become more distant, by one unit, in the
neighboring directions,
\bea
\nu_{i-1} &\to& \nu_{i-1} \, + \, 1 ;
\nonumber\\
\nu_{i+1} &\to& \nu_{i+1} \, + \, 1 .
\eea
This phenomenon is represented in fig.$\,$\ref{fig_pRMRM} 
for $N=3$.
In general, if the lattice size $N$ is large, 
there are many different directions 
one can take in the space 
$\mathcal{I}$, so that one can move
quite a lot in the course of the reduction 
process of a correlator having even with a small 
one-distance from $\mathrm{Hc}$,
before reaching primitive correlators.
%
%%%%%%%%%%%%%%%%%%% FIGURA %%%%%%%%%%%%%%%%%%%%
%%%%%%%%%% FINE FIGURA %%%%%%%%%%%%%%%%%%%%%%%%
%
\subsection{Evolution Equations for Primitive Correlators}

The derivation of a primitive correlator $P(\nu)$
with respect to $w_{i,i+1}$ can also be viewed 
geometrically in the space $\mathcal{I}$. 
With the derivation, the point $Q$ in $\mathcal{I}$ 
corresponding to $P(\nu)$ is translated by the vector
\beq
\theta^{(i)} \, \equiv \, 
\left(0 \, ; \cdots \, ; 0 \, ; 1_i \, ; 1_{i+1} \, ; 0 \, ; \cdots \, ; 0 \right),
\eeq
i.e.
\beq
\frac{\partial}{\partial w_{i,i+1}}:
\,\, Q \,\,\, \to \,\,\, Q \, + \, \theta^{(i)}. 
\eeq
Now, a primitive correlator is represented by a point
lying inside $\mathrm{Hc}$.
If the shifted point lies again inside $\mathrm{Hc}$, 
then the derivation has produced another primitive
correlator and we have a one-term differential equation.
Otherwise, the derivation has produced a reducible
correlator.
If a primitive correlator lying on the boundary of $\mathrm{Hc}$
is derived with respect to $w_{i,i+1}$, 
independent correlators with a long reduction chain 
are in general generated.

%%%%%%%%%%%%%%%%%%%%%%%%%%%%%%%%%%%%%%%%%%%%%%%%%%%%%%%%%%

\section{Inclusion of Lattice Symmetry Equations}
\label{sect11}

Let us now discuss the effects of Lattice Symmetry Equations $(LSE)$
on the reduction of kinematically-independent correlators 
$\left\{G(\nu)\right\}$ to primitive correlators 
$\left\{P(\nu)\right\}$, i.e. to dynamically-independent correlators.
Such a reduction has been made in the previous section by disregarding the symmetries of the theory.
It is clear that $LSE$ reduce, in general, the number
of primitive (i.e. non zero and independent) 
correlators existing at a given lattice size $N < \infty$.

In general, if the theory at lattice size $N$ has a symmetry described
by a finite group $G_N$, then the number
of primitive correlators satisfies the lower bound
\beq
\label{general_lower_bound}
\mathrm{Card}\left[ \left\{ P(\nu) \right\}_N \right] 
\, \ge  \,
\frac{3^{N}}{\left|G_N\right|},
\eeq
where $\left| G_N \right| < \infty$ is the order of $G_N$,
i.e. the number of its elements.
The inequality above follows from the fact that any orbit of primitive 
correlators under $G_N$ contains at most $\left|G_N\right|$ 
(distinct) elements.

Let us now consider in detail the symmetries
of the anharmonic oscillator on a circular lattice considered
before. 
\begin{enumerate}
\item
{\it Even Lattice Action,} 
\beq
S[-\Phi] \, = \, S[\Phi].
\eeq
Since, as already discussed, correlators $G(\nu)$ with odd weight 
are zero, the number of primitive correlators is roughly
reduced by a factor two,
\beq
\mathrm{Card}\left[ \left\{ P(\nu) \right\}_N \right] 
\,\, = \,\, 
\frac{3^N + 1 }{2}
\,\, > \,\,
\frac{3^N}{2} \, = \, \frac{3^N}{\left|\ZZ_2\right|} .
\eeq
In this simple case, the combination of the LDS equations
with the LSE is rather trivial, as they are basically
independent.
The LDS equations indeed involve corrrelators
all having the same parity of the index $\mathcal{R}$,
i.e. all the correlators entering a given $LDS$ equation 
all have even $\mathcal{R}$ or odd $\mathcal{R}$;
%%%%%
\item
{\it Lattice Action $S[\Phi]$ invariant under 
the dihedral group $D_N$,}
\beq
S[g \cdot \Phi] \, = \, S[\Phi], \qquad g \, \in \, D_N.
\eeq
This case is more complicated than the previous one,
because now combining together the restrictions coming
from both the LDS equations and LSE is not trivial.

Since $D_N$ is a finite group with
\beq
\left| D_N \right| \, = \, 2 N,
\eeq
according to the general inequality (\ref{general_lower_bound}), 
it holds
\beq
\mathrm{Card}\left[ \left\{ P(\nu) \right\}_N \right] 
\, \ge  \,
\frac{3^{N}}{2N}.
\eeq
At most, a mild, power-like suppression in the number of primitive correlators
is then obtained,
\beq
\mathrm{Card}\left[ \left\{ P(\nu) \right\}_N \right] 
\quad \ge \quad
C \, \frac{3^N}{N^k},
\eeq
with $C=1/2$ and $k=1$.
\end{enumerate}
The basic exponential growth with the lattice size $N$
in the number of primitive correlators,
which we have found in QFT,
can be compared with the situation in classical
field theory (which can be considered the limit
of QFT for $\hbar \to 0$).
Indeed, one can put on the lattice also a classical
field theory.
If we consider for example a classical scalar field
$\varphi(x,t)$ in space-time dimension $d=2$, 
discretized on a space lattice
of size $N$, the state of the system at a given
time $t$ is represented by the values
of the fields $\varphi_i(t) \equiv \varphi\left(x_i;t\right)$ 
at the lattice points $x_i=i \, a$, $i=1,2,\cdots,N$,
\beq
       \varphi_1\left(t \right); 
\,\,\, \varphi_2\left(t \right);
\,\,\, \cdots;
\,\,\, \varphi_N\left(t \right).
\eeq
If we deal for instance with a Cauchy problem,
one has typically to integrate $N$ evolution equations
of second order in $t$ of the form:
\beq
\frac{d^2 \varphi_i}{dt^2} 
\, = \, F\left(\varphi_i,\varphi_{i+1},\varphi_{i+1}\right),
\qquad i=1,2,\cdots,N.
\eeq
Each one of the $N$ real numbers above, 
$\varphi_i\left(t \right)$,
can be well approximated by, let' say, a binary
expansion with $k$ digits with $k \gg 1$.%
%%%%%%%%%%
\footnote{
As well known, convergence is exponentially
fast in the number $k$ of the digits, as the relative 
error is $\mathcal{O}(1/2^k)$;
By adding one digit, one doubles the accuracy.
}
%%%%%%%%%%%
Now, if we double the lattice size, i.e. if we go from
$N$ to $2N$, the number of state variables 
$\phi_i$ and of evolution equations simply doubles, 
i.e. there is a linear (power-like) growth with $N$,
in contrast to the exponential growth $\approx 3^N$ 
which we have found in the quantum case.
The quantum field theory case is therefore intrinsically 
much more complicated than the classical field theory case.

%%%%%%%%%%%%%%%%%%%%%%%%%%%%%%%%%%%%%%%%%%%%%%%%%%%%%%%%%%%%%

\section{Reduction to Primitive Correlators for $N \to \infty$}
\label{sect12}

In the weak limit $N \to \infty$, the index space $\mathcal{I}$ becomes
an infinite-dimensional discrete space,
\beq
\mathcal{I} \, \equiv \, 
\left\{ \left(\cdots,\nu_{-1},\nu_0,\nu_1, \nu_2,\cdots \right) 
\, \in \, \NN^\ZZ; 
\,\,\, 
0 \, \le \, \sum_{i=-\infty}^{+\infty} \nu_i \, < \, \infty
\right\}.
\eeq
However, the reduction of an arbitrary correlator
\beq
G\left(\cdots,\nu_{-1},\nu_0,\nu_1, \nu_2,\cdots \right) 
\eeq
to a linear combination of primitive correlators
\beq
P\left(\cdots,\mu_{-1},\mu_0,\mu_1, \mu_2,\cdots \right),
\qquad \mu_i \, = \, 0, 1, 2, \quad i \, \in \, \ZZ,
\eeq
is always {\it finite}, i.e. it involves
a finite number of step in any case.
That is because we require the sum of all the indices
to be finite,
\beq
\label{usual_finite}
\sum_{i=-\infty}^{+\infty} \nu_i \, < \, \infty,
\eeq
and, at any reduction step, the above sum is decreased
at least by two units.
Since, at each reduction step, the number of correlators involved
grows at most by a factor four, it follows that the number
of primitive correlators appearing in the final decomposition
of any $G(\nu)$ is finite.
The following remarks are in order.
\begin{enumerate}
\item
While always finite, an arbitrarily large number of indices $\nu_i$ 
in $G(\nu)$ can be different from zero
while, at finite lattice size $N$, such a number
is obviously limited by $N$;
%%%%%
\item
In the case of the strong continuum limit, since
condition (\ref{usual_finite}) is not imposed,
the sum of the indices is generally
infinite, so that the reduction to primitive correlators 
involves an infinite number of steps.
\end{enumerate}

%%%%%%%%%%%%%%%%%%%%%%%%%%%%%%%%%%%%%%%%%%%%%%%

\subsection{Imposing Lattice Symmetry Equations}

By combining the LSE with the LDS equations,
two scenarios are in principle possible
in the weak limit $N \to \infty$:
\begin{enumerate}
\item
{\it Normally-Symmetric Case} --- or more simply Normal Case.
There can be many $LSE$,
but the reduction of the dimensions of $\left\{ P(\nu) \right\}_N$
at finite lattice sizes $N$, after imposing the $LSE$, 
is not so strong to be able to reduce
the cardinality of the  basis in the limit $N \to \infty$,
which remains the countable one
\beq
\mathrm{Card}\left[ \left\{ P(\nu) \right\}_\infty \right]
\, = \, \mathrm{Card}\left( \NN \right)
\, \equiv \, \aleph_0.
\eeq
In this case, the addition of $LSE$ to $LDS$ equations
does not play, as far as the reduction to primitive correlators 
is concerned, any critical role. 
There is not any "collapse" of the cardinality of
$\left\{ P(\nu) \right\}_\infty$ 
from the countable one to a finite one 
and the theory is classified as {\it unsolvable};
%%%%%%%%%%%%%%%%%%%%%%%%%%%%%%%%%%%
\item
{\it Exceptionally-Symmetric Case} --- or simply Exceptional Case.
There are so many LSE --- independent on each other,
as well as on the LDS equations --- that the cardinality of the
primitive correlator basis is diminished from the countable one
to a finite one:
\beq
\mathrm{Card}\left[ \left\{ P(\nu) \right\}_\infty \right]
\, < \, \infty. 
\eeq
That implies that all the primitive correlators 
of the theory can be expressed in terms of a finite number 
of them.
In other words, after imposing the Ward identities
of the theory, only a finite number of 
independent primitive correlators is found,
as in the case of the Gaussian theory.
In such an "infinitely symmetric" case, there is
a "collapse" in the size of the primitive correlators basis
and, according to our philosophy, the theory is classified 
as {\it solvable}.
\end{enumerate}

%%%%%%%%%%%%%%%%%%%%%%%%%%%%%%%%%%%%%%%%%%%%%%%%%%%%%%%%%%%

\subsubsection{Examples}

In this section we consider a few examples 
of systems with different symmetries.
\begin{enumerate}
\item
{\it Anharmonic Oscillator on an infinite lattice
with symmetry group $D_\infty$}, the dihedral group
of infinite order defined previously.
That is the continuum limit of our reference model.
The infinitely-many primitive correlators with $\tau=1$,
\beq
P\left(\cdots; \, 0 \, ; \cdots \, ; \, 0 \, ; \, \nu_i \, = \, 1,2 \, ; \, 0 \, ; \, \cdots ; \, 0 \, ; \, \cdots \right),
\qquad i \, \in \, \ZZ,
\eeq
can be reduced to the two primitive correlators
\beq
P\left(\cdots \, ; \, 0 \, ; \, \nu_0 = 1,2 \, ; \, 0 \, ; \, \cdots \right),
\eeq 
by means of the shift of the indices $j \to j - i $, $j \in \ZZ$.
The symmetry $D_\infty$ is then very "efficient"
in this case, as it reduces an infinite set
of primitive correlators to a finite one. 
However, the primitive correlators with $\tau=2$,
\beq
P\left(\cdots \, ; \, 0 \, ; \, \nu_i = 1, 2 \, ; \, 0 \, ; \, \cdots ; \, 0 \, ; \, \nu_j = 1, 2 \, ; \, 0 \, ; \, \cdots \right),
\qquad i \, < \, j \, \in \, \ZZ,
\eeq
can only be reduced to primitive correlators of the form
\beq
P\left(\cdots \,; \, 0 \, ; \, \nu_0 = 1, 2 \, ; \, 0 \, ; \, \cdots ; \, 0 \, ; \nu_{j-i} = 1, 2 \, ; \, 0 \, ; \, \cdots \right),
\eeq
by means of the shift of the indices $j \to j - i $, $j \in \ZZ$.
If the indices $\nu_0$ and $\nu_{j-i}$ are different
from each other, one can use the reflection symmetry
to render for example $\nu_0<\nu_{j-i}$, but
no further reduction is possible.
In the latter case, the symmetry $D_\infty$ is not so efficient:
the number of primitive correlators after the reduction 
is still infinite. 
The conclusion is that, in the limit $N \to \infty$,
the number of primitive correlators of the anharmonic
oscillator is countably-infinite so the model, 
according to our definition, is a normally-symmetric one 
and therefore is unsolvable;
%%%%%
\item
{\it Scalar $\lambda\,\phi^4$ theory for
$N \to \infty$ invariant under $S_\infty$},
the full symmetric group acting on $\NN$.
The action reads:
\beq
S_{\mathrm{sym}}[\Phi] 
\, \equiv \,
\frac{k}{2} \sum_{i=-\infty}^{+\infty} \phi_i^2
\, - \, \frac{w}{2} \, \sum_{i \ne j }^{-\infty,+\infty} \, \phi_i \, \phi_j
\, + \, \frac{\lambda}{4} \sum_{i=-\infty}^{+\infty} \phi_i^4.
\eeq
The latter can also be written as
\beq
S_{\mathrm{sym}}[\Phi] 
\, \equiv \,
\frac{k+w}{2} \sum_{i=-\infty}^{+\infty} \phi_i^2
\, - \, \frac{w}{2} \, \left( \sum_{i=-\infty}^{+\infty} \phi_i \right)^2
\, + \, \frac{\lambda}{4} \sum_{i=-\infty}^{+\infty} \phi_i^4.
\eeq
In a plane, we can have up to three points
at the same distance from each other,
by putting them at the vertices of an equilater
triangle. If we construct a lattice theory
on such a triangle, the field at each site
interacts in the same way with the fields
in the other two lattice points.
In the ordinary, 3-dimensional space, we can defined a
lattice theory on a regular tetrahedron, and so on. 
Therefore, the theory above can be naturally
constructed in a infinite-dimensional ambient space.

Since, as already discussed, the integration measure 
\beq
\mathcal{D} \Phi 
\, \equiv \, 
\sum_{i=-\infty}^{+\infty} d \phi_i
\eeq
is invariant under $S_\infty$, also the quantum theory has
the full $S_\infty$ group as its symmetry group.
Any primitive correlator $P(\nu)$
has a string of indices $\nu_i$ equal to
an infinite sequence of zero's, one's
and two's.
By using the symmetry group $S_\infty$,
one can reduce any primitive correlator $P(\nu)$,
with $\tau(\nu)=n=0,1,2,\cdots$, 
to primitive correlators with the first $k$ indices 
$\nu_1$, $\nu_2$, $\cdots$, $\nu_k$ equal to two, 
the next $n-k$ indices $\nu_{k+1}$, $\nu_{k+2}$, $\cdots$,
$\nu_n$ equal to one, with $0 \le k \le n$,
and all the remaining indices identically zero:
\bea
&& \left\{ P(\nu) ; \,\, \tau(\nu) \, = \, n \right\}
\,\, \Rightarrow \,\,
\\
&&
\quad
\left\{ 
P\left( \cdots; \, \nu_{-1}=0 \, ; 
\, \nu_1=2 \, ; \, \cdots; \, \nu_k=2 \, 
; \, \nu_{k+1} = 1; \cdots \, ; \, \nu_n=1; 
\, \nu_{n+1}=0 \,; \, \cdots \right)
\right\}.
\nonumber
\eea
For any given $\tau=n$, there are therefore $n+1$
independent primitive correlators.
Since the variable $n$ can take any integer value,
\beq
n \, = \, 0, 1, 2, \cdots,
\eeq
the theory, despite its large symmetry group $S_\infty$, 
has an infinite set of primitive correlators, 
so it is classified as unsolvable. 
Actually, we tried to explicitly solve this model,
but we did not succeed, so it seems, at least
relative to this example, that our classification
scheme is practically relevant;
%%%%%
\item
{\it Symmetric Random Field},
i.e. the scalar $\lambda\,\phi^4$ theory for 
$N \to \infty$ for $w=0$. 
As we have seen, this model is invariant under $S_\infty$. 
There is only one independent primitive correlator.
Therefore, according to our definition, the
model is solvable; it is also solvable in practice.
\end{enumerate}
It would be interesting to investigate 
the connection between known solvable models, 
regularized on various lattices, and the cardinality of the 
corresponding  primitive correlator bases.
Let us remark that, in our scheme, solvability is just {\it defined}
by looking at the cardinality of the primitive basis.

%%%%%%%%%%%%%%%%%%%%%%%%%%%%%%%%%%%%%%%%%%%%%%%%%%%%%%%%
%%%%%%%%%%%%%%%%%%%%%%%%%%%%%%%%%%%%%%%%%%%%%%%%%%%%%%%%
%%%%%%%%%%%%%%%%%%%%%%%%%%%%%%%%%%%%%%%%%%%%%%%%%%%%%%%%

\section{Evaluation of Primitive Correlators}
\label{sect13}

After the reduction of correlators to primitive correlators
--- a purely algebraic step --- one has to evaluate all
the primitive correlators $P(\mu)$ --- the "analytic part"
of quantum field theory.
The most efficient way to accomplish this task
involves the following steps:
\begin{enumerate}
\item
Generate evolution (differential) equations for the $P(\mu)$'s
with respect to some parameter entering the action, such
as $w_{i,i+1}$ or $\lambda_i$;
\item
Impose initial values for the primitive correlators,
at which the latter can be exactly evaluated;
\item
Integrate the Cauchy problem specified at the two 
previous steps. 
\end{enumerate}

%%%%%%%%%%%%%%%%%%%%%%%%%%%%%%%%%%%%%

\subsection{Initial Conditions}

By looking at the integral expression of a correlator,
eq.(\ref{G_integral}),
one easily convinces himself that initial conditions
for the parameter flow can only by provided in the
following two cases:
\begin{enumerate}
\item
{\it Gaussian (or Free) Theory},
\beq
\lambda_i \, = \, 0, \qquad i \, \in \, \ZZ_N . 
\eeq
The free theory is solved by means of the Discrete
Fourier Transform (DFT).
Standard perturbation theory is then made by
expanding the exponential of minus the euclidean action 
in powers of the $\lambda_i$;
%%%%%
\item
{\it Random Field},
\beq
w_{i,i+1} \, = \, 0, \qquad i \, \in \, \ZZ_N .
\eeq
\end{enumerate}
In the first case above, one obtains differential
equations with irregular singular points at $\lambda_i=0$,
because of well-known vacuum instability, so the
second possibility is the only viable one%
%%%%%%%%%%
\footnote{To tame the factorial divergence of the perturbative
expansion produced by vacuum instability, one can make the Borel 
transform of the primitive correlators with respect to 
$\lambda_i \to s_i$, and then write evolution equations
in the Borel variables $s_i$.
This strategy gives rise to a theory similar
to the one obtained by direct derivation with respect to 
the $w_{i,i+1}$'s.
}.

%%%%%%%%%%%%%%%%%%%%%%%%%%%%%%%%%%%%%%%%%%%%%%

\subsection{System of Ordinary Differential Equations}

Let us consider, for simplicity's sake, the symmetric theory,
with weight
\beq
\label{weight_to_der}
\exp(-S[\Phi])
\, = \,
\exp\left\{
\sum_{i=1}^{N} 
\left(
- \frac{1}{2} \bar{k} \, \phi_i^2
\, + \, \bar{w} \, \phi_i \, \phi_{i+1}
\, - \, \frac{1}{4} \, \bar{\lambda} \, \phi_i^4
\right)\right\}.
\eeq
We have put a bar over the couplings to indicate that
we are interested in the theory with that specific choice
of the parameters.
Now, let us consider $w$ as a variable, by introducing
the action
\beq
S[\Phi; \, w] 
\, \equiv \,
\sum_{i=1}^{N} 
\left(
\frac{1}{2} \bar{k} \, \phi_i^2
\, - \, w \, \phi_i \, \phi_{i+1}
\, + \, \frac{1}{4} \, \bar{\lambda} \, \phi_i^4
\right).
\eeq
We want to evolve the primitive correlators from $w=0$ up
to the chosen $\bar{w} \ne 0$:
\beq
w: 0 \, \to \, \bar{w}.
\eeq
Since the derivation of the weight in eq.(\ref{weight_to_der}) 
with respect to $w$ brings down the sum
\beq
\sum_{i=1}^{N} \phi_i \, \phi_{i+1},
\eeq
the derivative of the primitive correlator reads:
\beq
\frac{\partial}{\partial w}
P\left(\nu_1; \cdots; \nu_i; \nu_{i+1}; \cdots, \nu_N\right)
\, = \, 
\sum_{i=1}^N G\left(\nu_1; \cdots; 1 + \nu_i; 1 + \nu_{i+1}; \cdots,\nu_N\right).
\eeq
Since the weight $\mathcal{R}$ of $P$ is increased by two units
upon derivation with respect to $w$, reducible correlators 
are in general generated on the r.h.s. of the above equation.
By reducing each correlator on the r.h.s., to a linear
combination of primitive correlators,
as shown in previous section,
we generate a linear system of coupled Ordinary Differential 
Equations (ODE's) with variable coefficients of order 
$\mathcal{O}\left(3^N\right)$:
\beq
\frac{\partial P(\nu)}{\partial w } \, = \, 
\sum_{\left\| \mu \right\|_\infty \, \le \, 2} g_\nu (\mu) \, P(\mu),
\qquad \left\| \nu \right\|_\infty \, \le \, 2.
\eeq
There is one $ODE$ for each  primitive correlator.

%%%%%%%%%%%%%%%%%%%%%%%%%%%%%%%%%%%%%%%%%%%%%%%%%%%%%%

\subsection{System of Partial Differential Equations}

In this section we describe an alternative form of the evolution equations, 
consisting of a system of partial differential equations,
which will turn out to be the only possibility in the continuum limit 
$N \to \infty$.

We introduce a different variable $w_i \equiv w_{i,i+1}$ for each term
$\phi_i \, \phi_{i+1}$ in the above action, which then becomes:
\beq
S[\Phi; \, {\bf w}] 
\, \equiv \,
\sum_{i=1}^{N} 
\left(
\frac{1}{2} \bar{k} \, \phi_i^2
\, - \, w_i \, \phi_i \, \phi_{i+1}
\, + \, \frac{1}{4} \, \bar{\lambda} \, \phi_i^4
\right),
\eeq
where we have defined the $N$-dimensional vector
\beq
{\bf w} 
\, \equiv \,
\left( w_1, w_2, \cdots,w_N \right).
\eeq
We consider the $w_i$'s as independent variables,
to evolve from the initial point
\beq
w_1 \, = \, w_2 \, = \, \cdots \, = \, w_N \, = \, 0
\eeq
up to the final point
\beq
w_1 \, = \, w_2 \, = \, \cdots \, = \, w_N \, = \, \bar{w}.
\eeq
In addition to the multi-index $\nu$,
the primitive correlators are now also functions of 
the vector ${\bf w}$: 
\beq
P \, = \, P(\nu,{\bf w}).
\eeq
According to the chain rule,
\beq
\frac{\partial }{\partial w} P(\nu;w)
\, = \, \sum_{i=1}^N 
\left.\frac{\partial}{\partial w_i} P(\nu,{\bf w}) \right|_{w_k \to w}
\eeq
We therefore have to evaluate the partial derivatives
of each primitive correlator with respect to each one of the $w_i$'s:
\beq
\frac{\partial P}{\partial w_i} \, 
P\left( \nu_1, \cdots; \, \nu_i; \,  \nu_{i+1}; \, \cdots, \nu_N \right)
\, = \, 
G\left(\nu_1, \cdots; \, 1 \, + \, \nu_i; \, 1 \, + \, \nu_{i+1}; 
\, \cdots, \nu_N \right) ,
\quad i \in \ZZ.
\eeq
By reducing the r.h.s. of the above equation to a linear
combination of primitive correlators,
we obtain a linear system of coupled Partial Differential Equations 
on the primitive correlators $\left\{ P(\nu; \,{\bf w}) \right\}$
in the $N$ independent variables $w_i$:
\beq
\qquad\qquad\qquad\qquad
\frac{\partial P(\nu)}{\partial w_i } \, = \, 
\sum_{  \left\| \mu \right\|_\infty \le 2} g_\nu^{(i)} (\mu) \, P(\mu),
\qquad 
i \, = \, 1,2,\cdots,N, \quad \left\| \nu \right\|_\infty \, \le \, 2. 
\eeq
Note that, in this case, there are $N$ Partial Differential Equations
for each primitive correlator.

%%%%%%%%%%%%%%%%%%%%%%%%%%%%

\subsection{Commuting flows}

Since the $w_i$'s constitute a set of $N$ independent variables
on each other, one has $N$ commuting flows.
The consequence of such commutativity 
can be expressed in differential form by requiring
mixed partial derivatives to be equal
\beq
\frac{\partial^2 P(\nu) }{\partial w_i \, \partial w_j}
\, = \, 
\frac{\partial^2 P(\nu) }{\partial w_j \, \partial w_i};
\qquad 
\| \nu \|_\infty \, \le \, 2;
\quad
i \, < \, j \, = \, 1,2,\cdots,N .
\eeq
By explicitating the derivatives, using the
flow equations and taking into account
that $P(\mu)$'s with different $\mu$'s, 
are linearly independent on each other,
one obtains the following
compatibility conditions
\beq
  \frac{\partial g_\nu^{(j)}(\xi)}{\partial w_i}
- \frac{\partial g_\nu^{(i)}(\xi)}{\partial w_j}
\, + \,  
\sum_{\| \mu \|_\infty \, \le \, 2}
\left[
  g_\nu^{(j)}(\mu) \, g_\mu^{(i)}(\xi)
- g_\nu^{(i)}(\mu) \, g_\mu^{(j)}(\xi)
\right]
\, = \, 0;
\eeq
where
\beq
\qquad\qquad\qquad\qquad\qquad\qquad\qquad\quad
\| \xi \|_\infty \, \le \, 2;
\qquad
i \, < \, j \, = \, 1, 2, \cdots, N .
\eeq

%%%%%%%%%%%%%%%%%%%%%%%%%%%

\subsection{Classification}

It is natural to classify the evolution equations above
according to the number of terms appearing on their r.h.s.:
\begin{enumerate}
\item
{\it One-Term Equations.}
These are the simplest equations, on primitive
correlators having all indices
\beq
\nu_i \, \le \, 1; \qquad i \, \in \, \ZZ_N;
\eeq 
namely:
\beq
\frac{\partial}{ \partial w_{i,i+1} } 
\, P\left(\cdots; \, \nu_i \, \le \, 1; \, \nu_{i+1} \, \le \, 1; \, \cdots \right)
\, = \, 
P\left(\cdots; \, \nu_i \, + \, 1; \, \nu_{i+1} \, + \, 1; \, \cdots \right).
\eeq
Note that, in particular, the above equations
have a structure independent on the lattice size $N$.
They are trivially integrate by quadrature;
%%%%%%
\item
{\it Many-Term Equations.}
These are the PDE's on primitive correlators
having at least one index equal to two.
If we consider for example the highest-weight primitive
correlator --- having all its indices equal to two --- 
by differentiating one obtains
\beq
\frac{\partial}{\partial w_{i,i+1}} \, 
P\left(
2; \, \cdots; \, \nu_i \, = \, 2 ; \, \nu_{i+1} \, = \, 2; \, \cdots; \, 2 
\right)
\, = \,  
G\left(
2; \, \cdots; \, \nu_i \, = \, 3 ; \, \nu_{i+1} \, = \, 3; \, \cdots; \, 2 
\right) .
\eeq
The reduction of the correlator on the r.h.s. of the above equation
involves the "waves" propagating along the whole lattice
discussed in the previous section.
\end{enumerate}

%%%%%%%%%%%%%%%%%%%%%%%%%%%%

\subsection{Rigidity}

In this section we discuss a property
of the exact theory that we have decided to
call "rigidity" --- namely the fact that,
unlike approximate solutions, 
the exact knowledge of any primitive correlator
requires the knowledge of all of them.
In order to exhibit this phenomenon in an easy way,
let us consider the evaluation of the
simplest (i.e. lowest weight) primitive correlator, 
namely the vacuum one
\beq
V \, \equiv \, P(0,0,\cdots,0).
\eeq
The latter obeys the system of $N$ PDE's
\beq
\frac{\partial V}{\partial w_{i,i+1}} 
\, = \, P\left(0,\cdots,0,1_i,1_{i+1},0, \cdots,0\right),
\qquad i = 1, 2, \cdots, N.
\eeq
The above equations are integrated by quadrature,
but they involve known terms on their r.h.s.,
having $\mathcal{R}=2$, which need to be determined.
The latter primitive correlators obey
differential equations of the form
\beq
\frac{\partial}{\partial w_{i,i+1}} 
P\left(0, \cdots, 0, 1_j, 1_{j+1}, 0, \cdots, 0 \right)
= 
\left\{
\begin{array}{cc}
P\left(\cdots, 1_i, 1_{i+1},\cdots, 1_j, 1_{j+1}, \cdots \right) 
&
\mathrm{for} \,\,  i \, < \, j \, - \, 1 ;
\\
P\left(\cdots\cdots, 1_i, 2_{i+1},1_{i+2}, \cdots \cdots \right) \,\,\,\,
& \mathrm{for} \,\,  i \, = \, j \, - \, 1 ;
\\
P\left(\cdots\cdots, 2_i, 2_{i+1}, \cdots\cdots \right) \qquad\quad
& 
\!\!\!\!\!\!\!\!\!\! \mathrm{for} \,\, i \, = \, j  ;
\end{array}
\right.
\eeq
where the dots denote zero indices,
and similar equations for $i>j$.
Also the latter PDE's are solved by quadrature,
but require the knowledge of the $\mathcal{R}=4$
primitive correlators on their r.h.s.'s.
By writing evolution equations for the $\mathcal{R}=4$
primitive correlators, one brings into the system 
correlators with $\mathcal{R}=6$,
i.e. with six indices equal to one, 
four indices equal to one and one index
equal to two, ... 
By iterating this process in order to close the system, 
one finds that all the indices
initially set to zero by our choice
(vacuum correlator), are progressively filled
in with one's and then with two's.
In general, all the primitive correlator basis
explicitly enters the evaluation
of any individual correlator.

%%%%%%%%%%%%%%%%%%%%%%%%%%%%

\subsection{Examples}

In this section we provide a pair of explicit
examples of systems of evolution equations
for small lattices.
\begin{enumerate}
\item
{\it Lattice with one point \cite{Guralnik},}
\beq
N \,= \, 1.
\eeq
There are two (non-zero by definition) primitive
correlators, 
\beq
P(0) \quad \mathrm{and} \quad P(2).
\eeq
Since there are no correlations between different
points in this case, 
\beq
S(\phi) \, = \, \frac{k}{2} \phi^2 \, + \, \frac{\lambda}{4} \phi^4,
\eeq
there is no a $w$ parameter with respect to which we can 
differentiate.
We can however evolve with respect of one of the two
parameters entering $S$:
\begin{enumerate}
\item
$k$-{\it Evolution.} The system of evolution equations reads:
\bea
\frac{\partial P(0)}{\partial k} &=& - \, \frac{1}{2} \, P(2);
\nonumber\\
\frac{\partial P(2)}{\partial k} &=& 
+ \, \frac{k}{2\lambda} \, P(2) \, - \, \frac{1}{2\lambda} \, P(0). 
\eea
Initial conditions can be assigned at $k=0$ ($\lambda >0$).
The following remarks are in order.
\begin{enumerate}
\item
The number of primitive correlators is in agreement
with the general formula:
\beq
\frac{3^N+1}{2} \, \to \, 2 \quad \mathrm{for} \quad N \, \to \, 1 ;
\eeq
%%%%%
\item
As discussed in the general classification:

\noindent
$I^0$ equation above is a one-term equation, whose
derivation did not use any LDS equation;

\noindent
$II^0$ equation required a single use of the LDS equation,
as it involves coefficients $\propto 1/\lambda^n$ with $n=1$;

%%%%%%
\end{enumerate}
%%%%%
\item
$\lambda$-{\it Evolution.}
The system of (two) differential equations in $\lambda$ 
explicitly reads:
\bea
\frac{\partial P(0)}{\partial \lambda} &=& 
\frac{k}{4 \lambda} \, P(2)
\, - \, \frac{1}{4 \lambda} \, P(0) ;
\nonumber\\
\frac{\partial P(2)}{\partial \lambda} &=& 
- \frac{1}{4\lambda}
\left( \frac{k^2}{\lambda} + 3 \right) P(2)
\, + \, \frac{k}{4\lambda^2} P(0) . 
\eea
The basic observation here is that 
the coefficients on the rhs's
are proportional to $1/\lambda^n$, like in 
the previous case, but now we are evolving with 
respect to $\lambda$ itself.
The point $\lambda=0$ is therefore singular
and we cannot assign initial conditions
there.
That can be seen more explicitly by deriving
a second-order equation for $P(0)$ from the
above system:
\beq
\frac{\partial^2 P(0)}{\partial \lambda^2} 
\, + \, \left( \frac{k^2}{4 \lambda^2} + \frac{2}{\lambda} \right)
\frac{\partial P(0)}{\partial \lambda} 
\, + \, \frac{3}{16 \lambda^2} P(0) \, = \, 0.
\eeq
Since the first derivative of $P(0)$ has a coefficient
containing a double pole in $\lambda=0$ for $k \ne 0$
(i.e. in the massive case), the latter
is a {\it irregular singular point}%
\footnote{
A {\it regular singular point}
of a second-order ordinary differential equation in $\lambda$, 
say $\lambda=0$, is a singularity
in the equation such that the solution $f(\lambda)$
can be written in a neighborhood of the origin
as the product of a simple function (logarithm, real power, etc.), 
singular at $\lambda=0$,
times a convergent power series in $\lambda$.
For that to occur, the coefficient of 
the first derivative, $df/d\lambda$,
must contain at most a simple pole at $\lambda=0$,
while the second derivative
$d^2f/d\lambda^2$, must contain at most a double pole.

An {\it irregular singular point} is 
a singularity of the equation which is
not a regular singular point.
The singularity in this case is so strong 
that it is not anymore possible to write the 
solution in the factorized form described above.
}.
\end{enumerate}
%%%%%%%%%%%%%%%%%%%%%%%%%%%%%%%%%%%%%%%%%
\item
{\it Lattice with two points,}
\beq
N \,= \, 2.
\eeq
This is, as already observed,
the simplest non-trivial
case, because of the occurrence of
correlations at different points:

The primitive correlators 
\beq
P\left(\nu_1,\nu_2\right)
\eeq
have to satisfy the conditions
\beq
0 \le \nu_1, \, \nu_2 \, \le \, 2;
\qquad \nu_1 + \nu_2 = \mathrm{even};
\eeq
the last condition coming from the $\phi_i \to -\phi_i$ symmetry.
Explicitly:
\beq
\left\{ P(\nu) \right\}
\, = \, \left\{ P(0,0); \,\,\, P(2,0); \,\,\, P(1,1); \,\,\, P(0,2); \,\,\, P(2,2) \right\}.
\eeq
The $P$' obey the following system of ordinary linear
differential equations:
\bea
\frac{\partial P(0,0)}{\partial w} &=& P(1,1);
\nonumber\\
\frac{\partial P(1,1)}{\partial w} &=& P(2,2);
\nonumber\\
\frac{\partial P(2,0)}{\partial w} &=& 
- \frac{k_1}{\lambda_1} P(1,1) + \frac{w}{\lambda_1} P(0,2);
\nonumber\\
\frac{\partial P(0,2)}{\partial w} &=& 
- \frac{k_2}{\lambda_2} P(1,1) + \frac{w}{\lambda_2} P(2,0);
\\
\frac{\partial P(2,2)}{\partial w} &=& 
\frac{k_1 k_2 + w^2}{\lambda_1 \lambda_2} P(1,1) 
- \frac{k_1 w}{\lambda_1 \lambda_2} P(2,0)
- \frac{k_2 w}{\lambda_1 \lambda_2} P(0,2)
+ \frac{w}{\lambda_1 \lambda_2} P(0,0).
\nonumber
\eea
The following remarks are in order.
\begin{enumerate}
\item
The number of primitive correlators is in agreement
with the general formula
\beq
\frac{3^N+1}{2} \, \to \, 5 \quad \mathrm{for} \quad N \, \to \, 2 ;
\eeq
%%%%%
\item
As discussed in the general classification:
\begin{itemize}
\item
$I^0$ and $II^0$ eqs.
are one-terms equation, not
requiring any LDS equations;
\item
$III^0$ and $IV^0$ equations have required
a single use of the LDS equations;
\item
$V^0$ equation has required a double use of LDS;
\end{itemize}
%%%%%
\item
Exchanging $\nu_1$ with $\nu_2$ is equivalent to
exchange the couplings $k_1$ with $k_2$ and $\lambda_1$
with $\lambda_2$.
In the case of a symmetric Lagrangian,
\beq
k_1 \, = \, k_2 \, = \, k; \qquad \lambda_1 \, = \, \lambda_2 \, = \, \lambda;
\eeq
it holds
\beq
T(2,0) \, = \, T(0,2) ;
\eeq
so there are only four primitive correlators and the system above 
reduces to a fourth-order one.
\end{enumerate}
\end{enumerate}

%%%%%%%%%%%%%%%%%%%%%%%%%%%%%%%%%%%%%%%%%%

\subsection{General Comment}

The method discussed above allows in principle
the exact (numerical, but "deterministic") solution 
of an arbitrary local quantum field
theory at any finite lattice spacing $N < \infty$
summarized by the following steps:
\begin{enumerate}
\item
Reduction of any correlator $G(\nu)$ to a (finite)
linear combination of primitive correlators $P(\mu)$,
with known coefficients;
\item
Evaluation of all the primitive correlators
by solving partial differential equations in the
couplings, with initial conditions given by explicit
analytic formulas.
\end{enumerate}
However, due to the exponential increase with $N$ of the
order of the system to be solved, the method
is not practical at all: one has to stop in any case
at very small $N$.
In order to solve for example a $\lambda \phi^4$ theory on a 
lattice  of size $N=10^4$ --- a lattice size well below 
current Monte-Carlo simulation ones --- one should indeed
solve 
\beq
\mathcal{O} \left( 10^{4771} \right). 
\eeq
differential equations.
That is actually the path of thought that lead us to define
as unsolvable a theory with an exponential growth
with the lattice size $N$ in the number of its primitive 
correlators, giving rise in the weak limit $N \to \infty$
to a countably-infinite set of primitive correlators.

%%%%%%%%%%%%%%%%%%%%%%%%%%%%%%%%%%%%%%%%%%%%%%%%%%%%%%%%%%%%%

\section{Evaluation of Primitive Correlators for $N \to \infty$}
\label{sect14}

By definition, to solve a theory in the weak limit $N \to \infty$ 
means to know all its correlators,
\beq
G\left(\cdots,\nu_{-1},\nu_0,\nu_1,\nu_2,\cdots\right)
\, \equiv \,
\left\langle 
\cdots \, \phi_{-1}^{\nu_{-1}} \, \phi_{0}^{\nu_0}
\, \phi_{1}^{\nu_{1}} \, \phi_{2}^{\nu_{2}} \, \cdots
\right\rangle, 
\eeq
having a finite sum of all the occupation numbers $\nu_i$,
\beq
\mathcal{R} \, \equiv \, \sum_{i=-\infty}^{+\infty} \nu_i \, < \, \infty. 
\eeq
In the case of the anharmonic oscillator on an infinite lattice,
a primitive correlator is a correlator
with all its finitely-many non-zero indices $\nu_i$, $i \in\ZZ$, 
less than or equal to two:
\beq
P \, = \, P\left(\cdots, \nu_{-1}, \nu_0, \nu_1, \nu_2, \cdots \right)
\qquad \nu_i \, \le \, 2, \,\,\, i \, \in \, \ZZ.
\eeq
Note that the weight $\mathcal{R}(\nu)$ 
and the index $\tau(\nu)$ for a primitive correlator $P(\nu)$, 
though always finite, can become arbitrarily large.
This situation is to be compared with the finite-dimensional
case (finite $N$), where a primitive correlator
\beq
P\left(\nu_1,\cdots,\nu_N\right)
\eeq
has the  weight $\mathcal{R}$ bounded from above by $2N$, 
and the index $\tau$ bounded by $N$,
with $N$ the lattice size,
\bea
\mathcal{R}\left[P\left(\nu_1,\cdots,\nu_N\right)\right] &\le& 2N;
\nonumber\\
\tau\left[P\left(\nu_1,\cdots,\nu_N\right)\right] &\le& N.
\eea
Since the recursion weight $\mathcal{R}$ is finite, the reduction
to primitive correlators involves a finite number of steps.
We want to know "how many" primitive correlators
are there in the case of the anharmonic oscillator on an infinite lattice.
The set of all $P(\nu)$'s can be written as
the following disjoint union:
\beq
\left\{ P(\nu) \right\}
\, = \, \cup_{n=0}^\infty \left\{ P(\nu); \,\, \tau(\nu) \, = \, n \right\}.
\eeq
By a similar computation to the one made in the case of the
independent correlators $G(\nu)$, it can be shown
that the set of primitive correlators is
countable.

Our aim is to calculate the primitive correlators
for the action
\beq
S[\Phi] \, = \, \sum_{i=-\infty}^{+\infty}
\left(
\frac{\bar{k}}{2} \, \phi_i^2 \, + \, \frac{\bar{\lambda}}{4} \, \phi_i^4
\right)
\, - \, \bar{w} \sum_{i=-\infty}^{+\infty}  \phi_i \, \phi_{i+1} ,
\eeq
where we have put a bar over all the parameters
to indicate that we are interested to the theory 
with those specific values of the couplings.
By thinking to $w$ as a variable, let us write
the weight in the functional integral as
\beq
\label{S_sym_af}
\exp[-S(\Phi,w)] \, = \, 
\exp\left[
\, - \, \sum_{i=-\infty}^{+\infty}
\left(
\frac{\bar{k}}{2} \, \phi_i^2 \, - \, \frac{\bar{\lambda}}{4} \, \phi_i^4
\right)
\, + \, w \sum_{i=-\infty}^{+\infty}  \phi_i \, \phi_{i+1} 
\right],
\eeq
The first idea that comes to the mind is to write evolution 
equations for the primitive correlators $P(\nu; w)$
with respect to $w$:
\beq
\label{eqdiff_naive}
\frac{\partial P(\nu; w)}{\partial w} 
\, = \, 
\sum_{\|\mu \|_\infty\le 2} c_\nu(\mu; \, w) \, P(\mu; w),
\qquad \|\nu \|_\infty \, \le \, 2 .
\eeq
To solve the theory, we integrate all the above equations from $w=0$
(initial condition provided by the random field)
up to $w=\bar{w}$:
\beq
w: 0 \, \to \, \bar{w} \, \ne \, 0.
\eeq
The problem is that this way one brings down from the
exponent in eq.(\ref{S_sym_af}) an infinite number of terms, 
namely the series
\beq
\sum_{i=-\infty}^{+\infty}  \phi_i \, \phi_{i+1} .
\eeq
As a consequence,
the differential equation (\ref{eqdiff_naive})
contains an infinite number of terms
on its r.h.s., even before the reduction to
primitive correlators has been made:
\beq
\frac{\partial P(\nu; \, w)}{\partial w} 
\, = \, \sum_{i=-\infty}^{+\infty}
G\left(\cdots; 
\,\nu_{i-1}; \, 1 \, + \, \nu_i; \, 1 \, + \, \nu_{i+1}; 
\, \nu_{i+2}; \, \cdots ; \,\, w \right).
\eeq
Now, as well known, unless some notion of convergence 
is provided, an infinite sum such as the above one,
does not make any sense.  
Instead of introducing ad ad-hoc metric or, more generally,
topology in the vector space of the correlators,
it is more convenient to 
follow a "localization" strategy: we replace 
the single ordinary differential equation (\ref{eqdiff_naive})
with an {\it infinite} set of partial differential equations,
each one containing a {\it finite} number of terms.
To this aim, let us consider the following
generalized action, depending on the infinite 
set of variables $w_i$, $i \in \ZZ$:
\beq
S\left[ \Phi; \, \bf{w} \right]
\, = \, \sum_{i=-\infty}^{+\infty}
\left(
\frac{\bar{k}}{2} \, \phi_i^2 \, + \, \frac{\bar{\lambda}}{4} \, \phi_i^4
\right)
\, - \, \sum_{i=-\infty}^{+\infty} w_i \, \phi_i \, \phi_{i+1} ,
\eeq
where we have defined the doubly-infinite vector
\beq
{\bf w} \, \equiv \, 
\left( \cdots,w_{-2},w_{-1},w_0,w_1,w_2,\cdots,w_{n-1},w_n,w_{n+1},\cdots \right).
\eeq
The dots "$\cdots$", both at the beginning and at the end
of the string of the $w_i$'s, indicate that there is neither
a first component nor a last component in ${\bf w}$.
Also the primitive correlators, computed with the above generalized 
action, depend on all the $w_i$'s:
\beq
P \, = \, P\left( \nu; \, \bf{w} \right).
\eeq
According to a (formal) infinite-dimensional generalization
of the chain rule,
\beq
\frac{\partial P(\nu; w)}{\partial w}
\, = \, \sum_{i=-\infty}^{+\infty} 
\left.\frac{\partial P(\nu; \, \bf{w})}{\partial w_i}\right|_{w_j\to w; \, j\in\ZZ}.
\eeq
In order to write evolution equations with 
a finite number of terms on the r.h.s.,
we differentiate $\partial P\left( \nu; {\bf w} \right)$
with respect to any given $w_i$:
\beq
\frac{\partial P\left( \nu; \, {\bf w} \right)}{\partial w_i} 
\, = \, 
\sum_{\|\mu \|_\infty\le 2} c_\nu^{(i)}(\mu; \, {\bf w}) \, 
P\left(\mu; {\bf w} \right),
\qquad\qquad
i \, \in \, \ZZ, \qquad \|\nu\|_\infty \, \le \, 2.
\eeq
It holds indeed:
\beq
\frac{\partial P( \nu; \, {\bf w} )}{\partial w_i} 
\, = \, 
G\left(\cdots \, ; \, \nu_{i-1} \, ; \, 1 \, + \, \nu_i \, ; 
\, 1 \, + \, \nu_{i+1} \, ; \,
\nu_{i+2} \, ; \, \cdots \, ; \, {\bf w} \right),
\qquad i \, \in \, \ZZ,
\eeq
and, as we have shown above, the reduction of the correlator 
on the r.h.s always involves a finite number of steps,
so that a finite number of primitive correlators appears in the
final, i.e. complete, decomposition.
We then evolve each primitive correlator $P\left( \nu; {\bf w} \right)$
with respect to any variables $w_i$ and at the
end we set $w_i=\bar{w}$, for all $i \in\ZZ$.
Note that, for each primitive correlator $P( \nu; \, {\bf w} )$, 
we have a countable number of commuting flows,
one for each $w_i$ variable, $i \in \ZZ$.

%%%%%%%%%%%%%%%%%%%%%%%%%%%%%%%%%%%%%%%%%%

Let us end this section with a few comments.
\begin{enumerate}
\item
Since the generation of an evolution equation
for a primitive correlator
in the continuum limit may involve
an arbitrarily large number of reduction steps,
arbitrarily big powers of $\lambda$
do appear in the coefficients
\beq
\frac{1}{\lambda^n} \qquad n \,\, \mathrm{arbitrarily \,\, large}.
\eeq
These terms are related to the order-three branch point 
of the exact interacting theory in the free point 
$\lambda = 0$ \cite{ZinnJustin};
%%%%%
\item
For $\lambda_i \gg 1$ 
one can make truncate the series, obtaining
strong coupling expansions \cite{Marchesini,MartinelliParisi,Sharpe}.
One can also expand in powers of the $w_{i,i,+1}\ll 1$.
\end{enumerate}

%%%%%%%%%%%%%%%%%%%%%%%%%%%%%%%%%%%%%%%%%%%%%%%%%%%%%%

\section{Lattice Scalar Theory in $d=2$}
\label{sect15}

Let us briefly discuss the generalization
of the above theory to a "true" quantum field,
in space dimension $d_S = 1$, i.e.
in space-time dimension $d = d_S + 1 = 2$.
In the continuum, the theory is formulated
in a (flat) torus $T^2$, the direct product of two circles:
\beq
T^2 \, \equiv \, S^1 \, \times \, S^1.
\eeq
The symmetry group $G$ contains the direct product of the
symmetry groups of the space factors, namely $O(2) \times O(2)$.
To understand whether the inclusion is proper or not
is beyond the aim of this paper.
%
%%%%%%%%%%%%%%%%%%% FIGURA %%%%%%%%%%%%%%%%%%%%
\begin{figure}[ht]
\begin{center}
\includegraphics[width=0.5\textwidth]{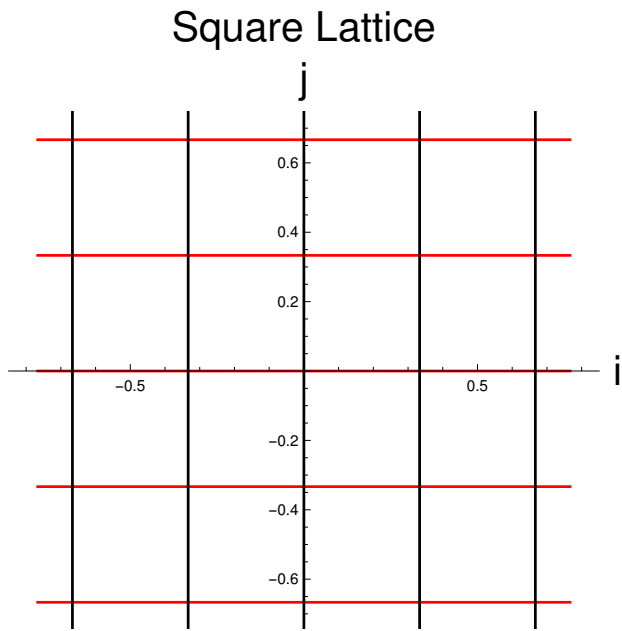}
\footnotesize
\caption{
\label{fig_platquad}
\it Square lattice $\Lambda^2$ in space-time dimension $d=2$.
Since $\Lambda^2$ is immersed in a two-dimensional (flat) torus
$T^2 \equiv S^1 \times S^1$, the points on the left-most vertical line
directly interact with the points in the right-most vertical line
at the same height.
Similarly for the points on the boundary horizontal lines.
}
\end{center}
\end{figure}
%%%%%%%%%% FINE FIGURA %%%%%%%%%%%%%%%%%%%%%%%%
%
We regularize the theory by means of a two-dimensional square lattice
$\Lambda^2$ immersed in $T^2$ (see fig.\ref{fig_platquad}). 
The symmetry group is a two-dimensional
generalization of the dihedral group $D_N$ found in $d=1$,
having as subgroup the direct product of the
symmetry groups of the space factors, $D_N \times D_N$.
The correlators to evaluate in this case read
\beq
G(\nu) \, = \, \int_{\RR^{N^2}} D \Phi \, \Phi^\nu \, e^{-S[\Phi]},
\eeq
where:
\begin{enumerate}
\item
The generalized euclidean action is defined as
\beq
S[\Phi] 
\, \equiv \,
\sum_{(i,j) \in I}
\left(
\frac{k_{i; \, j}}{2} \, \phi_{i; \, j}^2 
\, - \, w_{i + ; \, j} \, \phi_{i; \, j} \, \phi_{i+1; \, j} 
\, - \, w_{i ; \, j + } \, \phi_{i; \, j} \, \phi_{i; \, j+1} 
\, + \, \frac{\lambda_{i; \, j}}{4} \, \phi_{i; \, j}^4
\right);
\eeq
%%%%%
\item
The multi-index of two-subscript indices $\nu_{i;j}$ 
can be written as
\bea
\nu &\equiv&
\left(
\nu_{[-N/2]+1; \, [-N/2]+1} ; \,\,\, \nu_{[-N/2]+1; \, [-N/2]+2}; \,\, \cdots \,\,
\right.
\nonumber\\
&&\left. \,\, \cdots \,\, ;
\,\, \nu_{0; \, 0} \,\, ; \,\, \nu_{0; \, 1} \,\, ; \,\, \nu_{0; \, 2} \,\,; \,\, \cdots \,\, ;
\,\, \nu_{1; \, 0} \,\, ; \,\, \nu_{1; \, 1} \,\, ; \,\, \cdots \,\, ;
\,\, \nu_{[N/2]; \, [N/2]}   
\right) ;
\eea
%%%%
\item
The integration measure is the standard Lebesgue product measure in $\RR$,
\beq
\mathrm{D} \Phi 
\, \equiv \, 
\prod_{(i,j) \in I} d\phi_{i; \, j} .
\eeq
\end{enumerate}
We have defined the index set
\beq
I \, \equiv \, \ZZ_N \times \ZZ_N .
\eeq

Let us make a general remark.
In space-time dimension $d=1$, 
if we assume a single lattice spacing $a$,
there is only one kind of lattice.
It is not so in $d=2$.
We have assumed for simplicity's sake a square lattice, 
but one can also use a lattice composed
of equilater triangles or regular exagons.
%
%%%%%%%%%%%%%%%%%%% FIGURA %%%%%%%%%%%%%%%%%%%%
\begin{figure}[ht]
\begin{center}
\includegraphics[width=0.5\textwidth]{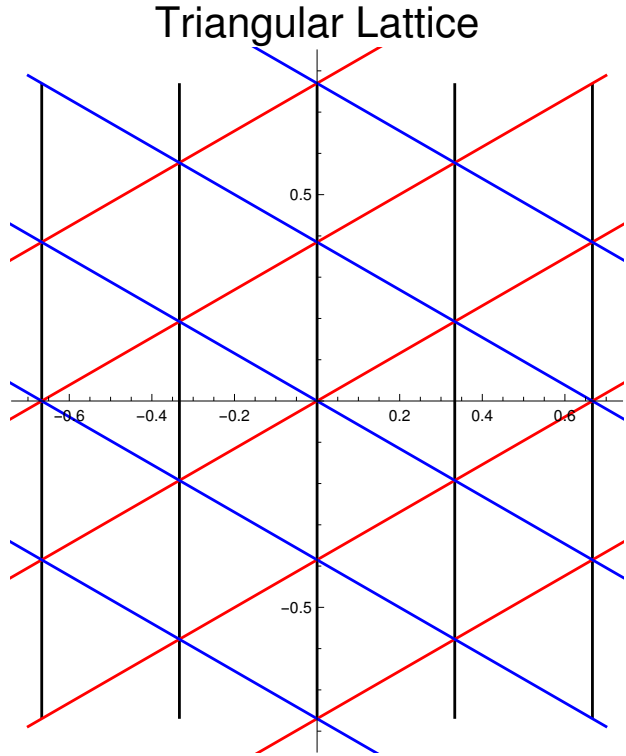}
\footnotesize
\caption{
\label{fig_platriang}
\it Plane lattice composed of equilater triangles.
Each point has six nearest neighbors.
By collecting together sets of six triangles
with one vertex in common, an hexagonal lattice
is formed. In the latter case, each point has
three nearest neighbors.
}
\end{center}
\end{figure}
%%%%%%%%%% FINE FIGURA %%%%%%%%%%%%%%%%%%%%%%%%
%
As we have seen, with a square lattice
each lattice point $p_i$ has
four nearest neighborhood,
while, on a triangular lattice, $p_i$ has
six nearest neighborhood and with an hexagonal
lattice $p_i$ has only three. 
The corresponding LDS equations are
therefore of different form for each kind 
of lattice.
Furthermore, the symmetry groups of the
triangular and hexagonal lattices
are different from the symmetry group 
of the square lattice, as well as from
each other.
In principle different conclusions can be 
reached, depending on the lattice
regularization chosen.
It is beyond the scope of this paper
to analyze different lattice regularizations
and compare the results.
On physical ground, however, we believe that
the results could differ at finite $N$,
but should become independent on the
chosen regularization when $N \to \infty$.
That is  because $a \ll 1$ for $N \gg 1$,
so one looks at fields $\phi_i, \phi_j,\phi_k,\cdots$ 
defined on lattice points $i,j,k,\cdots$ 
always at distances much greater than the
lattice spacing,
\beq
d_{i,j}, \,\, d_{i, k}, \,\, d_{j,k}, \cdots \, \gg \, a.
\eeq
As a consequence, regularization effects
should be small for $N \gg 1$ and 
vanish exactly in the limit $N \to \infty$.

%%%%%%%%%%%%%%%%%%%%%%%%%%%%%%%%%%%%%%

\subsection{Solution of LDS Equations}

By writing only the shifted indices, the
symbolic solutions of the LDS equations read:
\bea
G\left( \nu_{i;j} \, \ge \, 3 \right)
&\to&
\frac{\nu_{i;j} - 3}{\lambda_{i;j}} 
G\left( - \, 4 \, + \, \nu_{i;j} \right) \, +
\nonumber\\
&-& \frac{k_{ij}}{\lambda_{i;j}} 
G\left( - \, 2 \, + \, \nu_{i;j} \right) \, +
\nonumber\\
&+& \frac{w_{i-;j}}{\lambda_{i;j}} 
G\left( 1 \, + \, \nu_{i-1;j}; \,\, - \, 3 \, + \, \nu_{i;j} \right) \, +
\nonumber\\
&+& \frac{w_{i+;j}}{\lambda_{i;j}} 
G\left( 1 \, + \, \nu_{i+1;j}; \,\, - \, 3 \, + \, \nu_{i;j} \right) \, +
\nonumber\\
&+& \frac{w_{i;j-}}{\lambda_{i;j}} 
G\left( 1 \, + \, \nu_{i;j-1}; \,\, - \, 3 \, + \, \nu_{i;j} \right) \, +
\nonumber\\
&+& \frac{w_{i;j+}}{\lambda_{i;j}} 
G\left( 1 \, + \, \nu_{i;j+1}; \,\, - \, 3 \, + \, \nu_{i;j} \right) .
\eea
By solving the above system as in the one-dimensional case,
one obtains a basis of $3^{N^2}$ primitive correlators
on a lattice of $N\times N$ points.
Discrete symmetries of the two-dimensional lattice
are expected to produce an $N^2$ power-suppression,
so that the total number of primitive correlates is
\beq
\mathrm{Card}\left[ \left\{P(\nu) \right\}_{N^2} \right] 
\, \gsim \, \frac{3^{N^2}}{N^2} .
\eeq
As in the case of the an harmonic oscillator, 
the basis of primitive correlates has the cardinality 
of the integers in the weak limit $N \to \infty$.
   
%%%%%%%%%%%%%%%%%%%%%%%%%%%%%%%%%%%%%%%%%%%%%%%%%%%%%%%%%%%

\subsection{Evolution Equations for Primitive Correlates}

One has two families of two-indices
couplings to evolve on:
\bea
\frac{\partial P(\nu)}{\partial w_{i+; \, j}}
&=& \sum_{ 0 \le \mu_{k;l} \le 2; \, \mathrm{all} \, k,l} \, c_\nu^{(i;\, j)}(\mu) \, P(\mu) ;
\nonumber\\
\frac{\partial P(\nu)}{\partial w_{i; \, j+}}
&=& \sum_{ 0 \le \mu_{k;l} \le 2; \, \mathrm{all} \, k,l} \, d_\nu^{(i;\, j)}(\mu) \, P(\mu) ;
\qquad (i,j) \, \in \, \ZZ_N \times \ZZ_N.
\eea
Roughly speaking, with the first set of equations
one evolves "horizontally", while the second
set one evolves "vertically".
By evolving for example first with respect to the
$w_{i+; \, j}$'s and later with respect to the $w_{i; \, j+}$'s,
one is building up the torus drawing one parallel at a time.

%%%%%%%%%%%%%%%%%%%%%%%%%%%%%%%%%%%%%%%%%%%%

\subsection{Continuum Limit $N \to \infty$}

The continuum limit $N \to \infty$ is taken 
in weak sense, exactly as in the case of the 
an harmonic oscillator, i.e. $d=1$.

%%%%%%%%%%%%%%%%%%%%%%%%%%%%%%%%%%%%%%%%%%%%%%%%%%%%%%%%%%%%%%%%

\section{Lattice Scalar Theory in $d>2$}
\label{sect16}

The generalization to $d>2$ space-time dimensions 
is trivial, as it is actually rather trivial already
the extension from $d=1$ to $d=2$ sketched in
the previous section.
In general, on a (hyper-)cubic lattice in $d$ space-time dimensions, 
each point has $2d$ nearest neighbors, with
corresponding terms in the action
\bea
&& w_{i_1 +; \, i_2; \, \cdots; \, i_d} \,\,
\phi_{i_1; \, i_2; \, \cdots; \, i_d} \,\,
\phi_{1 \, + \, i_1; \, i_2; \, \cdots; \, i_d}
\nonumber\\
&+& w_{i_1 ; \, i_2 + ; \, \cdots; \, i_d} \,\,
\phi_{i_1; \, i_2; \, \cdots; \, i_d} \,\, 
\phi_{ i_1; \, 1 \, + \, i_2; \, \cdots; \, i_d}
\nonumber\\
&+& \,\, \cdots \cdots \cdots \cdots \cdots \cdots \cdots \cdots \cdots \cdots
\nonumber\\
&+& w_{i_1 ; \, i_2 ; \, \cdots; \, i_d + } \,\,
\phi_{i_1; \, i_2; \, \cdots; \, i_d} \,\, 
\phi_{ i_1; \,  i_2; \, \cdots; \, 1 \, + \, i_d}.
\eea
The LDS equations are still solved with respect to
the unique higher-weight term representing the
quartic interaction.

%%%%%%%%%%%%%%%%%%%%%%%%%%%%%%%%%%%%%%%%%%%%%%%%%%%%%%%%%%%%%%%%

\section{Further Generalizations}
\label{sect17}

The results derived in the previous sections
for the $\lambda \, \phi^4$ theory
can be generalized to arbitrary interacting bosonic 
theories, i.e. to theories involving interacting particles with
general integral spin, such as photons, gluons, intermediate
vector bosons, etc.
Admittedly, in the case of gauge theories, because of the geometrical 
structure, the technical implementation can be rather laborious:
one has to regularize the theory on a lattice by means
of links and plaquettes, write the corresponding Dyson-Schwinger
equations \cite{Migdal} and so on.
However, the general idea is that the arguments given
above apply to any bosonic degree of freedom, i.e. to any
physical polarization state.

%%%%%%%%%%%%%%%%%%%%%%%%%%%%%%%%%%%%%%%%%%%%%%%%%%%%%%%%%

\section{Conclusions}
\label{sect18}

We have considered euclidean $\lambda \, \phi^4$ scalar field theories
on lattices immersed in tori of different
space-time dimensions $T^d$,  with $d=1,2,3,\cdots$.
By writing the corresponding Dyson-Schwinger equations ---
called Lattice Dyson-Schwinger (LDS) equations for brevity ---
we have found that they close exponentially with the
lattice size $N$.
All the correlators $G(\nu)$ of the theory can be explicitly expressed, 
in purely algebraic way, in terms of a basis of 
$\mathcal{O}\left(3^N\right)$ correlators, which we have called primitive correlators:
\beq
G(\nu) \, = \, \sum_{\| \mu \|_\infty \le 2} c_\nu(\mu) \, P(\mu),
\qquad \| \nu \|_\infty \, \le \, 2 .
\eeq
In the weak continuum limit implying,
\beq
N \, \to \, \infty ,
\eeq
a countable basis of primitive correlators $\left\{ P(\nu) \right\}$
is obtained:
\beq
\mathrm{Card}\left[ \left\{ P(\nu) \right\}_\infty \right]
\, = \,  \mathcal{N}_0.
\eeq
In general, a bosonic quantum field theory having an infinite number 
of primitive correlators, after all Dyson-Schwinger equations and
Ward identities have been used, is defined as {\it unsolvable}.

Any conceivable exact analytic calculation
of the set of primitive correlators $\left\{ P(\nu) \right\}_\infty$
involves a linear system of coupled partial differential equations 
on the $P(\nu)$'s with respect to a countable set of independent variables

Let us remark that our results are restricted
to theories involving interacting bosons.
We do not know yet whether our arguments can be extended
to purely fermionic theories, or they do not.
In the latter case, as well known, the occupation numbers 
$n_i$ of the fermionic fields $\psi_i$ are restricted to
be less than one,
\beq
n_i \, = \, 0, 1,  
\eeq
so that, roughly speaking, a much smaller set
of independent correlators enters the game. 
Furthermore,  unlike bosonic theories, fermionic
theories with finitely-many degrees of
freedom --- described in the functional-integral
formalism by finite-dimensional Grassman algebras 
\cite{Berezin} --- 
are purely algebraic theories.
However, as discussed earlier, cardinal numbers, which 
are at the root of our considerations, do not need to 
change under such circumstances. 

The implications of our study are the following.
If our arguments are correct, no interacting bosonic
quantum field theory --- including the anharmonic oscillator ---
will ever be exactly solved.
Taking exactly into account anharmonic interactions
prevents indeed from any exact solution, as the system becomes
so correlated that couplings among its parts explode, 
in the cardinality sense explained above.
Conversely, if one day somebody will exactly solve
the anharmonic oscillator, our arguments
will be falsified and hope will raise again to 
exactly solve interacting quantum field theories 
(in space-time dimension $d>1$).

We also conjecture that only those
bosonic field theories which can be exactly transformed 
to Gaussian ones (via regular change of variables 
in the functional integral)
or possess an exceptionally large symmetry, can be possibly solved.
The relation between the two above possibilities
may also be worth investigating.

Let us end by saying that the exact solution of any truly-interacting  
bosonic quantum field theory --- if it exists --- lies at a 
transcendental distance from any regularized approximant.

\vskip 0.5truecm

\centerline{\bf Acknowledgments}

\vskip 0.3truecm

\noindent
I wish to express particular thanks to Prof. M. Testa
for various discussions.
I also acknowledge discussions with Prof. G. Parisi.

%%%%%%%%%%%%%%%%%%%%%%%%%%%%%%%%%%%%%%%%%%%%%%%%%%%%%%%%%%%%%%%%%%


\begin{thebibliography}{99}

\bibitem{BerrySimon} 
Standard references are:
B. Simon, {\it Functional Integration and Quantum Physics ---
Second Edition,}
AMS Chelsea Publishing, Providence, Rhode Island (2005);
%
{\it The $P(\Phi)_2$ Euclidean (Quantum) Field Theory,}
Princeton University Press, Princeton, New Jersey (1974);
%
A. Glimm and A. Jaffe, {\it Quantum Physics --- A Functional
Integral Point of View,} Springer-Verlag New-York, Inc (1981);

%%%%%%%%%%%%%%%%%%%%%%%%%%%%%%%%%%%%%%%%%%%%%%%%

\bibitem{Preparata} G. Preparata, oral tradition;
 
%%%%%%%%%%%%%%%%%%%%%%%%%%%%%%%%%%%%%%%%%%%%%%%% 
 
\bibitem{RemiddiGehrmann} 
T.~Gehrmann and E.~Remiddi,
Nucl.\ Phys.\ B {\bf 601} (2001) 248;
Nucl.\ Phys.\ B {\bf 601} (2001) 287; 
 
%%%%%%%%%%%%%%%%%%%%%%%%%%%%%%%%%%%%%%%%%%%%%%%% 

\bibitem{Remiddi} 
E.~Remiddi and L.~Tancredi, Nucl.\ Phys.\ B {\bf 907}, 
400 (2016);

%%%%%%%%%%%%%%%%%%%%%%%%%%%%%%%%%%%%%%%%%%%%%%%%%

\bibitem{noi} 
U.~Aglietti, R.~Bonciani, L.~Grassi and E.~Remiddi,
Nucl.\ Phys.\ B {\bf 789} (2008) 45.
   
%%%%%%%%%%%%%%%%%%%%%%%%%%%%%%%%%%%%%%%%%%%%%%%%   

\bibitem{Montvay}
For an introduction to quantum field theory
on the lattice, see for example:
I. Montvay, G. Munster,
{\it Quantum Fields on a Lattice}, Cambridge
University Press, Cambridge (1994);
 
%%%%%%%%%%%%%%%%%%%%%%%%%%%%%%%%%%%%%%%%%%%%%%%%%%%
 
\bibitem{DSeq}
F. Dyson, Phys. Rev. 75: 1736 (1949); 
J. Schwinger, PNAS. 37: 452–459 (1951);

%%%%%%%%%%%%%%%%%%%%%%%%%%%%%%%%%%%%%%%%%%%%%%%%%%%

\bibitem{ItzZuber}
For an introduction of the Dyson-Schwinger
equations in the continuum, see for example:
Claude Itzykson and Jean-Bernard Zuber,
{\it Quantum Field Theory}, McGraw-Hill (1980); 

%%%%%%%%%%%%%%%%%%%%%%%%%%%%%%%%%%%%%%%%%%%%%%%%% 
 
\bibitem{BenderWu} C.~M.~Bender and T.~T.~Wu,
Phys.\ Rev.\ D {\bf 7} (1973) 1620;
Phys.\ Rev.\ D {\bf 8} (1973) 3346;

%%%%%%%%%%%%%%%%%%%%%%%%%%%%%%%%%%%%%%%%%%%%%%%%%

\bibitem{Serre} 
See for example: J.P. Serre, {\it Linear Representations
of Finite Groups}, Springer-Verlag New-York, Inc (1977);

%%%%%%%%%%%%%%%%%%%%%%%%%%%%%%%%%%%%%%%%%%%%%%%%%

\bibitem{Guralnik} 
G.~Guralnik and Z.~Guralnik,
Annals Phys.\  {\bf 325}, 2486 (2010)
doi:10.1016/j.aop.2010.06.001
[arXiv:0710.1256 [hep-th]].
 
%%%%%%%%%%%%%%%%%%%%%%%%%%%%%%%%%%%%%%%%%%%%%%%%%

\bibitem{ZinnJustin}
See for example: 
J. Zinn-Justin, {\it Quantum Field Theory and
Critical Phenomena}, Oxford University Press Inc, New York
(1996);

%%%%%%%%%%%%%%%%%%%%%%%%%%%%%%%%%%%%%%%%%%%%%%%%%%

\bibitem{Marchesini}
P.~Butera, R.~Cabassi, M.~Comi and G.~Marchesini,
Comput.\ Phys.\ Commun.\  {\bf 44} (1987) 143.

%%%%%%%%%%%%%%%%%%%%%%%%%%%%%%%%%%%%%%%%%%%%%%%%%

\bibitem{MartinelliParisi}
R.~Benzi, G.~Martinelli and G.~Parisi,
Nucl.\ Phys.\ B {\bf 135} (1978) 429.

%%%%%%%%%%%%%%%%%%%%%%%%%%%%%%%%%%%%%%%%%%%%%%%%%

\bibitem{Sharpe}
C.~M.~Bender, F.~Cooper, G.~S.~Guralnik and D.~H.~Sharp,
Phys.\ Rev.\ D {\bf 19} (1979) 1865.

%%%%%%%%%%%%%%%%%%%%%%%%%%%%%%%%%%%%%%%%%%%%%%%%%

\bibitem{Migdal}
Y.~M.~Makeenko and A.~A.~Migdal,
Phys.\ Lett.\  {\bf 88B} (1979) 135
Erratum: [Phys.\ Lett.\  {\bf 89B} (1980) 437];
Nucl.\ Phys.\ B {\bf 188} (1981) 269
[Sov.\ J.\ Nucl.\ Phys.\  {\bf 32} (1980) 431]
[Yad.\ Fiz.\  {\bf 32} (1980) 838];

%%%%%%%%%%%%%%%%%%%%%%%%%%%%%%%%%%%%%%%%%%%%%%%%%

\bibitem{Berezin}
Berezin, {\it The Method of Second Quantization}, 
Academic Press Inc. (1966).
    
\end{thebibliography}
\end{document}